\documentclass[final,3p,times]{elsarticle}

\usepackage{amssymb}
\usepackage{amsmath}

\usepackage{tikz}
\usepackage{pgfplots}
\usepackage{nicefrac}
\usepackage[ruled]{algorithm2e}
\usepackage{appendix}
\usepackage{longtable}
\usepackage{url}
\usepackage{graphicx}
\usepackage{hyperref}

\usetikzlibrary{positioning, calc, shapes.misc, external}
\usepgfplotslibrary{external}
\pgfplotsset{compat=1.18}
\usepackage{nicefrac}
\usepackage{xcolor}
\usepackage{bm}

\usepackage{caption}
\usepackage{subcaption}
\usepackage[normalem]{ulem}

\renewcommand{\vec}[1]{\bm{#1}}
\newcommand{\ten}[1]{\bm{#1}}

\journal{Computer Methods in Applied Mechanics and Engineering}

\begin{document}

\begin{frontmatter}



\title{A manifold learning approach to nonlinear model order reduction of quasi-static problems in solid mechanics}


\author[label1]{Lisa Scheunemann\corref{cor1}\fnref{contr1}} 
\ead{lisa.scheunemann@mv.rptu.de}
\cortext[cor1]{Corresponding Author}
\author[label1]{Erik Faust\fnref{contr1}} 
\ead{erik.faust@mv.rptu.de}
\fntext[contr1]{These authors contributed equally to this work.}

\affiliation[label1]{organization={Chair of Applied Mechanics, Department of Mechanical and Process Engineering, RPTU Kaiserslautern-Landau},
            addressline={Postfach 3049}, 
            city={Kaiserslautern},
            postcode={67653}, 
            state={Rhineland-Palatinate},
            country={Germany}}

\begin{abstract}
The proper orthogonal decomposition (POD) -- a popular projection-based model order reduction (MOR) method -- may require significant model dimensionalities to successfully capture a nonlinear solution manifold resulting from a {parameterised} quasi-static solid-mechanical problem. The local basis method by Amsallem et al.~\cite{AmsZahFar:2012:nmo} addresses this deficiency by introducing a locally, rather than globally, linear approximation of the solution manifold. However, this generally successful approach comes with some limitations, especially in the data-poor setting.
In this proof-of-concept investigation, we instead propose a {graph-based} manifold learning approach to nonlinear projection-based MOR which uses a global, continuously nonlinear approximation of the solution manifold. Approximations of local tangents to the solution manifold, which are necessary for a Galerkin scheme, are computed in the online phase. {As an example application for the resulting nonlinear MOR algorithms, we consider simple representative volume element computations. On this example, the manifold learning approach Pareto-dominates the POD and local basis method in terms of the error and runtime achieved using a range of model dimensionalities.}
\end{abstract}



\begin{keyword}
Model Order Reduction \sep Manifold Learning \sep Multiscale Solid Mechanics \sep Nonlinear Solid Mechanics \sep Quasi-Static Solid Mechanics


\end{keyword}

\end{frontmatter}



\section{Introduction}

In solid mechanics, a demand for repeated evaluations of a parameterised simulation model frequently arises, for example, in optimisation~\cite{GanZab:2004:dls,ChaAntBuf:2023:lrb,ChaAntBuf:2024:fpa}, uncertainty quantification, inverse problems~\cite{Bha:2017:rom,BolBul:2011:ect,GhaWil:2021:lpm,BulMai:2011:pod,GarMaiNov:2012:coe}, and multiscale problems~\cite{FriBoh:2013:rbh,DenSodApe:2022:rmm,GuoRokVer:2024:rom,BhaMat:2016:nmr}.  
Engineers and scientists can use parametric model order reduction (p-MOR) methods to considerably accelerate simulation workflows in such applications (see e.g.~\cite{GuoRokVer:2024:rom}).
Crucially, the reductions in simulation time achieved by parametric reduced order models (ROMs) in an online application phase must offset the cost incurred in training these models in the offline stage~\cite{BenGugWil:2015:spm,BhaMat:2020:ndr,GouAmsBor:2016:asl}. At the same time, these methods should sacrifice as little as possible in the way of accuracy. 
Furthermore, whether an engineer or scientist can profitably apply a p-MOR method to a given problem will depend on that method's generality and user-time intensiveness.

In this publication, we are primarily concerned with broadly applicable methods which already yield suitable speed-ups in online simulation time with comparatively little training data, and thus comparatively little computational offline cost. At the same time, we aim to retain relatively high levels of accuracy and to minimise the investment of human effort required for applications to new problems.
In particular, we propose an implementationally simple projection-based MOR framework which uses graph-based manifold learning techniques to obtain a general, globally nonlinear approximation of the solution manifold in which to search for solutions. 
As motivated in the following, such a method improves upon some shortcomings of existing projection-based MOR methods in handling nonlinearities in the solution manifold, particularly in the data-poor context.

To the best of our knowledge, this is the first application of graph-based manifold learning techniques to projection-based MOR in the field of quasi-static solid mechanics. The simple local linearisation methodology proposed in Section~\ref{s:loclin} and its combination with a two-stage MOR scheme as outlined in Section~\ref{s:two_stage} have, as far as we know, not previously been proposed in the context of MOR.
In this work, we consider primarily p-MOR applications to
multiscale problems, although the methods discussed here are intended to be applicable to {parameterised} quasi-static problems in nonlinear solid mechanics more generally.

The multiscale Finite Element Method (FEM) or `Finite Element squared' (FE\textsuperscript{2}) method has made concurrent multiscale simulations tractable in principle, using representative volume elements (RVEs) via which the microscale mechanics are considered~\cite{SmiBreMei:1998:pmb,MieSchSch:1999:cmm,Sch:2014:nth}. For many relevant problems, however, fully coupled high-fidelity multiscale simulations remain prohibitively expensive with currently available computational resources \cite{YvoHe:2007:rmm,MonYvoHe:2008:chn,FriHod:2016:fes,FriFerLar:2019:oan,BraDavMer:2019:rmh,GuoRokVer:2024:rom}; advanced domain decomposition and parallelisation strategies on state-of-the-art supercomputers are required to tackle problems with comparatively simple macroscopic geometries~\cite{KlaKohLan:2020:chm}. Fortunately, several avenues are open to researchers looking to accelerate such a quasi-static, solid mechanical multiscale problem; a comparative, high-level overview can for example be found in~\cite{RocKerVan:2020:msm}.

One approach to accelerating parameterised multiscale problems in particular and quasi-static solid mechanical problems more generally is to construct simplified physical models to replace the more complex original models. In the context of the FE\textsuperscript{2} method, statistically similar RVEs (SSRVEs), which replace a complex microstructure with a simplified geometry while conserving essential statistical characteristics (see e.g.~\cite{SchBalBra:2015:d3s}), and simplified truss models of special microstructures (see e.g.~\cite{DycHub:2023:dmm}) could, for example, be classed among such methods. Where applicable, simplified physical models yield an appealing compromise between accuracy, acceleration of online simulation times, and frugal use of data~\cite{SchBalBra:2015:d3s,DycHub:2023:dmm}. However, formulating a simplified physical model for one problem or a class of problems can be user-time intensive. Furthermore, simplified physical models might still require a solution procedure which leads to non-negligible computational cost in the online phase, meaning that further reduction is necessary.

Rather than formulating an alternative, simplified model, some p-MOR methods instead aim to accelerate the solution procedure for a given model based on data gathered in the offline phase, often by introducing an approximation space for the primary unknown variables.
There are several p-MOR methods which utilise physics- and structure-
exploiting approximation spaces to yield tailored reduced equations in the multiscale context, such as the nonuniform transformation field analysis (NTFA)~\cite{MicSuq:2003:ntf}, potential-based reduced basis model order reduction (pRBMOR)~\cite{FriLeu:2013:rbh}, mixed transformation field
analysis (MxTFA)~\cite{CovDeFri:2018:cro}, self-consistent clustering analysis (SCCA)~\cite{LiuBesLiu:2016:sca}, or the proper generalised decomposition (PGD)~\cite{ElGonChi:2013:fml,NirAlfGon:2013:mor}. 
As the performance of an algorithm often scales with the extent to which it exploits problem characteristics~\cite{KerHooNeu:2019:aas}, the balance between accuracy, low offline cost, and efficiency achieved by these methods is very attractive. 
On the other hand, specialising algorithms to problems or extending them to novel fields of application is user-time-intensive; this investment may not be recouped in every application.
The decision as to which p-MOR method may be used in a given context should thus generally be informed by the trade-offs between relevant performance metrics~\cite{KerHooNeu:2019:aas}: the optimal, currently realisable compromise between accuracy, computational efficiency in the online phase, data-hungriness in the offline phase, generality, and user-time intensiveness is application-dependent. 
Accordingly, further developments in more general p-MOR methods, such as those explored in this investigation, usefully complement specialised strategies.

An appealing approach to reduce a quasi-static solid-mechanical problem in a general way is to approximate the set of all possible solutions based on as little training data as possible, and to confine the search for solutions 
to this approximation.
For a well-posed parameterised problem, each possible parameter combination uniquely defines a solution and the solution is a continuous function of these parameters~\cite{DanCasAkk:2022:pca}. Thus, the set of all possible solutions defines a solution manifold: a connected set of solutions which locally resembles an Euclidean space~\cite{DanCasAkk:2020:mor}. The dimensionality of the solution manifold is determined by the parameterisation; in the case of  a 
quasi-static RVE problem with given geometry, this may involve the macroscopic strain and, potentially, history variables. Since there are generally far fewer parameters than unknown variables in such a problem, the space in which to search for solutions can potentially be reduced considerably. In addition to the robustness that comes with the retention of the underlying solution scheme~\cite{FriFerLar:2019:oan}, such a projection-based MOR framework generalises well to a range of physical behaviours, with moderate theoretical and implementational effort. 

The proper orthogonal decomposition (POD)~\cite{Pea:1901:llp,Wei:2019:tpo}, which uses a linear subspace to approximate a superset of the solution manifold, is currently the most popular projection-based MOR method for RVE applications~\cite{YvoHe:2007:rmm,MonYvoHe:2008:chn,RadBedSti:2016:dmm}. It can significantly reduce the cost incurred in solving linear systems of equations resulting from a Newton-Raphson scheme
, provided the solution manifold can be successfully captured in the linear subspace obtained by the POD.
In conjunction with the POD, hyper-reduction methods such as those proposed in~\cite{Ryc:2009:hrm,ChaSor:2010:nmr,CarFar:2011:lcg,NegManAms:2015:emr,JaiTis:2019:hnm} are used to also circumvent the need to assemble these equation systems fully, meaning that the RVE problem can be approached with an algorithmic complexity no longer tied to the number of the degrees of freedom of an FE discretisation~\cite{GouKerBor:2014:bac,HerOliHue:2014:hmr,RadRee:2016:pbm,BonManQua:2017:mdt,GhaTisSim:2017:pdm,SolBraZab:2017:nsd,ZahAveFar:2017:mpb,BraDavMer:2019:rmh,RasLloHue:2021:hpr,AgoArgBer:2022:prm,GuoRokVer:2024:rom,LanHutKie:2024:mhr}. 
A limitation of the linear approximation space provided by the POD, however, is that substantial model {dimensionalities} might be required to successfully {approximate} the solution manifold~\cite{AmsZahFar:2012:nmo}. 

To address this issue, Amsallem et al.~\cite{AmsZahFar:2012:nmo,AmsZahWas:2015:flr} suggested a strategy which uses a clustering step to construct multiple local POD approximation spaces for distinct regions in parameter or solution space, reducing the model sizes and thus the simulation times required to yield simulation results with a desired level of accuracy.
In~\cite{PenMoh:2016:nmr}, a similar approach using an adaptive weighting of snapshots was proposed. {In contrast,~\cite{DieMuiZlo:2021:ndr} select among local POD bases defined via neighbourhood relationships in a reduced space defined by a kernel principal components analysis.} Such local basis methods 
were applied to quasi-static solid mechanical problems in~\cite{DanCasAkk:2020:mor,DanCasAkk:2022:pca,DanCasAkk:2022:uqi,ChaAntBuf:2023:lrb,ChaAntBuf:2024:fpa} and a multiscale problem in~\cite{HeAveFar:2020:sar}. 

These methods generally yield encouraging improvements over comparable POD ROMs~\cite{DanCasAkk:2020:mor,DanCasAkk:2022:pca,DanCasAkk:2022:uqi,ChaAntBuf:2023:lrb,ChaAntBuf:2024:fpa}. They are not excessively data-hungry and, due to the locally linear approximation, a tangent, which is required for projection-based MOR, is available by construction~\cite{AmsZahFar:2012:nmo}.
However, limitations remain, especially in the data-poor context. Firstly, sufficient data is required in each cluster to construct an accurate POD ROM. In applications and regions with sparse data, sufficiently large clusters may yield locally linear approximations which do not parameterise the solution manifold very closely. Additionally, properly handling cluster transitions is challenging~\cite{AmsZahWas:2015:flr}, and cluster transitions may yield instabilities or inaccuracies in the solution scheme~\cite{IdeCar:1985:rmn}. 
Finally, the quality of the obtained ROM is highly contingent on clustering quality and assuring this quality is far from trivial, though there has been some promising research to this end~\cite{DanCasAkk:2020:mor,DanCasAkk:2022:pca,DanCasAkk:2022:uqi}.

Other researchers have since sought a more continuous, nonlinear approximation space in which to search for solutions. These include operator inference approaches using Ansatz manifold parametrisations via e.g., separable polynomials \cite{JaiTisRut:2017:qmm,JaiTis:2019:hnm,GeeWil:2022:lnr,GeeBalWri:2024:lpr}, neural network parameterisations such as autoencoders~\cite{KimChoWid:2022:fap,FreDedMan:2021:cdl,FreMan:2022:ped}, and manifold learning methods combined with an interpolation scheme~\cite{MilArr:2013:nml}. These advances in projection-based nonlinear MOR (NLMOR) have almost universally been applied to dynamical problems, with the evolution of a transient solution being constrained to the nonlinear approximation space. Often, this evolution is driven by constant or weakly nonlinear system matrices; this allows quantities appearing in the reduced equations to be pre-computed efficiently in the offline phase via operator inference techniques~\cite{GeeWil:2022:lnr,GeeBalWri:2024:lpr}. A formal framework for NLMOR using continuous, nonlinear approximation spaces has been proposed for the dynamic case in~\cite{BucGlaHaa:2024:mrm}.

In the quasi-static case, in contrast, it is the {search for} one static solution that is constrained to the approximation space. This search is often driven by strongly varying tangent matrices arising from a Newton-Raphson scheme. In this context, local basis methods have so far been the method of choice to consider nonlinearity~\cite{DanCasAkk:2020:mor,DanCasAkk:2022:pca,DanCasAkk:2022:uqi,ChaAntBuf:2023:lrb,ChaAntBuf:2024:fpa}. 
Alternative solution manifold parameterisations using artificial neural networks such as the autoencoders used in~\cite{FreMan:2022:ped,KimChoWid:2022:fap} might yield {a} closer, continuously nonlinear approximation of the solution manifold, but are {considerably more data-hungry and require more training effort}. Although a continuous tangent is available, its evaluation is costly~\cite{KimChoWid:2022:fap}. Ansatz parameterisations of the solution manifold~\cite{JaiTisRut:2017:qmm,JaiTis:2019:hnm,GeeWil:2022:lnr,GeeBalWri:2024:lpr}, on the contrary, are comparatively data-prudent and deliver low-cost tangents a priori, but are limited to approximating solution manifolds with specific structural features. Graph-based manifold learning methods such as that used by Millan et al.~\cite{MilArr:2013:nml} in the context of fluid dynamics or those used in~\cite{BhaMat:2016:nmr,BhaMat:2020:ndr,KimShi:2024:dmf} in constructing surrogate models for RVE problems, meanwhile, generalise {comparatively} well and work with little data, but do not yield a tangent a priori. Provided a suitable tangent can be constructed, a manifold learning-based p-MOR scheme might, however, yield a smoother, closer approximation of solution manifold for RVE problems than alternative methods. If global, nonlinear trends in these solution manifolds can be exploited efficiently this way, a manifold learning-based NLMOR scheme prove a viable alternative to local basis methods and their alternatives in accelerating RVE computations based on limited data.

In this work, we propose a proof-of-concept for such a method. Firstly, Section~\ref{s:FEM} recalls some essentials of FE modelling for quasi-static, parameterised solid-mechanical problems. Section~\ref{s:problem} then outlines the framework of projection-based MOR for quasi-static solid-mechanical problems while emphasising the role of the chosen approximation spaces.
Then, Section~\ref{s:state_of_art} recalls the state of the art, i.e. the POD and the local basis method. In Section~\ref{s:ManLMOR}, we outline our proof-of-concept for a manifold-learning-based nonlinear MOR scheme: Subsections~\ref{s:ManL},~\ref{s:LEM}, and~\ref{s:LLE} cover the reduction step performed via manifold learning techniques and Subsection~\ref{s:loclin} discusses a tangent computation scheme. 
While we have not implemented a hyper-reduction methodology for this proof-of-concept investigation, a two-stage MOR scheme which enables a scaling behaviour independent of the dimension of the original FE-RVE problem in principle is proposed in Section~\ref{s:two_stage}.
Finally, Section~\ref{s:examples} features a simple numerical example and Section~\ref{s:parameter_study} an ad-hoc parameter study. Pseudocode for relevant algorithms used in the scope of this work is provided in the appendix.

\section{Parameterised quasi-static solid mechanical problems}\label{s:FEM}

In this work, we investigate nonlinear projection-based MOR methods to be applied to quasi-static, parametrised solid mechanical problems discretised via the FEM. As an example application, we consider computations on an RVE in the context of computational homogenisation.
For this class of problems, the local balance of linear momentum at a point $\Vec{X}$ in the reference configuration can be written as 
\begin{equation}
    \text{Div} \ten{P}(\Vec{X}) - \Vec{b}(\Vec{X}) = \Vec{0}\,, \label{eq:momentumbalance}
\end{equation}
where $\ten{P}$ is the first Piola-Kirchhoff stress tensor, $\Vec{b}$ the bulk force per volume element in the reference configuration, and $\text{Div}$ the divergence operator~\cite[p.146]{Hol:2002:nsm}. Boundary conditions, e.g. in the displacement $\Vec{\underline{u}}(\Vec{X})$ and the nominal traction vector $\Vec{t}(\Vec{X})=\ten{P}(\Vec{X})\Vec{N}$
\begin{equation*}
    \Vec{\underline{u}}(\vec{X})=\Vec{\underline{u}}^*\,, \Vec{X} \in \partial \Omega^D\,, \quad \text{and} \quad \ten{P}(\Vec{X})\Vec{N} = \Vec{t}^*\,, \Vec{X}\in \partial \Omega^N
\end{equation*}
then determine a solution $\Vec{\underline{u}}(\Vec{X}), \Vec{X}\in \Omega$ on a domain $\Omega$
with boundary $\partial \Omega = \partial \Omega^D \cup \partial \Omega^N$~\cite[p.146]{Hol:2002:nsm}. Here, $\Vec{\underline{u}}^*$ and $\Vec{t}^*$ are boundary values in the displacement and nominal traction, respectively, and $\Vec{N}$ denotes the outward normal at $\Vec{X}\in \partial \Omega^N$. In the context of RVEs, periodic boundary conditions, which are described in Section~\ref{s:RVE}, further play a crucial role.

To model the material behaviour at a point $\Vec{X}$, a constitutive law describing the stress-strain relationship $\ten{P}(\ten{F})$, e.g. in terms of the deformation gradient $\ten{F}$, is required~\cite[p.40]{Wri:2001:nf}.
In this proof-of-concept investigation, we consider a simple compressible hyperelastic neo-Hooke material law, which postulates a reversible behaviour derived from a stored energy function 
\begin{equation}
    W = \frac{\mu}{2} ( I_c - 3 ) + \frac{\kappa}{4} ( J^2 - 1 - 2 \ln{J} )\,,\label{eq:nH}
\end{equation}
such that $\ten{P}=\dfrac{\partial W}{\partial \ten{F}}$~\cite[p.207]{Hol:2002:nsm}. Here, $I_c$ is the first invariant of the right Cauchy Green strain tensor, $J=\text{det}(\ten{F})$ is the volumetric strain, and $\kappa$ and $\mu$ are the bulk and shear moduli~\cite[p.247,254]{Hol:2002:nsm}.

Note that solutions $\Vec{\underline{u}}(\Vec{X})$ of Eq.~\eqref{eq:momentumbalance} on $\Omega$ are minimisers of a potential $\Pi$~\cite[p.157,206]{Hol:2002:nsm}, where
\begin{equation}
    \Pi = \int_\Omega W(\Vec{X}) dV - \int_\Omega \Vec{b}(\Vec{X}) dV - \int_{\partial \Omega^N} \Vec{t}(\Vec{X}) dA\,.
\end{equation}
Taking the first variation of this potential, which must vanish at a minimum of $\Pi$~\cite[p.382]{Hol:2002:nsm}, or multiplying Eq.~\eqref{eq:momentumbalance} by a kinematically admissible test function and integrating over $\Omega$ yields the weak form of the balance of linear momentum~\cite[p.82]{Wri:2001:nf}
\begin{equation}
    \int_{\Omega} \ten{P}(\vec{X}) : \delta \ten{F}(\vec{X}) dV  - \int_{\Omega} \Vec{b}(\vec{X}) \cdot \delta \Vec{\underline{u}}(\vec{X}) dV - \int_{\partial \Omega} \Vec{t}(\vec{X}) \cdot \delta \Vec{\underline{u}}(\vec{X}) dA = 0\,. \label{eq:weakform}
\end{equation}
Here, $\delta\Vec{\underline{u}}(\vec{X})$ and $\delta \ten{F}(\vec{X})$ denote the kinematically admissible variations in the displacement and the deformation gradient, respectively~\cite[p.82]{Wri:2001:nf}. 

The discretisation of Eq.~\eqref{eq:weakform} via a standard Galerkin FE formulation which uses the same Ansatz functions for $\Vec{\underline{u}}(\Vec{X)}$ and $\delta \vec{\underline{u}}(\vec{X})$ yields
\begin{equation}
    \delta \Vec{u}^T \Vec{g}(\Vec{u};\Vec{p}) = 0\,,\label{eq:variational}
\end{equation}
and, since Eq.~\eqref{eq:variational} must vanish for any kinematically admissible $\delta \vec{u}$, 
\begin{equation}
    \Vec{g}(\Vec{u};\Vec{p}) = \Vec{0}\,.\label{eq:eq_system}
\end{equation}
Here, $\Vec{u} \in \mathbb{R}^D$ denotes the nodal displacement vector via which the continuous displacement $\vec{\underline{u}}(\vec{X})$ is discretised, $\delta \Vec{u} \in \mathbb{R}^D$ the discrete counterpart of the variation $\delta \vec{\underline{u}}(\vec{X})$, $\vec{g}(\Vec{u};\Vec{p})\in \mathbb{R}^D$ the nodal residua, and $\Vec{p}\in\mathbb{R}^\delta$ the parameters which determine a solution~\cite[p.125]{Wri:2001:nf}. The parameter vector $\Vec{p}$ might for example contain values necessary to specify boundary conditions. This is the case in the example RVE computation outlined in Section~\ref{s:examples}, which is parameterised in terms of the macroscopic displacement gradient $\ten{\bar{H}} = \ten{\bar{F}} - \ten{I} \in\mathbb{R}^{3\times 3}$. 

The $D$-dimensional vector $\Vec{u} \in \mathbb{R}^D$ in Eq.~\eqref{eq:eq_system} contains the primary unknowns of the system. 
A Newton-Raphson scheme can be used as a solver for these unknown variables, in which case Eq.~\eqref{eq:eq_system} is linearised around the current iterate $\vec{u_\text{cur}}$ to yield an equation system for the displacement increment $\Delta \Vec{u}$~\cite[p.148]{Wri:2001:nf}
\begin{equation}
    \ten{K}(\vec{u_\text{cur}};\Vec{p}) \Delta {\Vec{u}} = - \Vec{g} (\vec{u_\text{cur}};\Vec{p})\,. \label{eq:newton}
\end{equation}
 The tangent appearing in the above defines a classical finite element stiffness matrix~$\ten{K} \in \mathbb{R}^{D\times D}$~\cite{SaeSteJav:2016:ach}
\begin{equation}
    \ten{K}(\vec{u_\text{cur}};\Vec{p}) = \frac{\partial \Vec{g}(\vec{u_\text{cur}};\Vec{p})}{\partial {\Vec{u}}}\,, \label{eq:K_glob}
\end{equation}
which is also assembled from nodal values on the element level~\cite[p.127]{Wri:2001:nf}. 

Dirichlet, Neumann, or periodic boundary conditions (BCs) necessitate modifications to the linear equation system in Eq.~\eqref{eq:newton}~\cite{SaeSteJav:2016:ach}. In the following, we will assume these changes to have been made, such that Eq.~\eqref{eq:newton} defines a linear system of $D$ equations with $D$ unknown independent variables. Following the solution of this linear system, the previous value of the discretised displacement vector may be updated with the increment $\Delta \vec{u}$ such that 
\begin{equation}
    \vec{u_\text{cur}} \gets \vec{u_\text{cur}} + \Delta \vec{u}\,.\label{eq:newton_update}
\end{equation}
Pseudocode for an example implementation of a full FE solution scheme (for an RVE problem with periodic boundary conditions) is provided in Alg.~\ref{alg:FEM}.

\section{Nonlinear projection-based model order reduction for quasi-static solid-mechanical problems}\label{s:problem}

The number of unknowns $D$ of quasi-static solid-mechanical problems discretised via the FEM (e.g. in the case of RVE computations) is often significant: in Section~\ref{s:examples} we tackle a simple example with $D=17,151$ independent degrees of freedom; values of $D>1,000,000$ are readily encountered in real-world problems (see e.g.~\cite{FriHod:2016:fes,BalSchBra:2014:ctt,BraBalSch:2016:cmd}).
The computational cost incurred in computations like these can be considerable: in the best case, the assembly of the linear system in Eq.~\eqref{eq:newton} scales with $\mathcal{O}(D)$ and its solution with $\mathcal{O}(D^2)$~\cite{FriHaaRyc:2018:ach}. 
Applications like 
multiscale modelling often necessitate thousands of solutions to such 
problems. The resulting computational cost precludes treatment using full FE simulations with currently available computational resources.

{To achieve practicable overall simulation times {in such a multi-query setting}, the computational cost of {individual simulations} must be reduced. This can be realised if, rather than the costly original problem, a reduced problem with $d\ll D$ unknown degrees of freedom is considered.} As motivated in the introduction, parametric projection-based MOR offers an attractive framework for performing such a reduction. {Typically, projection-based MOR schemes define a reduced space $\mathbb{R}^d$ populated by reduced vectors $\vec{y}\in\mathbb{R}^d$, with the aim of searching for solutions within this reduced space~\cite{YvoHe:2007:rmm,MonYvoHe:2008:chn,RadBedSti:2016:dmm}. A reconstruction map $R:\mathbb{R}^d\rightarrow\mathbb{R}^D$ is required to define reconstructed (approximate) solutions 
$\vec{\bar{u}} = R(\vec{y}) \in \mathbb{R}^D$ in the original solution space. In solid-mechanical applications of projection-based MOR, the reconstruction mapping $R$ is commonly chosen to be linear, i.e. $\vec{\bar{u}} = \ten{\psi}\vec{y}$~\cite{YvoHe:2007:rmm,MonYvoHe:2008:chn,RadBedSti:2016:dmm} or piecewise linear~\cite{DanCasAkk:2020:mor,DanCasAkk:2022:pca,DanCasAkk:2022:uqi,ChaAntBuf:2023:lrb,ChaAntBuf:2024:fpa}. The columns of the matrices $\ten{\psi}\in\mathbb{R}^{D\times d}$ defining such a linear map, which might e.g. be obtained via the POD~\cite{Wei:2019:tpo}, can then be interpreted as modes for the solutions $\vec{u}\in\mathbb{R}^D$, while the reduced vectors $\vec{y}\in\mathbb{R}^d$ contain the corresponding mode activity coefficients.}

\begin{figure}[h!]
\centering
\includegraphics[scale=0.5]{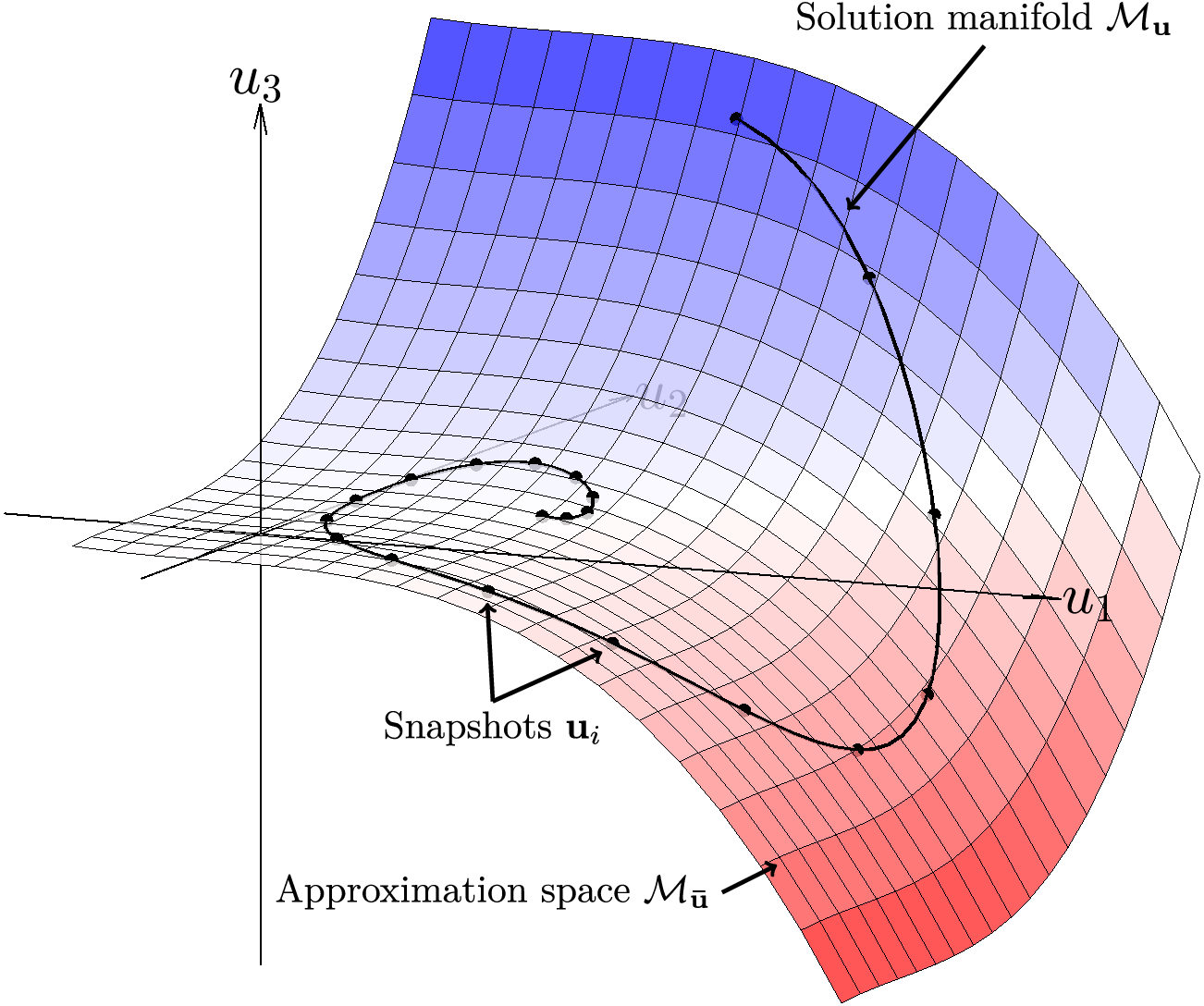}
\caption{Illustration of a $\delta=1$ dimensional solution manifold $\mathcal{M}_{\vec{u}}$ (black line) in a $D=3$ dimensional solution space, approximated via a $d=2$ dimensional approximation space $\mathcal{M}_{\vec{\bar{u}}}$ (surface with blue-red colour gradient). Snapshots are shown as black dots. Note that here, $\mathcal{M}_{\vec{\bar{u}}}\supset \mathcal{M}_{\vec{u}}$.}
\label{fig:approx_space}
\end{figure}

{Alternatively, MOR can be viewed from a manifold learning perspective.}
In the Introduction, we noted that the set of all possible solutions to a parameterised quasi-static solid mechanical problem defines a solution manifold $\mathcal{M}_{\vec{u}}=\{\Vec{u}\mid \exists \vec{p}: \vec{g}(\Vec{u};\Vec{p})=\Vec{0}\}$ with $\Vec{p}\in \mathbb{R}^\delta$, which is a connected set of solutions $\Vec{u}(\Vec{p})$ that locally resembles a Euclidean space~\cite{DanCasAkk:2020:mor,DanCasAkk:2022:pca}. For a well-posed parameterised problem, the solution manifold assumes the dimensionality $\delta$ of the parameter space $\mathbb{R}^\delta$, meaning that $\mathcal{M}_{\vec{u}}$ is much lower-dimensional than the solution space $\mathbb{R}^D$~\cite{DanCasAkk:2022:pca}. The visualisation in Fig.~\ref{fig:approx_space} illustrates a solution manifold with $\delta=1$ in a solution space with $D=3$. 
An appealing approach to reducing the original problem size $D$ then might be to approximate a superset of the solution manifold using an approximation space $\mathcal{M}_{\bar{\vec{u}}}$. Here, $\mathcal{M}_{\bar{\vec{u}}}$ is a $d$-dimensional manifold in the $D$-dimensional solution space, where $\delta<d\ll D$. An approximation space is also visualised in Fig~\ref{fig:approx_space} with $d=2$. {In the framework of projection-based model order reduction, $\mathcal{M}_{\bar{\vec{u}}}$ can be defined via the reconstruction map $R:\mathbb{R}^d\rightarrow\mathbb{R}^D$. The
solutions $\vec{\bar{u}}$ making up the approximation space can be defined as the images $\vec{\bar{u}} = R(\vec{y})$ of vectors in the reduced space $\vec{y}\in \mathbb{R}^d$, such that the $\vec{y}\in \mathbb{R}^d$ parameterise $\mathcal{M}_{\bar{\vec{u}}}$. Then, a ROM can seek for solutions in the approximation space $\mathcal{M}_{\bar{\vec{u}}}$, which is much lower dimensional than the solution space.} 
In the ideal case in which $\mathcal{M}_{\bar{\vec{u}}}$ is indeed a superset of the solution manifold, i.e. $\mathcal{M}_{\bar{\vec{u}}} \supset \mathcal{M}_{\vec{u}}$, all solutions in $\mathcal{M}_{\vec{u}}$ are also in $\mathcal{M}_{\bar{\vec{u}}}$. In this Section, we briefly sketch a framework for NLMOR along these lines. Note that, for the dynamic case, a formal framework for NLMOR using continuous Ansatz approximation spaces has been proposed in~\cite{BucGlaHaa:2024:mrm}.

{Practically, the reduced space and a reconstruction map $R$ can be constructed in a data-based manner.}
To this end, snapshot solutions $\Vec{u}_i\in \mathbb{R}^D, i \in 1,..,s$ to the full FEM problem in Eq.~\eqref{eq:eq_system} for $s$ parameter samples $\Vec{p}_i\in \mathbb{R}^\delta$ might be used, where $\Vec{g}(\Vec{u}_i;\Vec{p}_i)=\Vec{0}$~\cite{BenGugWil:2015:spm}.
{To construct the reduced space, a projection-based MOR scheme based on manifold learning techniques might then find an embedding $\vec{y}_i = M(\Vec{u}_i)$ for each $\Vec{u}_i$ with an embedding map $M:\mathbb{R}^D\rightarrow\mathbb{R}^d$, e.g. via a criterion that preserves {essential structural features}.}
The reconstruction $\Vec{\Bar{u}}_i=R(\Vec{y}_i)=M^{-1}(\Vec{y}_i)$ {might} also be found, e.g. by minimising the reconstruction error {$\|\vec{\bar{u}}_i-\vec{u}_i\|_2$ over all $i$}. {Note that this reconstruction error generally does not vanish, such that $\vec{\bar{u}} \neq \vec{u}$.} Specific schemes for finding suitable embeddings $\Vec{y}_i \in \mathbb{R}^d$ and approximation spaces $\mathcal{M}_{\bar{\vec{u}}}$ are discussed in Sections~\ref{s:POD},~\ref{s:LPOD},~\ref{s:LEM}, and~\ref{s:LLE}. Note, however, that many powerful dimensionality reduction techniques which might be used to this end only define the embedding {mapping} $M:\mathbb{R}^D\rightarrow\mathbb{R}^d$ and 
reconstruction mapping \mbox{$R:\mathbb{R}^d\rightarrow\mathbb{R}^D$} 
implicitly. {Instead, these techniques only provide an embedding $\Vec{y}_i \in \mathbb{R}^d$ corresponding to previously gathered solution data $\Vec{u}_i \in \mathbb{R}^D$, since it is highly nontrivial to find explicit mappings between points in $d$ and $D$-dimensional spaces $\Vec{y}_i\in \mathbb{R}^d$ and $\Vec{u}_i\in\mathbb{R}^D$ in a general manner~\cite{LeeVer:2007:ndr}.}

It will prove useful in the following to locally approximate $R$ via a linearisation, i.e. an approximation of a map to the tangent space $\mathcal{T}_{\Vec{u}} \mathcal{M}_{\bar{\vec{u}}}$ to $\mathcal{M}_{\bar{\vec{u}}}$. At a point with current reduced coordinates $\Vec{y}_\text{cur}$ and current displacement $\vec{u_\text{cur}}$, such a linearisation might be expressed as
\begin{equation}
    \Vec{\Bar{u}}-\vec{u_\text{cur}} \approx \ten{\varphi} (\Vec{y}-\Vec{y}_\text{cur})\,,\label{eq:lin}
\end{equation}
where $\Vec{\Bar{u}}\in \mathbb{R}^D$ denotes the reconstructed, discretised displacement corresponding to $\Vec{{y}}\in \mathbb{R}^d$.
The tangent $\ten{\varphi}$, which will turn out to be very useful in the following, is defined as
\begin{equation}
    \ten{\varphi} = \frac{\partial R(\Vec{y})}{\partial \Vec{y}}\Bigr|_{\Vec{y}_\text{cur}}\,.\label{eq:tangent}
\end{equation}
If $R$ is only given implicitly, $\ten{\varphi}$ can of course not be calculated directly and must be estimated. In the framework of a Galerkin scheme (see e.g.~\cite{PytAbe:2016:nmr}), the columns of $\ten{\varphi}$ can be interpreted as discrete local Ansatz functions for the increments $\Vec{\Bar{u}}-\vec{u_\text{cur}}$, while $\Vec{y}-\Vec{y}_\text{cur}$ are the corresponding coefficients.
Note that the familiar POD~\cite{Wei:2019:tpo}, which is outlined in Section~\ref{s:POD}, uses a $d$-dimensional linear subspace of $\mathbb{R}^D$ as an approximation space $\mathcal{M}_{\bar{\vec{u}}}$, such that $M:\mathbb{R}^D\rightarrow\mathbb{R}^d$ and $R:\mathbb{R}^d\rightarrow\mathbb{R}^D$ are globally linear by construction. 
Alternative methods for the construction of a linearisation as in Eq.~\eqref{eq:lin} will be discussed in Sections~\ref{s:LPOD} and~\ref{s:loclin}.

In the framework of a Galerkin scheme, a projection-based MOR method can then search for solutions in the approximation space $\mathcal{M}_{\bar{\vec{u}}}$, i.e. search for solutions to
\begin{equation*}
    \delta \vec{\bar{u}}^T \Vec{g}(\Vec{\bar{u}};\Vec{p})=0\,,
\end{equation*}
where $\Vec{\bar{u}} \in \mathcal{M}_{\bar{\vec{u}}}$ and $\delta \vec{\bar{u}}$ denotes the corresponding variation, which lies in the tangent space to the manifold at $\Vec{\bar{u}}$, i.e. $\delta \vec{\bar{u}} \in \mathcal{T}_{\Vec{u}} \mathcal{M}_{\bar{\vec{u}}}$. With $\mathcal{M}_{\bar{\vec{u}}}$ parameterised via $\Vec{y}\in\mathbb{R}^d$ {through the reconstruction $R$}, we can search for solutions in the much lower-dimensional reduced space $\mathbb{R}^d$, and approximate the variation as $\delta \vec{\bar{u}} = \ten{\varphi} \delta \vec{y}$ to yield
\begin{equation*}
    \delta \vec{y}^T \ten{\varphi}^T \Vec{g}(R(\Vec{y});\Vec{p})=0\,.
\end{equation*}
This must, again, vanish for any admissible $\delta \vec{y}$, which implies that the reduced residual
\begin{equation}
    \Vec{g}_r(\Vec{y};\Vec{p}) = \ten{\varphi}^T \Vec{g}(R(\Vec{y});\Vec{p})=\Vec{0}\,,\label{eq:redres}
\end{equation}
which is now also $d$- rather than $D$-dimensional, must vanish.

In analogy to Eq.~\eqref{eq:newton}, we can search for solutions to Eq.~\eqref{eq:redres} via a Newton scheme
\begin{equation}
    \ten{K}_r(\Vec{y}_\text{cur};\Vec{p}) \Delta \vec{y} = - \Vec{g}_r(\Vec{y}_\text{cur};\Vec{p})\,,\label{eq:newton_red}
\end{equation}
where $\ten{K}_r$ defines a reduced stiffness matrix
\begin{equation*}
    \ten{K}_r(\Vec{y};\Vec{p}) = \frac{\partial \Vec{g}_r(\Vec{y};\Vec{p})}{\partial \Vec{y}} = \frac{\partial (\ten{\varphi}^T \Vec{g}(R(\Vec{y});\Vec{p}))}{\partial \Vec{y}} = \ten{\varphi}^T \frac{\partial \Vec{g}(\Vec{\bar{u}};\Vec{p})}{\partial \Vec{\bar{u}}} \frac{\partial \Vec{\Bar{u}}}{\partial \Vec{y}}\,,
\end{equation*}
i.e.
\begin{equation}
    \ten{K}_r(\Vec{y};\Vec{p}) = \ten{\varphi}^T \ten{K}(R(\Vec{y});\Vec{p}) \ten{\varphi}\,. \label{eq:K_red}
\end{equation}
As in Eq.~\eqref{eq:newton_update}, the reduced variables $\vec{y}$ may be incremented following the solution of Eq.~\eqref{eq:newton_red} via
\begin{equation}
    \vec{y_\text{cur}} \gets \Vec{y_\text{cur}} + \Delta \vec{y}\,.\label{eq:newton_update_red}
\end{equation}

\section{State of the art projection-based model order reduction methods}\label{s:state_of_art}

\subsection{Proper Orthogonal Decomposition: robust linear MOR}\label{s:POD}

\begin{figure}[h!]
\centering
\begin{subfigure}[t]{.45\textwidth}
  \centering
  \includegraphics[scale=0.15,trim={15cm 5.5cm 15cm 5.5cm},clip]{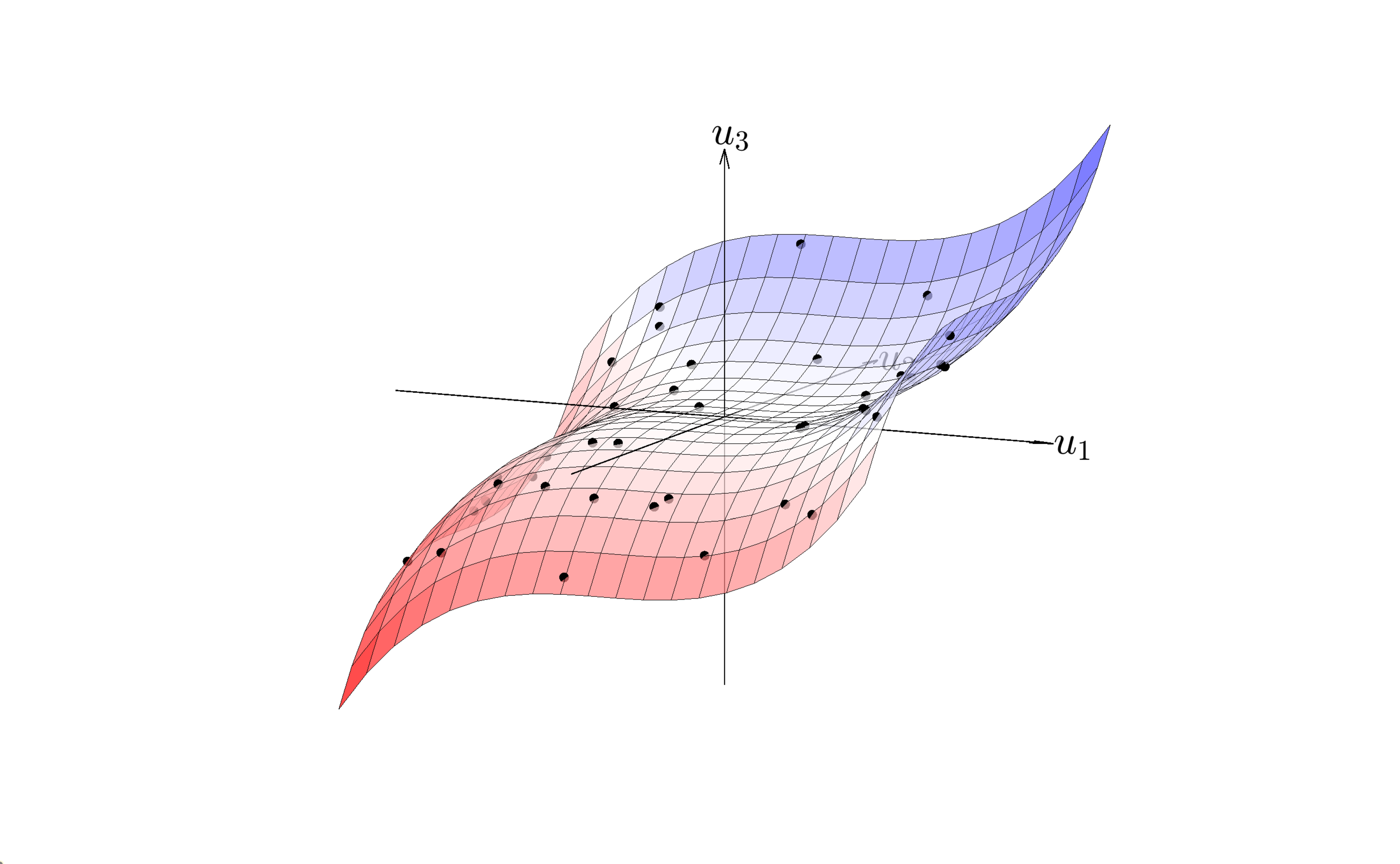}
  \caption{Snapshot vectors on solution manifold in high-dimensional solution space $\Vec{u} \in \mathbb{R}^D$.}
  \label{fig:POD1}
\end{subfigure}%
\hspace{0.5cm}
\begin{subfigure}[t]{.45\textwidth}
  \centering
  \includegraphics[scale=0.15,trim={15cm 5.5cm 15cm 5.5cm},clip]{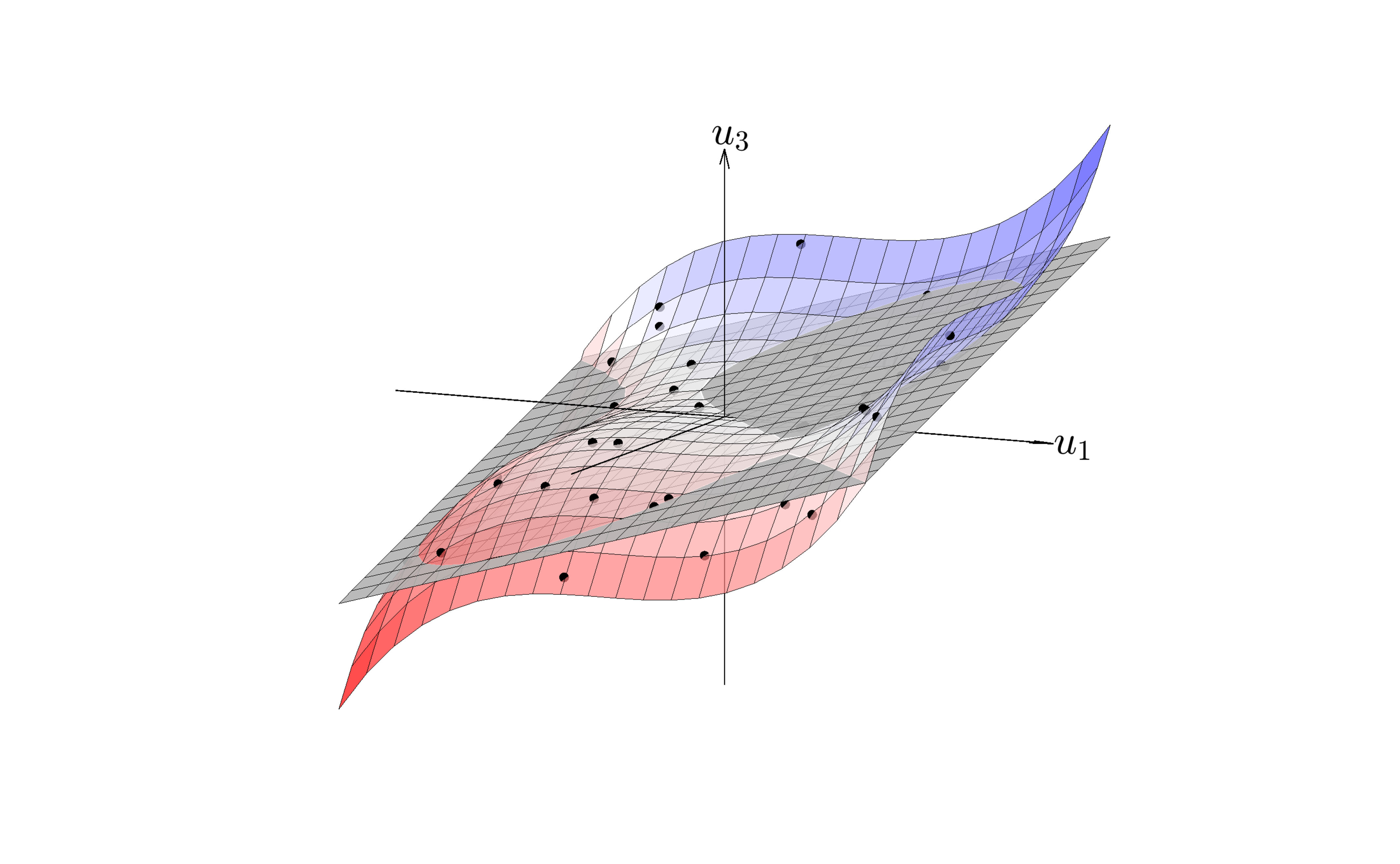}
  \caption{Linear approximation of solution manifold using the POD based on snapshot vectors $\Vec{U} \in \mathbb{R}^{D\times s}$. }
  \label{fig:POD2}
\end{subfigure}
\caption{Illustration of dimensionality reduction using the POD. Note that low-dimensional visualisations of manifold learning techniques obscure some of the nuances of dimensionality reduction.}
\label{fig:POD}
\end{figure}

The proper orthogonal decomposition (or POD; also known as principal components analysis, Karhunen-Loeve decomposition, or singular value decomposition) is a simple, yet robust linear projection-based MOR method~\cite{RadRee:2014:mre}. It works on the principle of projecting the snapshot data $\ten{U} = [ \Vec{u}_1,..,\Vec{u}_s] \in \mathbb{R}^{D\times s}$ from the original solution space $\mathbb{R}^D$ to a $d>\delta$-dimensional linear subspace $\mathbb{R}^d$ in which the captured variance of $s$ snapshots $\ten{U}$ is maximised. A low-dimensional visualisation of this approach is provided in Fig.~\ref{fig:POD}. Note that such visualisations can be somewhat misleading. In the case of the POD, the dimensionality of the reduced model $d$ is generally significantly larger than the intrinsic manifold dimension $\delta$. This allows for nonlinearities in the solution manifold to be captured to the extent that the $\delta$-dimensional solution manifold is captured in the $d$-dimensional linear subspace obtained by the POD. Note that both $\delta$ and appropriate model sizes $d$ can be estimated based on the snapshot data: in ~\ref{s:corr_dim}, we outline how the correlation dimension can be used to estimate $\delta$, while the Eigenvalue decay considered in Fig.~\ref{fig:EV_decay} in Section~\ref{s:RVE_prob} can be utilised to choose $d$.

The POD produces the mapping to and from the low-dimensional linear subspace of dimension $d$ which minimises the reconstruction error~\cite{Wei:2019:tpo}. The implementational approach outlined here, which is known as the snapshot POD, is sensible in case there are fewer snapshots than dimensions in the original solution space, i.e. $s<D$. 
The snapshot POD may be implemented as follows:
first, the covariance matrix $\ten{C}\in \mathbb{R}^{s\times s}$ of the snapshot data is computed~\cite{RadRee:2014:mre}
\begin{equation*}
    \ten{C} = \frac{1}{s-1} \ten{U}^T \ten{U}\,.
\end{equation*}
The linear embedding $M:\mathbb{R}^D\rightarrow\mathbb{R}^d$ from the solution space to the reduced space which maximises retained variance is then defined via the $d$ Eigenvectors of the Eigenvalue problem which correspond to the $d$ largest eigenvalues~\cite{RadRee:2014:mre}
\begin{equation}
    (\ten{C}-\lambda_i \ten{I}) \tilde{\Vec{v}}_i = \Vec{0}\,.\label{eq:POD_EV}
\end{equation}
Orthonormalisation of the modes $\tilde{\Vec{v}}_i$ with the snapshot matrix $\ten{U}$
\begin{equation*}
    \Vec{v}_i = \frac{\ten{{U}} \tilde{\Vec{v}}_i}{\|\ten{{U}} \tilde{\Vec{v}}_i\|_2}
\end{equation*}
yields the columns of the mapping matrix~\cite{RadRee:2014:mre}
\begin{equation*}
    \ten{{\psi}} = [\Vec{v}_1,..,\Vec{v}_d] \in \mathbb{R}^{D\times s}\,.
\end{equation*}
$\ten{\psi}$ defines a globally linear approximation space $\mathcal{M}_{\bar{\vec{u}}}$, as well as globally linear embedding $M:\mathbb{R}^D\rightarrow\mathbb{R}^d$ and reconstruction $R:\mathbb{R}^d\rightarrow\mathbb{R}^D$ maps, i.e.
\begin{equation}
    {\Vec{y}} = \ten{{\psi}}^T \Vec{u}\,, \quad \text{and}\quad \vec{\bar{u}} = \ten{{\psi}} \Vec{y}\,,\label{eq:POD_map}
\end{equation}
where $\ten{\psi}$ assumes the role of the tangent $\ten{\varphi}$ in Eq.~\eqref{eq:lin}. 
Algorithm~\ref{alg:POD_offline} provides pseudocode for the computational operations involved in the offline phase of the POD.

Once $\ten{\psi}$ has been identified, solutions can be sought in the reduced space $\mathbb{R}^d$ via a reduced Newton scheme as outlined in Eqs.~\eqref{eq:redres},~\eqref{eq:newton_red}, and~\eqref{eq:K_red}, i.e.~\cite{RadRee:2014:mre}
\begin{equation}
    \ten{K}_r \Delta \Vec{y} = - \Vec{g}_r\,,\quad \text{where} \quad \ten{K}_r = \ten{\psi}^T \ten{K} \ten{\psi}\,, \quad \text{and} \quad
    \Vec{g}_r = \ten{\psi}^T \Vec{g}\,.\label{eq:POD_red}
\end{equation}
After the $d$-dimensional linear system in Eq.~\eqref{eq:POD_red} has been solved in an iteration, the reduced displacement can be updated via Eq.~\eqref{eq:newton_update_red}~\cite{RadRee:2014:mre}.

Crucially, the solution of the linear equation system Eq.~\eqref{eq:POD_red} now scales with $\mathcal{O}(d^2)$ in the worst case, rather than $\mathcal{O}(D^2)$ as in Eq.~\eqref{eq:newton}~\cite{FriHaaRyc:2018:ach}. If $d\ll D$, this accelerates the Newton-Raphson step immensely. At the same time, the assembly of the stiffness matrix $\ten{K}$ and the residuum $\Vec{g}$ still scale with $\mathcal{O}(D)$, which proves a bottleneck for moderately sized problems not dominated by the solution of Eq.~\eqref{eq:newton} and problems featuring expensive operations within the element routine. A range of hyper-reduction techniques have been proposed to achieve improved scaling~\cite{Ryc:2009:hrm,ChaSor:2010:nmr,CarFar:2011:lcg,NegManAms:2015:emr,JaiTis:2019:hnm}. 
Due to the linearity of the approximation spaces obtained via the POD
, the reduced quantities $\ten{K}_r$ and $\Vec{g}_r$ can be estimated efficiently based on few entries of their unreduced counterparts $\ten{K}$ and $\Vec{g}$ (see e.g.~\cite{GouKerBor:2014:bac,HerOliHue:2014:hmr,RadRee:2016:pbm,BonManQua:2017:mdt,GhaTisSim:2017:pdm,SolBraZab:2017:nsd,ZahAveFar:2017:mpb,BraDavMer:2019:rmh,RasLloHue:2021:hpr,AgoArgBer:2022:prm,GuoRokVer:2024:rom,LanHutKie:2024:mhr}).
Algorithm~\ref{alg:POD_online} in \ref{s:code} provides pseudocode for the computational operations involved in the online phase of the POD for an RVE problem with periodic boundary conditions (omitting hyper-reduction). 

For some problems, the POD can reduce computation times on RVE problems considerably, especially in tandem with a hyper-reduction scheme. However, for a highly nonlinear solution manifold $\mathcal{M}_{\vec{u}}$ to be successfully captured in the linear approximation space $\mathcal{M}_{\bar{\vec{u}}}$, model dimensionalities significantly in excess of the manifold dimensionality may be required, i.e. $d_{\text{POD}}\gg \delta$. This limitation, which is known as the Kolmogorov barrier in fluid mechanics~\cite{DanCasAkk:2022:pca,Peh:2022:bkb}, limits the reduction in simulation time which may be achieved via the POD for highly nonlinear problems.



\subsection{Local bases: nonlinearity via locality}\label{s:LPOD}

\begin{figure}[h!]
\centering
\begin{subfigure}[t]{.45\textwidth}
  \centering
  \includegraphics[scale=0.15,trim={15cm 5.5cm 15cm 5.5cm},clip]{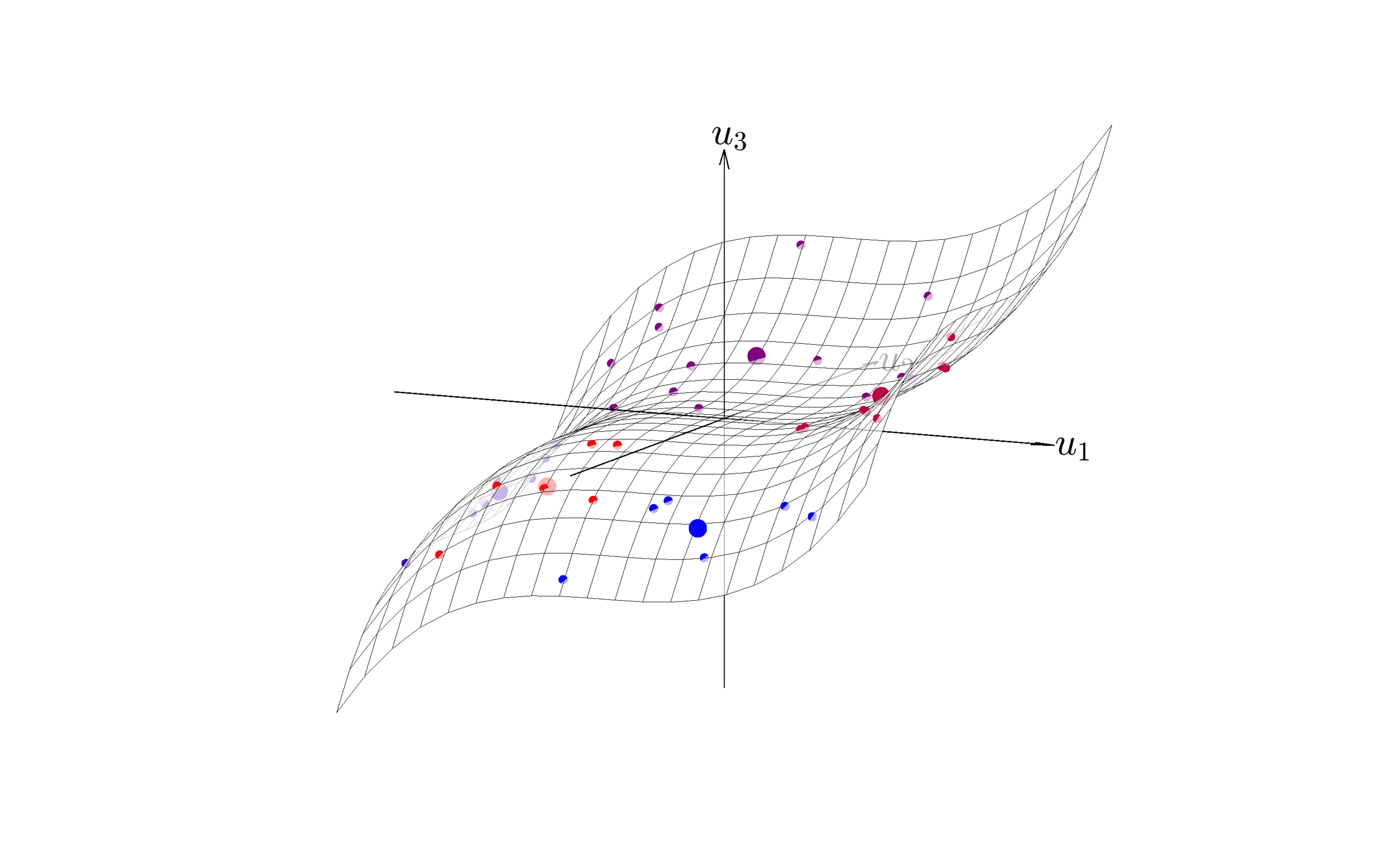}
  \caption{Clustered snapshot vectors on solution manifold in high-dimensional solution space $\Vec{u} \in \mathbb{R}^D$. }
  \label{fig:LPOD1}
\end{subfigure}%
\hspace{0.5cm}
\begin{subfigure}[t]{.45\textwidth}
  \centering
  \includegraphics[scale=0.15,trim={15cm 5.5cm 15cm 5.5cm},clip]{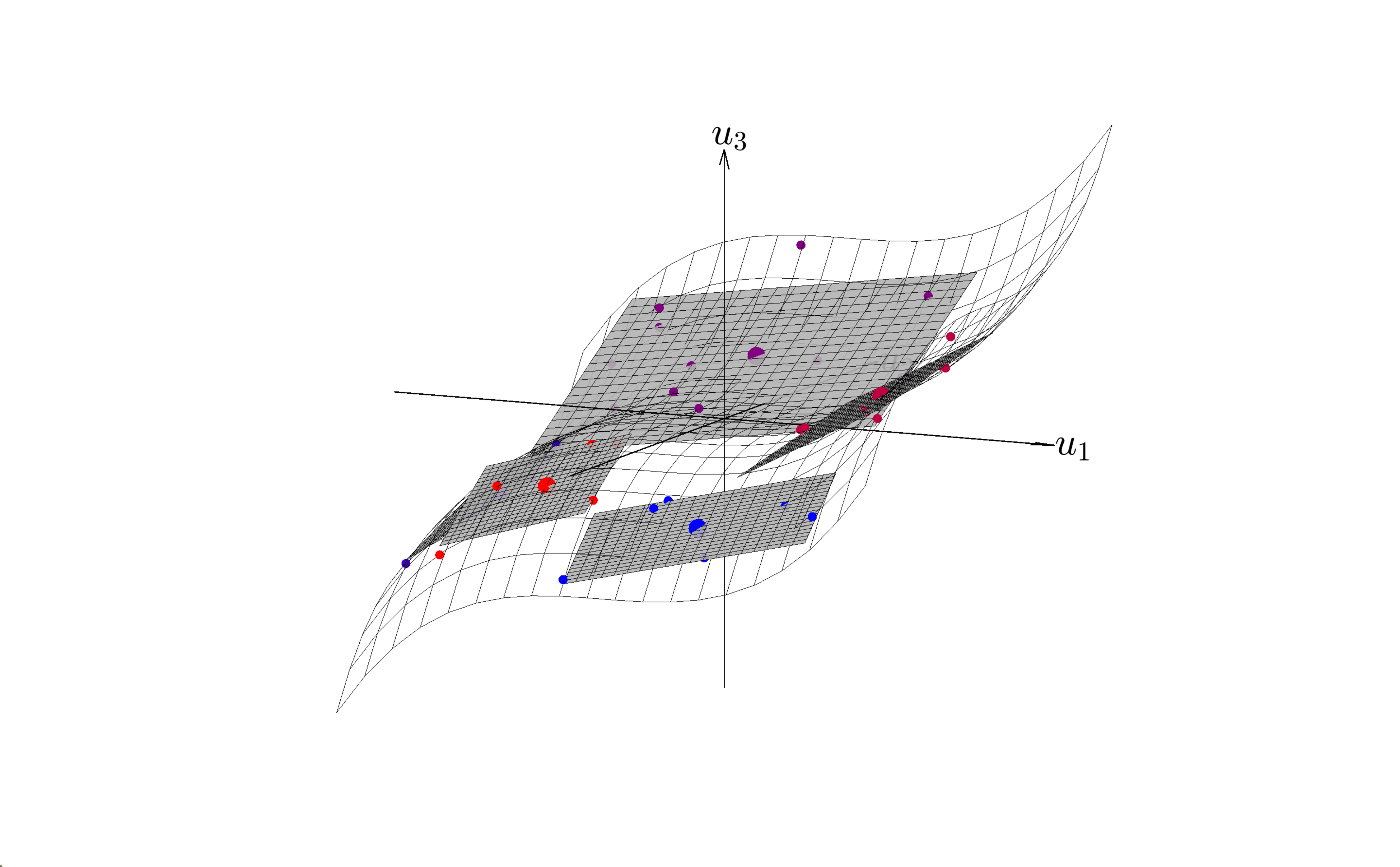}
  \caption{Piecewise linear approximation of solution manifold using the LPOD based on clustered snapshot vectors $\Vec{U} \in \mathbb{R}^{D\times s}$. }
  \label{fig:LPOD2}
\end{subfigure}
\caption{Illustration of dimensionality reduction using the local basis method. Cluster centroids shown as larger spheres. Note that low-dimensional visualisations of manifold learning techniques obscure some of the nuances of dimensionality reduction.}
\label{fig:LPOD}
\end{figure}

As discussed in Section~\ref{s:POD}, the POD may require model sizes $d_\text{POD}$ significantly in excess of the dimensionality $\delta$ of the solution manifold $\mathcal{M}_{\vec{u}}$, which limits the achievable reductions in simulation time.
The local basis method (LPOD) attempts to circumvent this obstacle by localising the POD and using an approximation space $\mathcal{M}_{\bar{\vec{u}}}$ which consists of multiple locally linear approximation spaces $\mathcal{M}_{\bar{\vec{u}}}^j$, each of which is valid in a cluster or subdomain $j$ in solution $\mathbb{R}^D$ or parameter space $\mathbb{R}^d$.
An illustration of this approach can be found in Fig.~\ref{fig:LPOD}. During the online solution procedure, the most appropriate of these local approximation spaces $\mathcal{M}_{\bar{\vec{u}}}^c$ for cluster $c$ is selected for reduction in each iteration.
Originally, the local basis method was proposed in the context of a Gauss-Newton scheme and a scheme from~\cite{CarFar:2011:lcg} was used for hyper-reduction~\cite{AmsZahFar:2012:nmo}, but the application to a Newton-Raphson scheme is straightforward  (see e.g.~\cite{ChaAntBuf:2023:lrb}).

As the POD, the LPOD constructs its reduced spaces from a snapshot matrix $\ten{U}$ of solutions to a full FE problem. 
In the offline phase, a $k$-means clustering step is performed on this data via Lloyd's algorithm~\cite{AmsZahFar:2012:nmo}. $k$ cluster centroids $\Vec{u}_j^c, j=1,...,k$ are initialised as randomly selected snapshots. In each iteration of the clustering scheme, the cluster association $\Vec{c}^u \in \mathbb{Z}^s$ for each snapshot is updated. Each snapshot is associated with the cluster to the centroid of which it is closest, i.e.
\begin{equation*}
    c_i^u=\text{argmin}_j d(\vec{u}_i,\Vec{u}_j^c)\,,
\end{equation*}
where $d(\vec{u}_i,\Vec{u}_j^c)$ is an appropriate metric. Here, a classical L2 norm is used for simplicity, i.e. ${d(\vec{u}_i,\Vec{u}_j^c)=\|\vec{u}_i-\Vec{u}_j^c\|_2}$.
The centroid is then updated as the algorithmic mean of the snapshots now belonging to that cluster, i.e.
\begin{equation*}
    \Vec{u}_j^c = \frac{1}{|\mathcal{C}_j|} \sum_{i \in \mathcal{C}_j} \Vec{u}_i\,,
\end{equation*}
where $\mathcal{C}_j = \{ i: c_i^u = j\}$ 
is the set of snapshots in the cluster and $|\mathcal{C}_j|$ its cardinality (the number of snapshots in the cluster). The procedure is repeated until no further change arises in the cluster association update. To avoid degenerate clusters, we only accept the output of Lloyd's algorithm if each cluster's size exceeds $|\mathcal{C}|_{\text{core,min}}$.
Note that alternative clustering schemes utilising a physics- or problem-based distance metric are also possible, and often advantageous~\cite{DanCasAkk:2020:mor,DanCasAkk:2022:pca,DanCasAkk:2022:uqi}.

After convergence has been achieved with Lloyd's algorithm, the next $\text{ceil}(r|\mathcal{C}_j|)$ nearest snapshots $\Vec{u}_i$ to cluster centroid $\Vec{u}_j^c$ are furthermore added to each cluster $j$ to introduce an overlap. Here, $r>0$ defines the relative amount by which the cardinality of each cluster is to be increased, and $\text{ceil}$ rounds to the nearest larger whole number. This improves the stability of the solution scheme in basis switches, as in the ideal case, transition regions between clusters are captured by all adjacent local POD bases~\cite{AmsZahFar:2012:nmo}.

We introduce a slight modification of this clustering scheme for robustness: it seems helpful, especially in the low-data context, to introduce a minimum cluster size $|\mathcal{C}|_{\text{min}}$ to which clusters must be and a maximum cluster size $|\mathcal{C}|_{\text{max}}$ beyond which they may not be enlarged. In the cluster enlargement step, a cluster is then enlarged until its cardinality reaches {$\max \big (|\mathcal{C}|_{\text{min}},\min(|\mathcal{C}_j|+\text{ceil}(r|\mathcal{C}_j|),|\mathcal{C}|_{\text{max}}) \big)$}. 

Following the clustering step, linear local approximation spaces $\mathcal{M}_{\bar{\vec{u}}}^c$ can be constructed for the snapshots $\Vec{u}_i, i \in \mathcal{C}_j$ in each of the $j = [1,...,k]$ clusters~\cite{AmsZahFar:2012:nmo}. Firstly, the snapshots within a cluster are centered; in~\cite{AmsZahFar:2012:nmo}, this is done using the initial condition $\vec{u}_0$
for the snapshots in all clusters. To obtain approximations of tangents $\mathcal{T}_{\Vec{u}} \mathcal{M}_{\vec{u}}$ to the solution manifold, we instead centre using the cluster centroids
\begin{equation*}
    \Vec{u}_i^s = \Vec{u}_i - \Vec{u}_j^c\,, \quad i \in \mathcal{C}_j, j = [1,...,k]\,,
\end{equation*}
which ensures that the cluster centroids are contained in the local approximation spaces $\mathcal{M}_{\bar{\vec{u}}}^c$ associated with that cluster (see Fig.~\ref{fig:LPOD}). Finally, a snapshot POD can be performed for each of the cluster snapshots matrices $\ten{U}_j^c$, where
\begin{equation*}
    \ten{U}_j^c = [\vec{u}_i^s]_{i\in \mathcal{C}_j}\,,
\end{equation*}
which yields the local POD bases $\ten{\psi}_j^c$, as described in Section~\ref{s:POD}. 
See Algorithm~\ref{alg:LPOD_offline} in \ref{s:code} for pseudocode outlining the offline computations required for the LPOD.

In the iterative online solution procedure, the cluster $c$ with the closest centroid to the current solution $\Vec{\bar{u}_\text{cur}}$ is selected, i.e.
\begin{equation*}
    c =\text{argmin}_j d(\vec{\bar{u}_\text{cur}},\Vec{u}_j^c)\,.
\end{equation*}
The POD basis $\ten{\psi}_c^c$ associated with this cluster is then chosen to define the local approximation space $\mathcal{M}_{\bar{\vec{u}}}^c$.
In~\cite{AmsZahFar:2012:nmo}, the local bases and reduced increments saved at previous time/load steps are exploited such that the distance computation $d(\cdot,\cdot)$ does not scale with the dimensionality of the original problem $D$. Since performance optimisation and runtime comparisons are not within our scope currently, this is not done here; the distance computation is performed in the solution space.
Additionally, a basis realignment scheme is outlined in~\cite{AmsZahWas:2015:flr} to account for the point of transition from one cluster $a$ to the next $b$ in the affine subspace spanned by the reference initial condition $\Vec{u}_0$ and the basis $\ten{\psi}_b^c$. 
Note that the cluster enlargement which produces overlapping clusters already goes some way toward ensuring that possible points of transition between subspaces are contained in the linear approximation spaces $\mathcal{M}_{\bar{\vec{u}}}^c$ associated with the neighbouring clusters.
Finally, a simple projection could ensure that after a transition from cluster $a$ to cluster $b$, the displacement $\Vec{\bar{u}}_\text{cur}$ is contained in the linear approximation space spanned by the local basis $\ten{\psi}_b^c$ around the cluster centroid $\Vec{u}_b^c$
\begin{equation*}
    \Vec{\bar{u}}_{\text{cur}} \gets \ten{\psi}_b^c {\ten{\psi}_b^c}^T ( \Vec{\bar{u}}_{\text{cur}} - \Vec{u}_b^c ) + \Vec{u}_b^c\,.
\end{equation*}
Such a projection does not, however, seem to be required for the examples considered in the scope of this investigation.

Once an appropriate linear approximation space $\mathcal{M}_{\bar{\vec{u}}}^c$ defined by basis $\ten{\psi}_c^c$ has been identified, incremental solutions can be sought in this subspace as in the case of the POD, i.e.
\begin{equation*}
    \ten{K}_r \Delta \Vec{y} = - \Vec{g}_r\,,\quad \text{where} \quad \ten{K}_r = \ten{\psi}_c^{cT} \ten{K} \ten{\psi}_c^c\,, \quad \text{and} \quad
    \Vec{g}_r = \ten{\psi}_c^{cT} \Vec{g}\,,
\end{equation*}
whereupon the current iterate can be incremented
\begin{equation*}
    \Delta {\Vec{\bar{u}}} = \ten{\psi}_c^{c} \Delta \Vec{y}, \quad {\Vec{\bar{u}_\text{cur}}} \leftarrow {\Vec{\bar{u}_\text{cur}}} + \Delta {\Vec{\bar{u}}}\,.
\end{equation*}
Algorithm~\ref{alg:LPOD_online} in \ref{s:code} features pseudocode for the online phase of the LPOD for an RVE problem with periodic BCs.
As in the case of the POD, the linearity of the approximation spaces $\mathcal{M}_{\bar{\vec{u}}}^c$ obtained via the LPOD means that a tangent is available by construction and hyper-reduction schemes can be applied without undue effort. The desired scaling behaviour independent from the problem size $D$ may be achieved~\cite{AmsZahFar:2012:nmo,ChaAntBuf:2023:lrb}.

By identifying locally rather than globally linear structures in solution space, the LPOD achieves a closer approximation of the solution manifold than the POD. In the trade-off between accuracy and computational cost in the online phase, this has enabled researchers to obtain overall performance improvements ~\cite{DanCasAkk:2020:mor,DanCasAkk:2022:pca,DanCasAkk:2022:uqi,ChaAntBuf:2023:lrb,ChaAntBuf:2024:fpa}. As outlined in the introduction, however, this generally successful approach is subject to limitations, especially in the data-poor context: locally linear approximations might not parameterise the solution manifold as closely as possible, properly handling cluster transitions is challenging~\cite{AmsZahWas:2015:flr}, cluster transitions may yield instabilities or inaccuracies in the solution scheme~\cite{IdeCar:1985:rmn}, and the quality of the obtained ROM scales with the quality of the underlying clusters (the assurance of which is intricate).

\section{Projection-based MOR by manifold learning: global nonlinearity}\label{s:ManLMOR}

As motivated in the introduction, we {instead pursue a manifold learning approach to nonlinear MOR}: graph-based manifold learning methods~\cite{LeeVer:2007:ndr,BelNiy:2003:led,RowSau:2000:ndr} are used to exploit global, nonlinear trends in the solution manifold based on the snapshot data. This may yield a closer, smoother approximation $\mathcal{M}_{\bar{\vec{u}}}$ of the solution manifold $\mathcal{M}_{\vec{u}}$ than established methods, especially in the data-poor context, meaning that a manifold learning-based NLMOR scheme might outperform local basis methods and their alternatives in accelerating RVE computations based on limited data. Note that manifold learning methods do not yield a tangent $\ten{\varphi}$ (see Eq.~\eqref{eq:tangent}) by construction; a linearisation needs to be constructed a posteriori~\cite{LeeVer:2007:ndr}.

\subsection{Manifold learning for dimensionality reduction}\label{s:ManL}

\begin{figure}[h!]
\centering
\begin{subfigure}[t]{.45\textwidth}
  \centering
  \includegraphics[scale=0.15,trim={15cm 5.5cm 15cm 5.5cm},clip]{figures/manifold_visualisations/manifold_samples.eps}
  \caption{Snapshot vectors on solution manifold in high-dimensional solution space $\Vec{u} \in \mathbb{R}^D$.}
  \label{fig:manl1}
\end{subfigure}%
\hspace{0.5cm}
\begin{subfigure}[t]{.45\textwidth}
  \centering
  \includegraphics[scale=0.15,trim={15cm 5.5cm 15cm 5.5cm},clip]{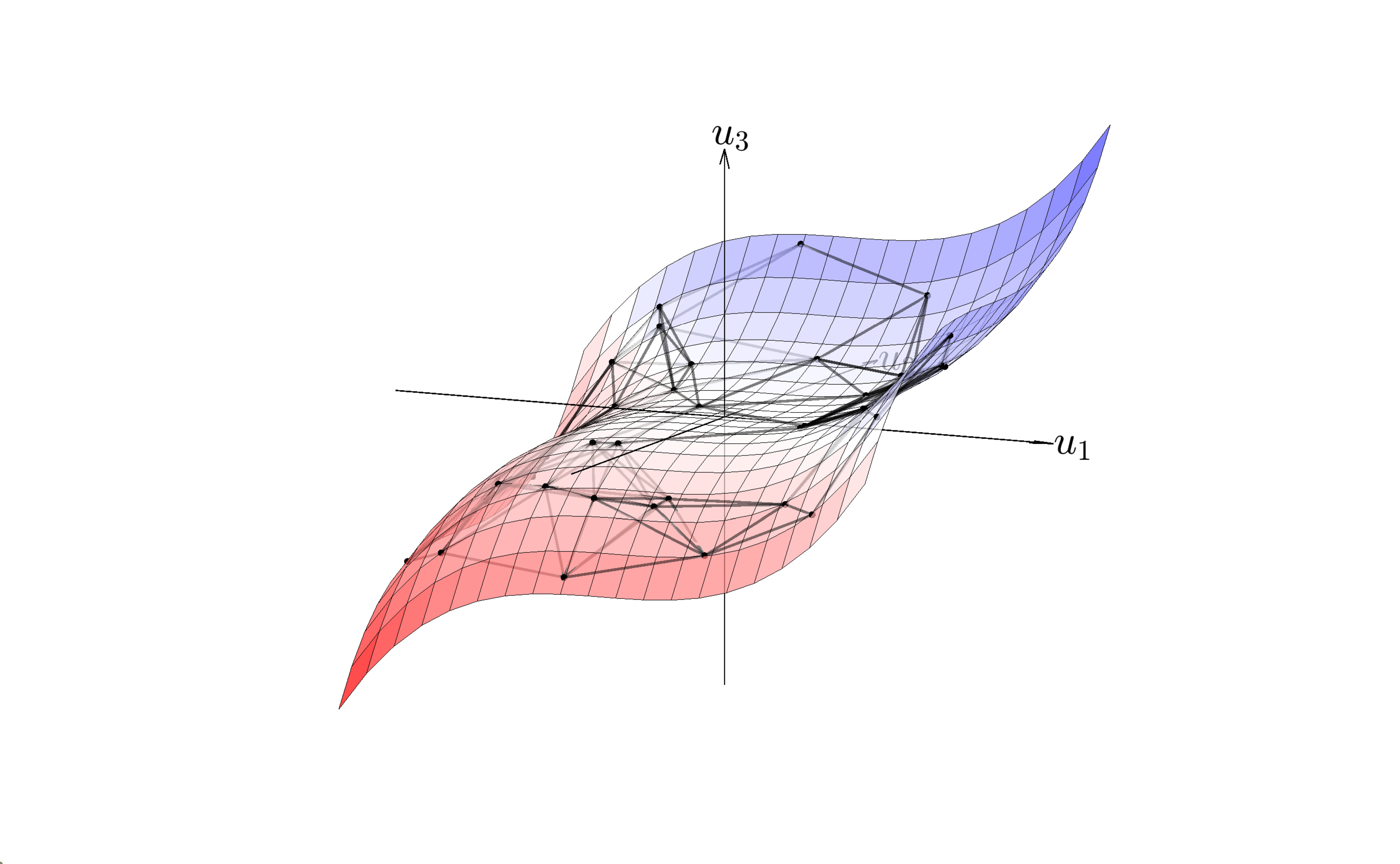}
  \caption{k-nearest-neighbours graph between snapshot vectors in high-dimensional solution space $\Vec{u} \in \mathbb{R}^D$.}
  \label{fig:manl2}
\end{subfigure}
\par
\begin{subfigure}[t]{.45\textwidth}
  \centering
  \includegraphics[scale=0.12,trim={7cm 2cm 7cm 2cm},clip]{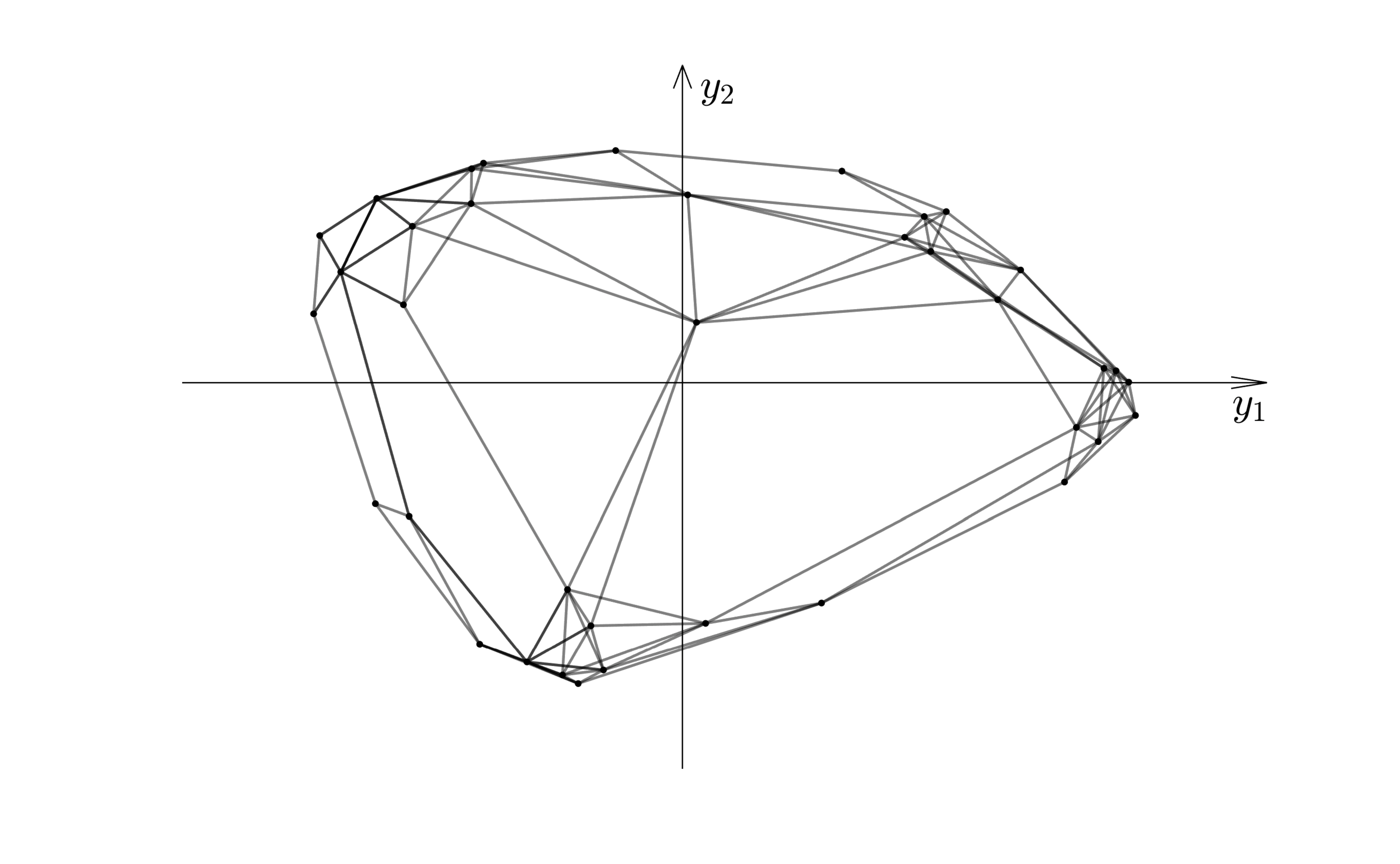}
  \caption{Low-dimensional embedding $\Vec{y} \in \mathbb{R}^d$ of snapshot vectors $\Vec{u} \in \mathbb{R}^D$ (using Laplacian Eigenmaps).}
  \label{fig:manl3}
\end{subfigure}
\caption{Illustration of dimensionality reduction using Laplacian Eigenmaps or Locally Linear Embedding. Note that low-dimensional visualisations of manifold learning techniques obscure some of the nuances of dimensionality reduction.}
\label{fig:manl}
\end{figure}

As discussed in Section~\ref{s:problem}, a sensible, generalisable approach to nonlinear projection-based MOR is to aim to parameterise a region in solution space, i.e. the approximation space $\mathcal{M}_{\bar{\vec{u}}}$, with reduced variables $\Vec{y} \in \mathbb{R}^d$ to approximate a superset of the solution manifold $\mathcal{M}_{\vec{u}}$. Here, the dimension of the approximation $d$ exceeds the dimensionality of the solution manifold $\delta$ but is far smaller than the dimension of the original solution space, i.e. $\text{dim}(\mathcal{M}_{\vec{u}})=\delta<\text{dim}(\mathcal{M}_{\bar{\vec{u}}})=d \ll D$. As the problem parameterisation $\vec{p}\in\mathbb{R}^\delta$ determines the dimensionality of the solution manifold, $\delta$ is often known a priori. When it is not, for example when plastic behaviour introduces a load path dependency, $\delta$ may be estimated from the snapshot data $\ten{U}\in \mathbb{R}^{D\times s}$, as outlined in ~\ref{s:corr_dim}. 

The dimensionality reduction problem underlying such a projection-based MOR scheme could be formulated as follows: 
use snapshot data $\ten{U}$ to find a nonlinear embedding map $M$ from the solution space to a reduced space, i.e. $M: \mathbb{R}^D\rightarrow \mathbb{R}^d$. The embedded coordinates in the reduced space $\Vec{y}_i=M(\Vec{u}_i) \in\mathbb{R}^d$ should be constructed so as to retain as much as possible of the {information contained within} the snapshots $\Vec{u}_i \in \mathbb{R}^D$, i.e. capture as much as possible of the structure of the solution manifold.

An approach to tackle this nonlinear dimensionality reduction problem 
are graph-based manifold learning techniques (see e.g.~\cite{LeeVer:2007:ndr}), which are popular in image and speech processing. Some researchers have also successfully applied these methods to surrogate modelling in fluid dynamics and solid statics~\cite{FraZimGor:2014:irm,BhaMat:2016:nmr,Bha:2017:rom} as well as to projection-based MOR in fluid dynamics~\cite{PytAbe:2016:nmr,PytMeySch:2017:snr,Pyt:2018:muo,MilArr:2013:nml}.
In this proof-of-concept investigation, we consider the Laplacian Eigenmaps (LEM)~\cite{BelNiy:2003:led} and the Locally Linear Embedding (LLE)~\cite{RowSau:2000:ndr,LeeVer:2007:ndr} as examples of graph-based manifold learning methods.

Conceptually, these methods find an embedding $\ten{Y}\in \mathbb{R}^{d\times s}$ for snapshot data $\ten{U} \in \mathbb{R}^{D\times s}$ in a reduced space $\mathbb{R}^d$ which conserves certain structural information (e.g., local distances in the case of LEM~\cite{BelNiy:2003:led} or local linearity in the case of LLE~\cite{RowSau:2000:ndr}).
Firstly, a graph $\ten{G} \in \mathbb{B}^{s \times s}$ (where $\mathbb{B}$ indicates the Boolean domain $\mathbb{B}\in \{ 0,1 \}$) is created for the snapshot solution data $\ten{U}$, e.g. via a mutual or symmetric $k$-nearest neighbour approach. Here, $G_{ij}=1$ indicates that snapshots $\Vec{u}_i$ and $\Vec{u}_j$ are connected via an edge, while $G_{ij}=0$ indicates unconnected snapshots. The graph thus contains information about local connectivity within the data and acts as a data-based model of the structure of the manifold $\mathcal{M}_{\vec{u}}$~\cite{BelNiy:2003:led}. Based on the structural data contained in the graph adjacency matrix and some additional, method-specific data, an embedding $\ten{Y} \in \mathbb{R}^{d\times s}$ in the reduced space $\mathbb{R}^{d}$ is then found.
Figure~\ref{fig:manl} visualises this conceptual approach on a low-dimensional example.

As motivated in Section~\ref{s:problem}, a reconstruction mapping $R:\mathbb{R}^d \rightarrow\mathbb{R}^D$ from the reduced space to the solution space as well as its derivative $\ten{\varphi} = \frac{\partial R(\Vec{y})}{\partial \Vec{y}}$, or at least approximations thereof are required for the online phase of a projection-based Galerkin MOR scheme.
The embedding $M$, reconstruction $R$, and derivative $\nicefrac{\partial R(\Vec{y})}{\partial\vec{y}}$, however, are only defined implicitly by graph-based manifold learning schemes and approximations need to be identified separately~\cite{BelNiy:2003:led}. Doing so in a general fashion is nontrivial, as the implicitly defined embedding mapping 
$M: \mathbb{R}^D\rightarrow \mathbb{R}^d$ is highly nonlinear in general, and must be so in order for the approximation space $\mathcal{M}_{\bar{\vec{u}}}$ to closely approximate the solution manifold $\mathcal{M}_{\vec{u}}$.

\subsection{Laplacian Eigenmaps}\label{s:LEM}

The Laplacian Eigenmaps (LEM) by Belkin et al.~\cite{BelNiy:2003:led} is a manifold learning method which can be shown to optimally preserve local distance information.
A graph of the snapshot data $\ten{U} \in \mathbb{R}^{D\times s}$ is first constructed via one of three methods:     
\begin{itemize}
    \item in the $\epsilon$-ball approach, an edge in the graph $G_{ij}=1$ is introduced whenever the distance between samples $\Vec{u}_i$ and $\Vec{u}_j$ is below a threshold, $\| \Vec{u}_i - \Vec{u}_j \| < \epsilon$~\cite{BelNiy:2003:led};
    \item in the symmetric $k$-nearest neighbour approach, an edge in the graph $G_{ij}=1$ is introduced whenever sample $\Vec{u}_i$ is among the $k$ nearest neighbours of sample $\Vec{u}_j$, \textbf{or} vice versa~\cite{BelNiy:2003:led};
    \item in the mutual $k$-nearest neighbour approach, an edge in the graph $G_{ij}=1$ is introduced whenever sample $\Vec{u}_i$ is among the $k$ nearest neighbours of sample $\Vec{u}_j$, \textbf{and} vice versa~\cite{BelNiy:2003:led}.
\end{itemize}
As suggested in~\cite{PytAbe:2016:nmr} and borne out in our preliminary numerical investigations, a symmetric $k$-nearest neighbour approach might be most promising as it is relatively robust in the face of varying data densities.

Weights $\ten{W}\in\mathbb{R}^{s\times s}$ are then assigned to each graph edge. This can for example be done via a Gaussian kernel $W_{ij} = E_{ij}\exp \left ( - \frac{1}{t} \| \Vec{u}_i- \Vec{u}_j \|^2 \right )$~\cite{BelNiy:2003:led}. An unweighted graph with $W_{ij}=E_{ij}$ is recovered in the limit as $t\rightarrow\infty$.
It can also be advantageous to weight the distance function with some variability metric of the underlying problem~\cite{PytMeySch:2017:snr}. A Gaussian kernel eventually yields an embedding $\ten{Y}\in\mathbb{R}^{d\times s}$ which optimally preserves local distance information since minimising distance changes in the embedding $M:\mathbb{R}^D\rightarrow\mathbb{R}^d$ corresponds to a search for Eigenfunctions of the manifold Laplace-Beltrami operator, the Green's function of which is the Gaussian~\cite{BelNiy:2003:led}.

Following~\cite{BelNiy:2003:led}, a solution is found to the generalised Eigenvalue problem in terms of the graph Laplacian $\ten{L}$
\begin{equation*}
    \ten{L} \Vec{v}_i = \lambda_i \ten{D} \Vec{v}_i\,, \quad \text{where}\quad \ten{L} = \ten{D} - \ten{W}\,,
\end{equation*}
with weight matrix $\ten{W}$ and diagonal weight matrix $\ten{D}$
\begin{equation*}
    D_{(ii)} = \sum_j W_{ij} = \sum_i W_{ij}\,.
\end{equation*}
$D_{(ii)}$ measures the sum total of the weights of the edges of the graph that are adjacent to sample $\Vec{u}_i$, and can be seen as an importance measure of sample $\Vec{u}_i$ \cite{BelNiy:2003:led}. Alternatively, the scaled Eigenvalue problem 
\begin{equation*}
    \ten{D}^{-1} \ten{L} \Vec{v}_i = \lambda_i \Vec{v}_i
\end{equation*}
may be solved instead~\cite{Che:2022:le}. In either case, Eigenvalues must be sorted in ascending order~\cite{BelNiy:2003:led}.

The lowest Eigenvalue vanishes (i.e., $\lambda_1=0$), and a $d$-dimensional embedding ($d \ll D$) of the samples $\Vec{u}_i, i \in [1,..,s]$ is defined by the next lowest Eigenvectors $\Vec{v}_i, i \in [2,..,d+1]$~\cite{BelNiy:2003:led}. In a rough analogy to the POD, these Eigenvalues contain the components of an embedding in decreasing order of relevance to the LEM's embedding criterion~\cite{BelNiy:2003:led}. The new coordinates $\Vec{y}_i$ in the reduced space corresponding to $\Vec{u}_i$ are given by row $i$ of the matrix~\cite{Che:2022:le}
\begin{equation*}
    \ten{Y}^T = [ \Vec{v}_2,..,\Vec{v}_{d+1}] \,, \quad \text{i.e.} \quad
    \Vec{y}_i = [ v_{2,i},..,v_{d+1,i}]^T\,.
\end{equation*}
Pseudocode for the LEM algorithm is provided in Algorithm~\ref{alg:LEM} in \ref{s:code}.

As noted in Section~\ref{s:ManL}, LEM does not, by construction, yield the reconstruction \mbox{$R:\mathbb{R}^d\rightarrow\mathbb{R}^D$} and the tangent $\ten{\varphi}=\frac{\partial R(\Vec{y})}{\partial \Vec{y}}$ 
which are required for projection-based MOR in the Galerkin framework (see Eqs.~\eqref{eq:redres},~\eqref{eq:newton_red}, and~\eqref{eq:K_red} in Section~\ref{s:problem}).
In~\cite{Pyt:2018:muo}, a global linearisation is suggested, as outlined in Section~\ref{s:globlin}. Furthermore, in~\cite{He:2002:lei} a suggestion for a linear Laplacian Eigenmap is made, which, however, is badly conditioned in the case that $D>s$.
Finally, Kernel regression is suggested in~\cite{CarLu:2007:lel}, which addresses the reconstruction but not the tangent computation.

Note furthermore that, as visualised in the examples in~\cite{LeeVer:2007:ndr}, Laplacian Eigenmaps can behave problematically for MOR in some cases, as points which are far apart in the original space might be mapped closely in the reduced space. This occurs because the LEM only considers the distances between data points which are connected by an edge of the underlying graph, i.e., points which are close, explicitly. Furthermore, only local distance information -- rather than, for example, information on local angles and topology -- are used.

\subsection{Locally Linear Embedding}\label{s:LLE}

The Locally Linear Embedding (LLE) by Roweis et al.~\cite{RowSau:2000:ndr} is a manifold learning method which preserves local topology. The summary in this Section is based mainly on the description in~\cite{LeeVer:2007:ndr}.

Firstly, LLE constructs a graph $\ten{G}\in \mathbb{B}^{s\times s}$ via a $k$-nearest neighbour approach, as already outlined in Section~\ref{s:LEM}: an edge in the graph $G_{ij}=1$ is introduced, whenever sample $\Vec{u}_i$ is among the $k$ nearest neighbours of sample $\Vec{u}_j$, either \textbf{and} vice versa (mutual $k$-nearest neighbours) or \textbf{or} vice versa (symmetric $k$-nearest neighbours), else $G_{ij}=0$~\cite{LeeVer:2007:ndr}.

Then, a weight matrix $\ten{W}\in \mathbb{R}^{s\times s}$ for the graph edges is constructed based on the principle of locally linear reconstruction.
All weights $W_{ij}$ for nodes $i$ and $j$ which are not connected are set to $0$.
The weights $W_{ij}$ for nodes $i$ and $j$ which are connected by an edge in Graph $\ten{G}$ are constructed on the principle that the displacement vector corresponding to node $i$ in solution space $\Vec{u}_i \in \mathbb{R}^D$ is optimally reconstructed via a linear combination of the displacement vectors $\Vec{u}_j \in \mathbb{R}^D, j \in N_i$ of its neighbours $N_i=\{j\mid G_{ij}=1\}$ with (reconstruction) weights $\ten{W}$, i.e. the minimisation problem
\begin{equation*}
    \min_{\ten{W}} \sum_i \| \Vec{u}_i - \sum_{j \in N_i} W_{ij} \Vec{u}_j \|^2
\end{equation*}
is solved~\cite{LeeVer:2007:ndr}. 

Practically, this can be implemented as follows~\cite{LeeVer:2007:ndr}:
iterate through all nodes $i$ and assemble the local covariance matrix of the distances between the point $\Vec{u}_i$ and its neighbours $\Vec{u}_j, j \in N_i$. The centered distances from $\vec{u}_i$ to each of its neighbours $\Vec{u}_j$ are collected into a matrix $\Delta \ten{U}^{N_i}$
and the local covariance matrix is computed as
\begin{equation*}
    \ten{G}_i =  \Delta \ten{U}^{N_i T} \Delta \ten{U}^{N_i} \,, \quad \text{where} \quad \Delta \ten{U}^{N_i} = [ \vec{u}_i - \vec{u}_j ]_{j \in N_i}\,.
\end{equation*}
Note that in the above, the indices $i$ and $j$ denote snapshot vector $i$ and its neighbours $j \in N_i$. 
The entries in the weight matrix $\ten{W}$ corresponding to the columns $j \in N_i$ in row $i$ can then be computed via the normalised solution of the (local) linear equation system
\begin{equation*}
    \ten{G}_i \Vec{w}_i = \vec{\bar{1}}\,,
\end{equation*}
where $\vec{\bar{1}}$ denotes a vector of ones and
\begin{equation*}
    W_{i,j \in N_i} = \Vec{w}_i\,.
\end{equation*}
When $\ten{G}_i$ is badly conditioned, 
the solution can be determined by modifying the diagonal of $\ten{G}_i$ with a small factor $\Delta$
\begin{equation*}
    \ten{G}_i = \ten{G}_i + \frac{\Delta^2}{N_i} \text{tr} (\ten{G}_i) \ten{I}\,.
\end{equation*}

Finally, the embedding $\ten{Y}\in\mathbb{R}^{d\times s}$ is constructed such that the locally linear interpolation can be performed with minimum error using the same weight matrix $\ten{W}$ in the reduced space~\cite{LeeVer:2007:ndr}. In other words, the reduced vectors $\Vec{y}_i$ are determined as solutions to the minimisation problem
\begin{equation*}
    \min_{\ten{Y}} \sum_i \| \Vec{y}_i - \sum_{j \in N_i} W_{ij} \Vec{y}_j \|^2\,.
\end{equation*}
As in the case of the Laplacian Eigenmaps, the $d$ entries of the $s$ reduced vectors $\ten{Y}\in \mathbb{R}^{d\times s}$ can be computed as the $d$ Eigenvectors corresponding to the second to $d+1$-th lowest Eigenvalues of a matrix $\ten{M \in \mathbb{R}^{s\times s}}$, in this case
\begin{equation*}
    \ten{M} = ( \ten{I}- \ten{W} )^T (\ten{I}- \ten{W} )\,,
\end{equation*}
i.e.
\begin{equation*}
    \ten{M} \Vec{v}_i = \lambda_i \Vec{v}_i\,.
\end{equation*}
The new coordinates $\Vec{y}_i$ in the reduced space $\mathbb{R}^d$ corresponding to $\Vec{u}_i$ are again given by row $i$ of the matrix~\cite{Che:2022:le}
\begin{equation*}
    \ten{Y}^T = [ \Vec{v}_2,..,\Vec{v}_{d+1}] \,, \quad \text{i.e.} \quad
    \Vec{y}_i = [ v_{2,i},..,v_{d+1,i}]^T\,.
\end{equation*}
Pseudocode for the LLE algorithm can be found in Algorithm~\ref{alg:LLE} in \ref{s:code}.

As outlined in Section~\ref{s:LEM} for the case of the LEM, LLE also provides no reconstruction $R:\mathbb{R}^d\rightarrow\mathbb{R}^D$ or tangent $\ten{\varphi}=\frac{\partial R(\Vec{y})}{\partial \Vec{y}}$  by construction. An approximation of the reconstruction operation $R:\mathbb{R}^d\rightarrow\mathbb{R}^D$ can be defined intuitively by locally linear interpolation with reconstruction weights~\cite{SauRow:2003:tgf}, but a tangent is still required for the reduction operation in Eq.~\eqref{eq:POD_red}.

\subsection{Global linearisation}\label{s:globlin}

\begin{figure}[h!]
    \centering
    \includegraphics[scale=0.15,trim={15cm 5.5cm 15cm 5.5cm},clip]{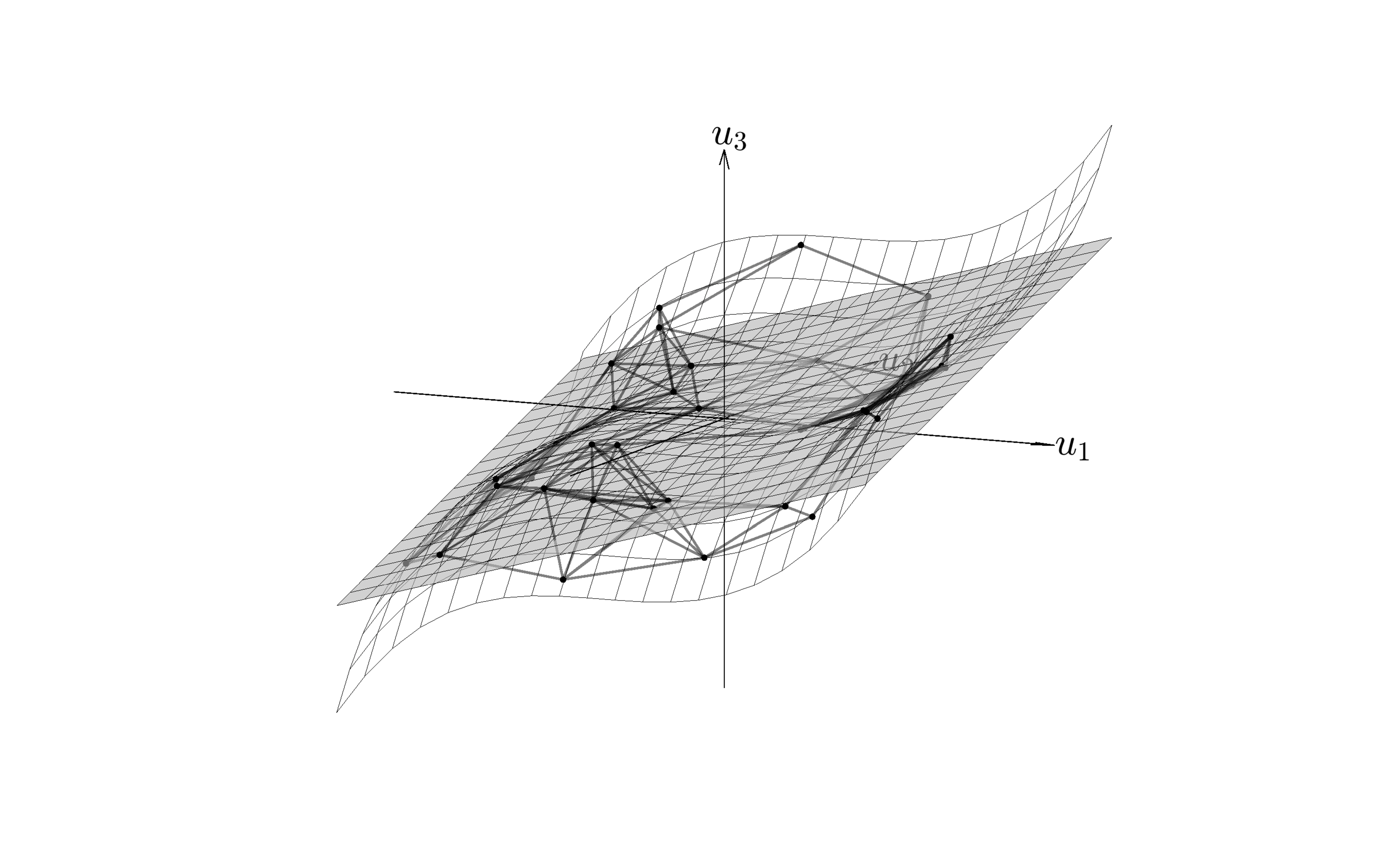}
    \caption{Illustration of global linearisation by \cite{Pyt:2018:muo}: linear approximation of solution manifold based on reduced and original coordinates of neighbouring snapshot vectors $\ten{Y}_n \in \mathbb{R}^{d \times n}$ and $\ten{U}_n \in \mathbb{R}^{D \times n}$. Note that low-dimensional visualisations of manifold learning techniques obscure some of the nuances of dimensionality reduction.}
    \label{fig:globlin}
\end{figure}

As outlined in Section~\ref{s:problem}, Galerkin schemes for MOR require approximations of the reconstruction mapping \mbox{$R:\mathbb{R}^d\rightarrow\mathbb{R}^D$} as well as its tangent $\frac{\partial R}{\partial \Vec{y}}$. The LEM and LLE which were outlined in Sections~\ref{s:LEM} and~\ref{s:LLE}, however, do not yield these a priori.

In the context of projection-based MOR for dynamic problems, Pyta et al. globally linearised an LEM embedding to yield a linear reconstruction $R:\mathbb{R}^d \rightarrow \mathbb{R}^D$ as that obtained via the POD (see Eq.~\eqref{eq:POD_map}), i.e. $ \vec{\bar{u}} = \ten{{\psi}} \Vec{y}$
~\cite{PytAbe:2016:nmr,PytMeySch:2017:snr,Pyt:2018:muo}.
Fig.~\ref{fig:globlin} features a low-dimensional visualisation of the linearisation scheme. 
Note, however, that this visualisation might obscure some of the nuances of dimensionality reduction: since the embedding may be performed to a reduced space $\Vec{y} \in \mathbb{R}^d$ with $d>\delta$, some of the nonlinearity in the solution manifold may still be captured by a global linearisation, provided the manifold can be successfully captured in the reduced space $\Vec{y} \in \mathbb{R}^d$~\cite{LeeVer:2007:ndr}. 
Note that in~\cite{Pyt:2018:muo}, such linear ROMs were assembled for individual points $\Vec{p}_i\in \mathbb{R}^\delta$ in parameter space from snapshot solutions obtained via simulations in time $\Vec{u}(t;\Vec{p}_i)$ at that parameter point. ROMs for unseen parameter points were constructed by interpolation on the Grassmann manifold using the methodology of Amsallem et al.~\cite{AmsFar:2008:ima,AmsCorCar:2009:mim}.

To ensure a consistent embedding of the zero vector, we first centre the reduced snapshot data around the embedding of the zero vector, i.e. $\Vec{y}_i \leftarrow \Vec{y}_i- \Vec{y}_0$. Then, a linear projection matrix $\ten{\psi}$ is sought to optimally map between the original snapshot matrix $\ten{U}$ and the corresponding reduced coordinates $\ten{Y}$ in the least-squares sense. As shown in~\cite{Pyt:2018:muo} and outlined in ~\ref{a:globlin}, $\ten{\psi}$ is given by
\begin{equation*}
    \ten{\psi} = \ten{U} \ten{Y}^T ( \ten{Y} \ten{Y}^T )^{-1}\,.
\end{equation*}
$\ten{\psi}$ can be orthonormalised via a Gram-Schmidt process or a reduced QR decomposition $QR(\cdot)$~\cite{Pyt:2018:muo}
\begin{equation*}
    {\ten{\psi}}_\perp, \ten{R} = QR \left (\frac{1}{D} \ten{\psi} \right )\,.
\end{equation*}
Pyta et al. then introduce a modified embedding defined via this map, such that~\cite{Pyt:2018:muo}
\begin{equation*}
    {\Vec{y}}_i^\perp = {\ten{\psi}}_\perp^T \Vec{u}_i\,, \quad \text{and} \quad \vec{\bar{u}}_i = {\ten{\psi}}_\perp \Vec{y}_i^\perp  
    \,.
\end{equation*}
See Alg.~\ref{alg:globlin} in \ref{s:code} for the construction of the mapping. The online phase of a projection-based MOR scheme utilising a globally linearised embedding is generally equivalent to that of the POD in Alg.~\ref{alg:POD_online} in \ref{s:code}.

\subsection{Local linearisation}\label{s:loclin}

\begin{figure}[h!]
\centering
\begin{subfigure}[t]{.45\textwidth}
  \centering
  \includegraphics[scale=0.12,trim={7cm 2cm 7cm 2cm},clip]{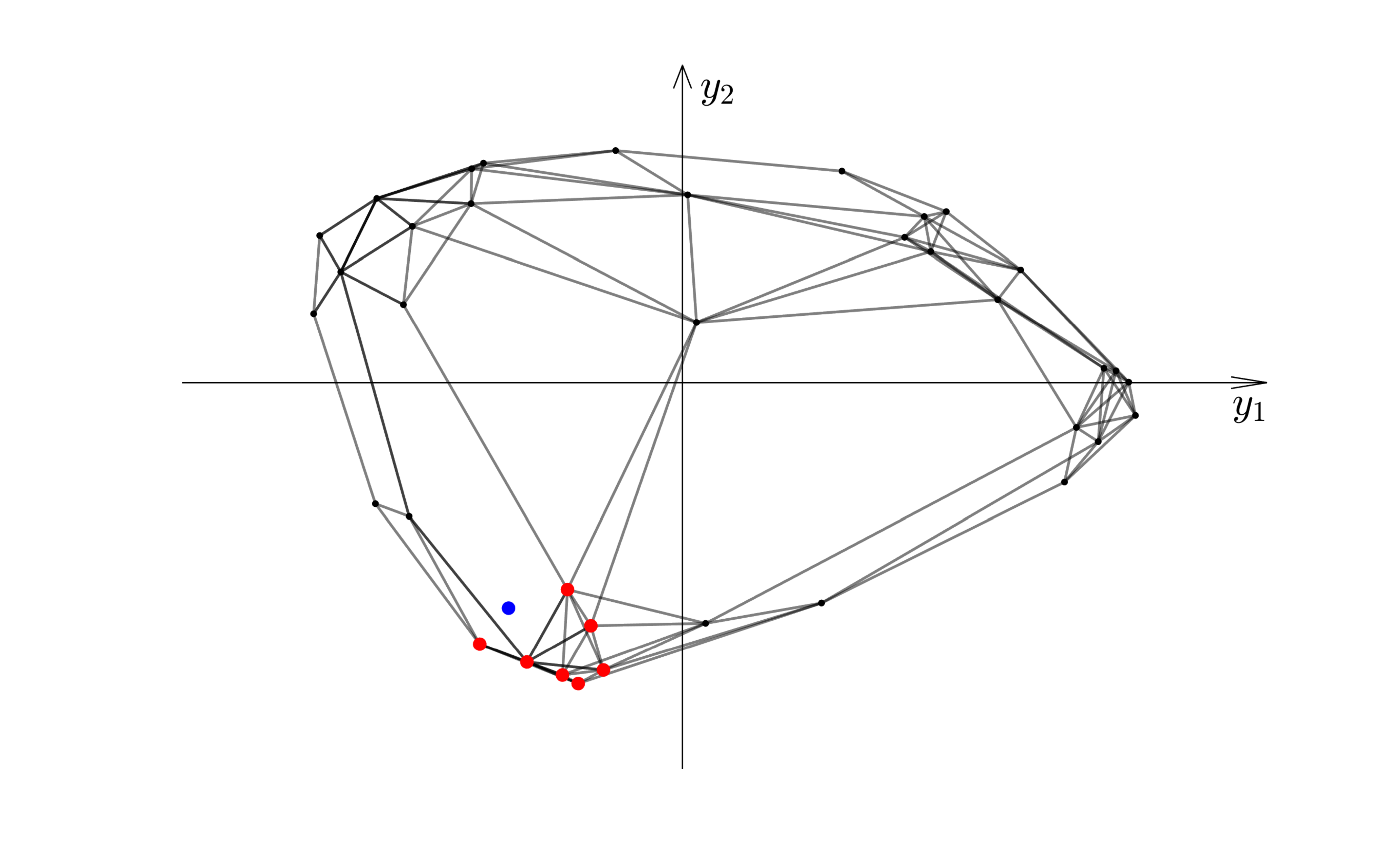}
  \caption{Neighbouring snapshots $\ten{Y}_n \in \mathbb{R}^{d\times n}$ {(red dots)} of point $\vec{y}$ {(blue dot)} in reduced solution space $\Vec{y} \in \mathbb{R}^d$.}
  \label{fig:loclin1}
\end{subfigure}%
\hspace{0.5cm}
\begin{subfigure}[t]{.45\textwidth}
  \centering
  \includegraphics[scale=0.15,trim={15cm 5.5cm 15cm 5.5cm},clip]{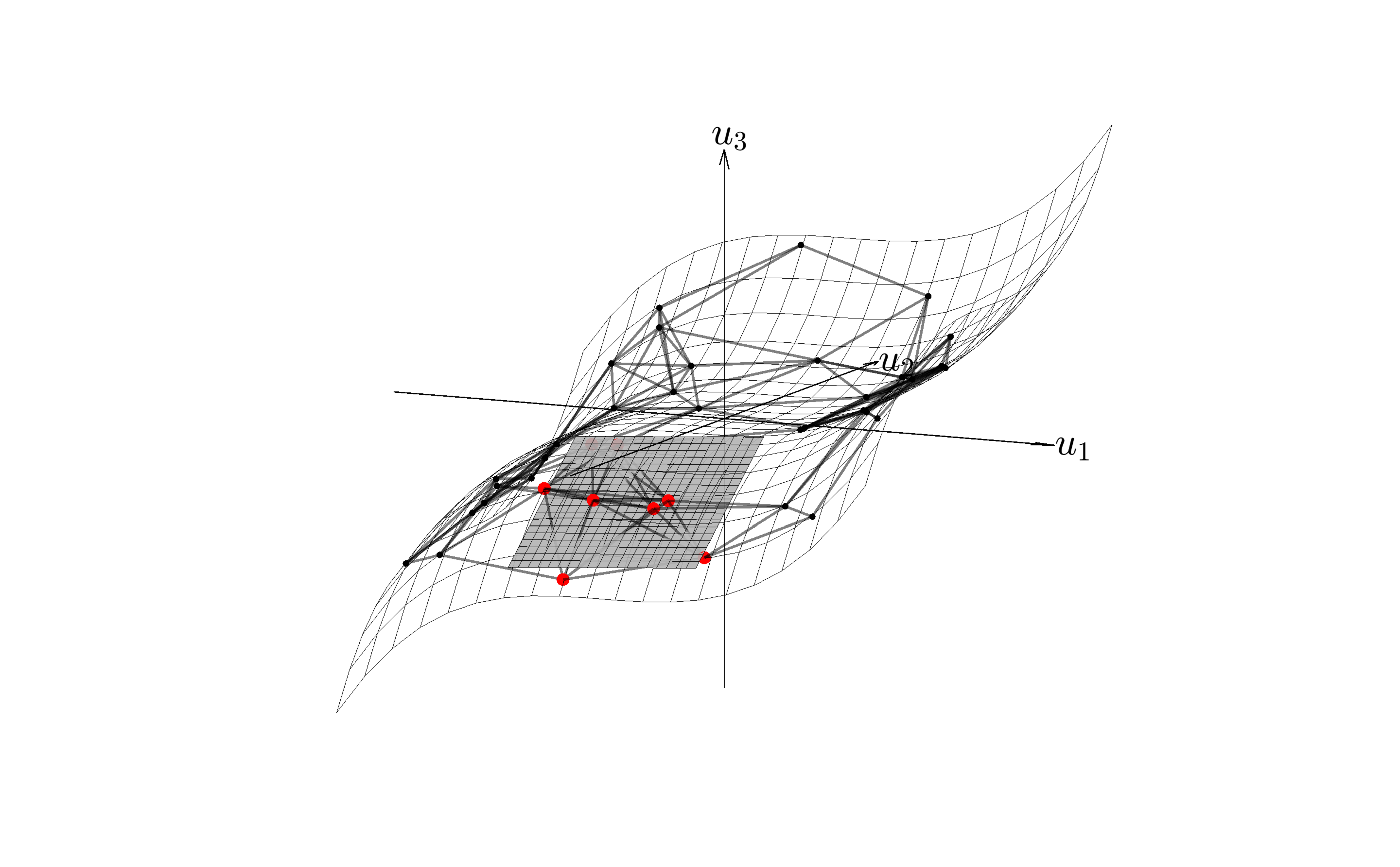}
  \caption{(Mapping to) approximation of tangent to solution manifold based on coordinates of neighbouring snapshot vectors {(red dots)} in reduced $\ten{Y}_n \in \mathbb{R}^{d \times n}$ and original $\ten{U}_n \in \mathbb{R}^{D \times n}$ solution spaces.}
  \label{fig:loclin2}
\end{subfigure}
\caption{Illustration of local linearisation: approximation of (mapping to) tangent to solution manifold based on reduced and original coordinates of neighbouring snapshot vectors $\ten{Y}_n \in \mathbb{R}^{d \times n}$ and $\ten{U}_n \in \mathbb{R}^{D \times n}$. Note that low-dimensional visualisations of manifold learning techniques obscure some of the nuances of dimensionality reduction.}
\label{fig:loclin}
\end{figure}


In this work, we instead propose a nonlinear projection-based MOR scheme which directly utilises the embedding $\Vec{y}_i\in\mathbb{R}^d$ and reduced space $\mathbb{R}^d$ obtained via manifold learning techniques to parameterise the approximation $\mathcal{M}_{\bar{\vec{u}}}$ of the solution manifold $\mathcal{M}_{\vec{u}}$. As motivated in Section~\ref{s:problem}, such an approach enables the use of lower-dimensional reduced spaces $\mathbb{R}^d$, as the solution manifold can be approximated more closely, meaning that higher levels of accuracy can be reached with a given model size $d$.
Instead of linearising globally, we compute a locally linear approximation $\Delta \Vec{\bar{u}} = \ten{\varphi} \Delta \Vec{y}$ (cf. Eq~\eqref{eq:loclin}) of the reconstruction mapping $R: \mathbb{R}^d \rightarrow \mathbb{R}^D$
around a point $\Vec{y}_{\text{cur}}$ in the reduced space. The mapping matrix $\ten{\varphi}$ approximates the tangent to the reconstruction $R:\mathbb{R}^d \rightarrow \mathbb{R}^D$ according to Eq.~\eqref{eq:tangent}.
This is tantamount to mapping from the reduced space $\mathbb{R}^d$ to an approximation of the tangent to the approximation space $\mathcal{T}_{\Vec{u}} \mathcal{M}_{\bar{\vec{u}}}$ at the corresponding point in the solution space $\Vec{u}_{\text{cur}}$. 

The linearisation around ${\Vec{y}}_\text{cur}$ is performed in a data-based manner in the online phase, based on the reduced $\Vec{y}_i \in \mathbb{R}^d, i \in N_y$ and the original coordinates $\Vec{u}_i \in \mathbb{R}^D, i \in N_y$ of the $N$ snapshots $N_y$ nearest to ${\Vec{y}}_\text{cur}$. The search for the set of nearest neighbours $N_y$ is performed in the reduced space; pseudocode can be found in Algorithm~\ref{alg:neighbours} in \ref{s:code}. The linearisation scheme is visualised schematically in Fig.~\ref{fig:loclin}.

We are thus looking for the optimal linear mapping 
\begin{equation}
    \Delta \Vec{\bar{u}} = \ten{\varphi} \Delta \Vec{y}\label{eq:loclin}
\end{equation}
around $\vec{y}_\text{cur}$. 
Note that this linear mapping is defined with respect to the current value of the original and reduced degrees of freedom $\vec{\bar{u}}_\text{cur}$ and $\vec{y}_\text{cur}$ resulting from the previous Newton iteration, i.e.
\begin{equation*}
    \Vec{\bar{u}} - \vec{\bar{u}}_\text{cur} = \ten{\varphi} ( \Vec{y} - \vec{y}_\text{cur} )\,.
\end{equation*}
As the FEM problem is to be solved in the reduced space, $\vec{y}_\text{cur}$ is generally known while $\vec{\bar{u}}_\text{cur}$ is not.
Rearranging for $\Vec{\bar{u}}$ yields
\begin{equation}
    \Vec{\bar{u}} = \ten{\varphi} \Vec{y} - \ten{\varphi} \vec{y}_\text{cur} + \vec{\bar{u}}_\text{cur} = \ten{\varphi} \Vec{y} + \Vec{u}^0\label{eq:loclin_param}
\end{equation}
with (unknown) offset $\Vec{u}^0=\vec{\bar{u}}_\text{cur}- \ten{\varphi} {\Vec{y}}_\text{cur}$.
We now seek the $\ten{\varphi} \in \mathbb{R}^{D\times d}$ and $\Vec{u}^0 \in \mathbb{R}^D$ with which Eq.~\eqref{eq:loclin_param} optimally maps between the reduced $\Vec{y}_i \in \mathbb{R}^d, i \in N_y$ and original $\Vec{u}_i \in \mathbb{R}^D, i \in N_y$ coordinates of the $N$ nearest neighbours $N_y$ of $\vec{y}_\text{cur}$, in a least squares sense. 

For convenience, we introduce local snapshot matrices $\ten{U}_N\in \mathbb{R}^{D\times N}$ and $\ten{Y}_N\in \mathbb{R}^{d\times N}$ featuring the original and reduced coordinates of the identified neighbours $N_y$
\begin{equation*}
    \ten{U}_N = [\vec{u}_n]_{n\in N_y}, \quad \text{and} \quad
    \ten{Y}_N = [\vec{y}_n]_{n\in N_y}\,,
\end{equation*}
with which the least squares problem may be written as
\begin{equation*}
    \min_{\ten{\varphi}} f(\ten{\varphi},\Vec{u}^0) = \min_{\ten{\varphi},\Vec{u}^0} \sum_n \sum_i ( \sum_j \varphi_{ij} Y_{Njn} + u_i^0 - U_{Nin} )^2 \,.
\end{equation*}
The necessary conditions are given by
\begin{equation*}
    \frac{\partial f}{\partial \varphi_{kl}}= 0\,, \quad \text{and} \quad \frac{\partial f}{\partial u_k^0}= 0\,.
\end{equation*}
Differentiation yields
\begin{equation*}            
    \frac{\partial f}{\partial \varphi_{kl}} = \sum_n 2 Y_{Nln} (\sum_j \varphi_{kj} Y_{Njn} + u_k^0 - U_{Nkn} ) = 0
\end{equation*}
and
\begin{equation*}
    \frac{\partial f}{\partial u_k^0} = \sum_n 2 ( \sum_j \varphi_{kj} Y_{Njn} + u_k^0 - U_{Nkn} ) = 0\,.
\end{equation*}

The necessary conditions thus define a linear equation system with $D\times d + D$ equations and unknowns
\begin{align*}
    \sum_n \sum_j Y_{Nln} \varphi_{kj} Y_{Njn} + \sum_n Y_{Nln} u_k^0 - \sum_n Y_{Nln} U_{Nkn} &= 0\,,\\
    \sum_n \sum_j \varphi_{kj} Y_{Njn} + \sum_n u_k^0 - \sum_n U_{Nkn} &= 0\,,
\end{align*}
or, in matrix notation
\begin{align}            
    \ten{Y}_N \ten{Y}_N^T \ten{\varphi}^T + \ten{Y}_N \vec{\bar{1}}_N \Vec{u}^{0T}- \ten{Y}_N \ten{U}_N^T &= \ten{0}\,.\label{eq:NC1}\\
    \ten{\varphi} \ten{Y}_N \vec{\bar{1}}_N + N \Vec{u}^0 - \ten{U}_N \vec{\bar{1}}_N &= \Vec{0} \,,\label{eq:NC2}
\end{align}
where $\vec{\bar{1}}_N$ denotes a vector of ones with $N$ entries.
Elimination of $\Vec{u}^0$ via Eq.~\eqref{eq:NC2}
\begin{equation*}
    \Vec{u}^0 = \frac{1}{N} ( \ten{U}_N - \ten{\varphi} \ten{Y}_N ) \vec{\bar{1}}_N\,,
\end{equation*}
and substitution into Eq.~\eqref{eq:NC1} yield
\begin{equation*}
    \ten{Y}_N \ten{Y}_N^T \ten{\varphi}^T + \ten{Y}_N \vec{\bar{1}}_N \frac{1}{N} \vec{\bar{1}}_N^T ( \ten{U}_N^T - \ten{Y}_N^T \ten{\varphi}^T )- \ten{Y}_N \ten{U}_N^T = \ten{0}\,,
\end{equation*}
i.e.
\begin{equation*}
    \ten{Y}_N \ten{Y}_N^T \ten{\varphi}^T +  \ten{Y}_N \frac{1}{N} \ten{1}_N \ten{U}_N^T - \ten{Y}_N \frac{1}{N} \ten{1}_N \ten{Y}_N^T \ten{\varphi}^T - \ten{Y}_N \ten{U}_N^T = \ten{0}\,,
\end{equation*}
where $\ten{1}_N$ is an $N \times N$ matrix of ones.
By collecting terms, we obtain
\begin{equation*}
    ( \ten{Y}_N \ten{Y}_N^T - \ten{Y}_N \frac{1}{N} \ten{1}_N \ten{Y}_N^T) \ten{\varphi}^T = \ten{Y}_N \ten{U}_N^T - \ten{Y}_N \frac{1}{N} \ten{1}_N \ten{U}_N^T
\end{equation*}
and
\begin{equation*}
    \ten{Y}_N ( \ten{I}_N - \frac{1}{N} \ten{1}_N ) \ten{Y}_N^T \ten{\varphi}^T = \ten{Y}_N ( \ten{I}_N - \frac{1}{N} \ten{1}_N ) \ten{U}_N^T \,,
\end{equation*}
with the $N \times N$ identity matrix $\ten{I}_N$.
With the symmetric $N \times N$ weight matrix $\ten{W}_N$
\begin{equation*}
    \ten{W}_N = \ten{I}_N - \frac{1}{N} \ten{1}_N\,,
\end{equation*}
this becomes
\begin{equation*}
    \ten{Y}_N \ten{W}_N \ten{Y}_N^T \ten{\varphi}^T = \ten{Y}_N \ten{W}_N \ten{U}_N^T\,.
\end{equation*}
We can now finally solve for $\ten{\varphi}$
\begin{equation*}
    \ten{\varphi} \ten{Y}_N \ten{W}_N \ten{Y}_N^T = \ten{U}_N \ten{W}_N \ten{Y}_N^T\,,
\end{equation*}
to yield
\begin{equation}
    \ten{\varphi} = \ten{U}_N \ten{W}_N \ten{Y}_N^T (\ten{Y}_N \ten{W}_N \ten{Y}_N^T)^{-1}\,.\label{eq:loclin_mat}
\end{equation}
Note that $\ten{Y}_N \ten{W}_N \ten{Y}_N^T$ must be full-rank and invertible. To ensure that this condition is met, we must simply ensure that the number of (non-degenerate) neighbouring points exceeds the dimensionality of the reduced space, i.e. $N>d$. Pseudocode for the local linearisation can be found in Algorithm~\ref{alg:loclin} in \ref{s:code}.

The application of a locally linearised manifold learning method to projection-based MOR with the FEM is straightforward. In each solver iteration, the set $N_y$ of the $N$ nearest snapshots $\vec{y}_i \in \mathbb{R}^d, i \in N_y$ to the current point in reduced space $\Vec{y}\in \mathbb{R}^d$ are found and collected in the local snapshot matrices $\ten{U}_N = [ \vec{u}_i ]_{i\in N_y}$ and $\ten{Y}_N = [ \vec{y}_i ]_{i\in N_y}$. Then, we perform the data-based local linearisation defined by Eq.~\eqref{eq:loclin_mat}.
With the local tangent matrix $\ten{\varphi}$, incremental solutions can be sought in the approximated tangent space of the approximation space $\mathcal{M}_{\bar{\vec{u}}}$ as outlined in Eqs.~\eqref{eq:redres},~\eqref{eq:newton_red}, and~\eqref{eq:K_red}, i.e.
\begin{equation}
    \ten{K}_r \Delta \Vec{y} = - \Vec{g}_r\,,\quad \text{where} \quad \ten{K}_r = \ten{\varphi}^T \ten{K} \ten{\varphi}\,, \quad \text{and} \quad
    \Vec{g}_r = \ten{\varphi}^T \Vec{g}\,.\label{eq:loclin_red}
\end{equation}
The increment in solution space $\mathbb{R}^D$ can be approximated by the linearisation in Eq.~\eqref{eq:loclin}, whereupon the current iterate in reduced space and in solution space can be incremented via
\begin{equation}
     \Vec{y_\text{cur}} \leftarrow \Vec{y_\text{cur}} + \Delta \Vec{y}\,, \quad \text{and} \quad {\Vec{\bar{u}_\text{cur}}} \leftarrow {\Vec{\bar{u}_\text{cur}}} +  \ten{\varphi} \Delta \Vec{y}\,.\label{eq:loclin_inc}
\end{equation}

In the above procedure, the local tangent $\ten{\varphi}$ is obtained via a data-based linearisation procedure and is thus not orthonormal by construction, in contrast to the projection matrices obtained via the POD~\cite{Wei:2019:tpo}, the LPOD~\cite{AmsZahFar:2012:nmo}, and the globally linearised LEM~\cite{Pyt:2018:muo}. 
However, the mapping matrix $\ten{\varphi} \in \mathbb{R}^{D\times d}$ can be factorised into a $D \times d$ orthonormal matrix $\ten{\varphi}_\perp \in \mathbb{R}^{D\times d}$ and a $d \times d$ upper triangular matrix $\ten{R}_\perp \in \mathbb{R}^{d \times d}$ using a reduced QR-factorisation, assuming $\ten{\varphi}$ to be of rank $d$~\cite{TreBau:2022:nla}, i.e.
\begin{equation*}
    \ten{\varphi} = \ten{\varphi}_\perp \ten{R}_\perp\,.
\end{equation*}
The factorisation can be substituted into the reduced Newton system in Eq.~\eqref{eq:newton_red} 
to yield
\begin{equation*}
    \ten{R}_\perp^T \ten{\varphi}_\perp^T \ten{K} \ten{\varphi}_\perp \ten{R}_\perp \Delta \Vec{y} = \ten{R}_\perp^T \ten{\varphi}_\perp^T \Vec{g}\,.
\end{equation*}
Since the factor $\ten{R}_\perp^T$ is invertible, we can left multiply with $\ten{R}_\perp^{-T}$ to eliminate it. Furthermore, we could define a new (local) reduced increment $\Delta \Vec{y}_\perp$ such that
\begin{equation}
    \Delta \Vec{y}_\perp = \ten{R}_\perp \Delta \Vec{y}\,.\label{eq:inc_perp}
\end{equation}
to yield a new incremental problem
\begin{equation*}
    \ten{\varphi}_\perp^T \ten{K} \ten{\varphi}_\perp \Delta \Vec{y}_\perp = \ten{\varphi}_\perp^T \Vec{g}\,.
\end{equation*}
Here, $\ten{\varphi}_\perp$ can be interpreted as $d$ orthonormal (local) modes and $\Delta \Vec{y}_\perp$ as the $d$ corresponding (local) mode activity coefficient increments. These are mostly analogous to their equivalents in the case of the POD, but are only valid locally. The reduced coordinates $\Vec{y}$ obtained by nonlinear MOR then define the underlying reduced space in which neighbourhood relationships between these local orthonormal modes and their coefficients are defined. 

With the new local orthonormal modes $\ten{\varphi}_\perp$, the reduced equation system can be defined similarly to Eq.~\eqref{eq:loclin_red} as
\begin{equation}
    \ten{K}_r \Delta \Vec{y} = - \Vec{g}_r\,,\quad \text{where} \quad \ten{K}_r = \ten{\varphi}_\perp^T \ten{K} \ten{\varphi}_\perp\,, \quad \text{and} \quad
    \Vec{g}_r = \ten{\varphi}_\perp^T \Vec{g}\,.\label{eq:loclin_newton}
\end{equation}
The increment in the reduced space $\Vec{y} \in \mathbb{R}^{d}$ can be obtained from Eq.~\eqref{eq:inc_perp} via the inverse of $\ten{R} \in \mathbb{R}^{d \times d}$ which is invertible if $\ten{\varphi}$ is of rank $d$
\begin{equation}
    \Delta \Vec{y} = \ten{R}_\perp^{-1} \Delta \Vec{y}_\perp\,,\label{eq:loclin_qr_inc}
\end{equation}
and the reduced and original solutions incremented via Eq.~\eqref{eq:loclin_inc}.
Pseudocode for the resulting projection-based nonlinear MOR algorithm based on an orthonormalised, locally linearised manifold learning scheme can be found in Algorithm~\ref{alg:manl_online} in \ref{s:code}. 


\section{Two-stage projection: achieving problem size independence}\label{s:two_stage}

The nonlinear projection-based MOR scheme proposed in Section~\ref{s:loclin} satisfies many of the requirements set forth in the introduction: 
it reduces the search space in which solutions are sought to a relatively tight approximation $\mathcal{M}_{\bar{\vec{u}}}$ of the solution manifold $\mathcal{M}_{\vec{u}}$, i.e. $d \not \gg \delta$; as the manifold learning methods outlined in Sections~\ref{s:LEM} and~\ref{s:LLE} are not data-hungry, it may function with relatively few snapshots $s$; the scheme is sufficiently general to apply to a wide range of quasi-static parametric problems.
For a given desired level of accuracy, the model size $d_{\text{ManL}}$ required by a manifold learning-based NLMOR scheme, meanwhile, may be considerably smaller than that required by the POD, i.e. $d_{\text{POD}}>d_{\text{ManL}}$. The solution of the linear system of equations in Eq.~\eqref{eq:loclin_newton}, which scales with {$\mathcal{O}(d^3)$ in the worst case}, may therefore be much accelerated. 

\begin{figure}[h!]
    \centering
    \includegraphics[scale=0.8]{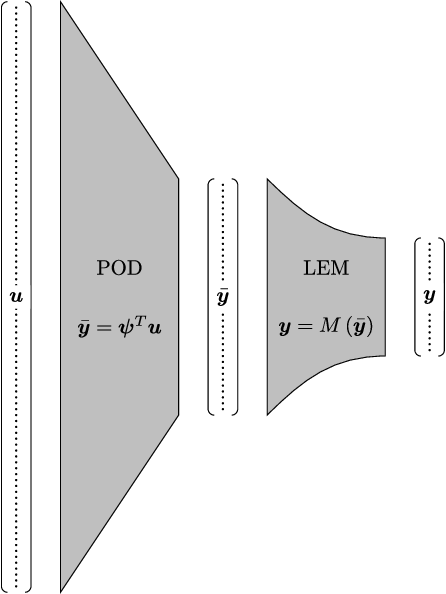}
    \caption{Schematic of two-stage reduction process using POD and manifold learning technique such as Laplacian Eigenmaps.}
    \label{fig:pod_manl}
\end{figure}

However, some components of the algorithm are still tied to the dimension of the original problem $D$, such as the local linearisation and the QR decomposition, as well as the assembly, each of which scale with $\mathcal{O}(D)$.
In this Section, we propose a two-stage MOR scheme with which independence from the problem size $D$ can be achieved. The approach is motivated by~\cite[p.231]{LeeVer:2007:ndr}: data on a nonlinear $\delta$-manifold $\mathcal{M}_{\vec{u}}$ in a $D$-dimensional space can often be linearly compressed nearly losslessly into a $\bar{d}$-dimensional intermediate space $\mathbb{R}^{\bar{d}}$, from which a nonlinear embedding can be performed to a reduced $d$-dimensional space $\mathbb{R}^d$, where $D\ll\bar{d}<d$. Thus, we use the POD to perform a linear reduction step $\Bar{M}:\mathbb{R}^D\rightarrow\mathbb{R}^{\bar{d}}$ from the $D$-dimensional solution space $\mathbb{R}^D$ to a $\bar{d}$-dimensional intermediate space $\mathbb{R}^{\bar{d}}$ and a nonlinear reduction $M:\mathbb{R}^{\bar{d}} \rightarrow \mathbb{R}^d$ (e.g., via the LEM) from the $\bar{d}$-dimensional intermediate space to the $d$-dimensional reduced space $\mathbb{R}^d$. This scheme is visualised schematically in Fig.~\ref{fig:pod_manl}. The local linearisation and QR decomposition must then no longer be performed with respect to the original solution space $\mathbb{R}^D$.
Beyond the improved scaling behaviour, the graph building, manifold learning, and local linearisation steps may also be more robust if performed with respect to the lower-dimensional intermediate space. {Of course, a hyper-reduction scheme is required to achieve the desired scaling behaviour for the assembly operation, but this is beyond the scope of this work.}

The application of such a two-stage MOR scheme is straightforward in principle: in the offline phase, a linear reduction is performed on the snapshot data $\ten{U}$ to the intermediate space $\mathbb{R}^{\bar{d}}$ via the POD, yielding a projection matrix $\ten{\psi}$ (see Section~\ref{s:POD}). 
$\ten{\psi}$ defines a mapping between the solution and intermediate spaces as well as a new snapshot matrix $\ten{\bar{Y}}$ in the intermediate space
\begin{equation*}
    \vec{\bar{y}} = \ten{\psi}^T \Vec{u}\,,\quad \text{and}\quad
    \vec{\bar{Y}}= \ten{\psi}^T \ten{U} \,.
\end{equation*}

Then, the linearly reduced solution manifold in the intermediate space $\mathcal{M}_{\Bar{\vec{y}}}=\{\Vec{\bar{y}} \mid \Vec{\bar{y}} = \ten{\psi}^T \vec{u}\,, \vec{g}(\Vec{u};\Vec{p})=\Vec{0}, \}$ is approximated with a graph; a graph adjacency matrix (and possibly weights) are computed as discussed in Section~\ref{s:LEM}, using the snapshot coordinates in the intermediate space $\ten{\bar{Y}}\in \mathbb{R}^{\bar{d}\times s}$ as nodes.
A graph-based nonlinear MOR technique is used to obtain an array of $s$ reduced $d$-dimensional snapshot vectors $\ten{Y} \in \mathbb{R}^{d \times s}$ for the snapshot coordinates in the intermediate space $\ten{\bar{Y}}\in \mathbb{R}^{\bar{d}\times s}$.

In each iteration in the online phase of such a two-stage nonlinear MOR scheme, a local linearisation can be performed as discussed in Section~\ref{s:loclin}, but with respect to the intermediate space $\mathbb{R}^{\bar{d}}$. The local snapshot matrices are now defined as
\begin{equation*}
    \ten{\bar{Y}} = [\vec{\bar{y}}_i]_{i \in N_y}\,, \quad \text{and} \quad
    \ten{Y}_N = [\Vec{y}_i]_{i \in N_y}\,,
\end{equation*}
and the local tangent $\ten{\varphi}$ from Eq.~\eqref{eq:tangent} becomes
\begin{equation*}
    \ten{\varphi} = \ten{\bar{Y}} \ten{W}_N \ten{Y}_N^T (\ten{Y}_N \ten{W}_N \ten{Y}_N^T)^{-1}\,.
\end{equation*}
As above, the mapping matrix can be orthonormalised via a reduced QR decomposition
\begin{equation*}
    \ten{\varphi}_\perp, \ten{R}_\perp \leftarrow \text{QR}(\ten{\varphi})\,.
\end{equation*}
Incremental solutions can now be sought in the approximated tangent space to the solution manifold in the intermediate space $\mathcal{T}_{\vec{\bar{y}}} \mathcal{M}_{\Bar{y}}$ via a reduced Newton scheme. The approach is analogous to that articulated in Section~\ref{s:loclin}, but with respect to the intermediate space $\mathbb{R}^{\bar{d}}$, i.e. the nonlinear reduction is performed on the reduced equations obtained via the POD as outlined in Section~\ref{s:POD}. This yields a doubly reduced Newton scheme
\begin{equation*}
    \ten{K}_r \Delta \Vec{y} = - \Vec{g}_r\,,\quad \text{where} \quad \ten{K}_r = \ten{\varphi}_\perp^T \ten{\psi}^T \ten{K} \ten{\psi} \ten{\varphi}_\perp\,, \quad \text{and} \quad
    \Vec{g}_r = \ten{\varphi}_\perp^T \ten{\psi}^T \Vec{g}\,.
\end{equation*}
The increment in reduced space can be obtained as in Eq.~\eqref{eq:loclin_qr_inc}, i.e. $\Delta \Vec{y}=\ten{R}_\perp^{-1} \Delta \Vec{y}_\perp$, and the linearised reconstruction of the increment in solution space $\mathbb{R}^D$ becomes
\begin{equation*}
    \Delta {\Vec{\bar{u}}} = \ten{\psi} \ten{\varphi} \Delta \Vec{y}\,.
\end{equation*}
Note that the local linearisation and the QR decomposition now scale with $\mathcal{O}(\bar{d})$, while the solution of the linear system scales with $\mathcal{O}(d^2)$. 
Hyper-reduction is not the focus of this proof-of concept investigation, and was not implemented for the examples considered in Section~\ref{s:examples}. See~\cite{Ryc:2009:hrm,ChaSor:2010:nmr,CarFar:2011:lcg,NegManAms:2015:emr,JaiTis:2019:hnm} for details about potential hyper-reduction schemes.

\section{Numerical Example}\label{s:examples}

\subsection{RVE computations for the FE\textsuperscript{2} method}\label{s:RVE}

In some applications, such as composite materials design~\cite{LLoGonMol:2011:mmc} and biomechanical research~\cite{BhaVic:2017:mmm}, it is desirable to model the mechanical behaviour of a macroscopic component as well as its microstructure.
Classical computational multiscale methods 
exploit a scale separation assumption to resolve macroscopic and microscopic behaviour separately; a microscale RVE is constructed to approximate the microstructure. 

In the case of the FE\textsuperscript{2} method~\cite{Sch:2014:nth}, the evaluation of a constitutive law on the macroscale is replaced by a volume averaging over the solution to the RVE problem. Other computational homogenisation schemes, such as the data-based methods proposed in \cite{LeYvoHe:2015:chn, BhaMat:2016:nmr, GuoRokVer:2021:lcm}, first build a considerable database of RVE solution data and use this data to train surrogate models for the constitutive behaviour $\ten{\bar{P}}(\ten{\bar{H}})$ exhibited by the RVE. In either case, a considerable number of computations on an RVE are required, which is why projection-based MOR methods such as those outlined in Sections~\ref{s:POD},~\ref{s:LPOD}, and~\ref{s:ManLMOR} are particularly attractive in this field.

This microscale RVE can be subjected to macroscale strain information using the macroscopic deformation gradient $\ten{\bar{F}}$ or displacement gradient $\ten{\bar{H}}$ via periodic boundary conditions: the displacement on the scale of the microstructure $\vec{\underline{u}}(\vec{X})$ is assumed to consist of a linear contribution due to the prescribed displacement gradient $\ten{\bar{H}}$ and a fluctuation contribution $\Tilde{\vec{\underline{u}}}(\vec{X})$ which is periodic on the unit cell, i.e.~\cite{Sch:2014:nth}
\begin{equation*}
    \vec{\underline{u}}(\vec{X}) = \ten{\Bar{H}}\vec{X}+\Tilde{\vec{\underline{u}}}(\vec{X})\,.
\end{equation*} 
The boundary $\partial \Omega$ of the RVE can be decomposed into two parts which are coupled due to periodicity. For the implementation, it is convenient to define an independent part {consisting of points $\vec{X}_-\in \partial \Omega_-$} and an associated dependent part {consisting of points $\vec{X}_-\in \partial \Omega_-$} on this boundary, where the behaviour of the dependent part is formulated in terms of the behaviour of the independent part as~\cite{Sch:2014:nth}
\begin{equation*}
    \Tilde{\vec{\underline{u}}}(\vec{X}_+) = \Tilde{\vec{\underline{u}}}(\vec{X}_-)\,, \quad \text{and} \quad \vec{t}(\vec{X}_+) = -\vec{t}(\vec{X}_-)\,.
\end{equation*}
Note that periodic boundary conditions satisfy Hill's micro-macro heterogeneity condition~\cite{Hil:1965:smc}.

With this formulation, the nodal degrees of freedom discretising $\Tilde{\vec{\underline{u}}}(\vec{X})$ become the primary unknowns to the RVE problem to be solved for using the FEM and a Newton-Raphson scheme as in Eq.~\eqref{eq:newton}.
The homogenised stress and stiffness response of the RVE to the load can then be used to obtain information about the effective constitutive behaviour of the inhomogeneous material at the considered point on the macroscale, as characterised by the first Piola-Kirchhoff stress $\ten{\bar{P}}$ and nominal constitutive modulus~$\ten{\bar{\mathbb{A}}}$~\cite{SaeSteJav:2016:ach}.

\subsection{Example RVE computation}\label{s:RVE_prob}

\begin{figure}[h!]
\centering
\begin{minipage}[t]{.45\textwidth}
    \centering
    \includegraphics[trim=16cm 2.5cm 14cm 6cm,clip,scale=0.2]{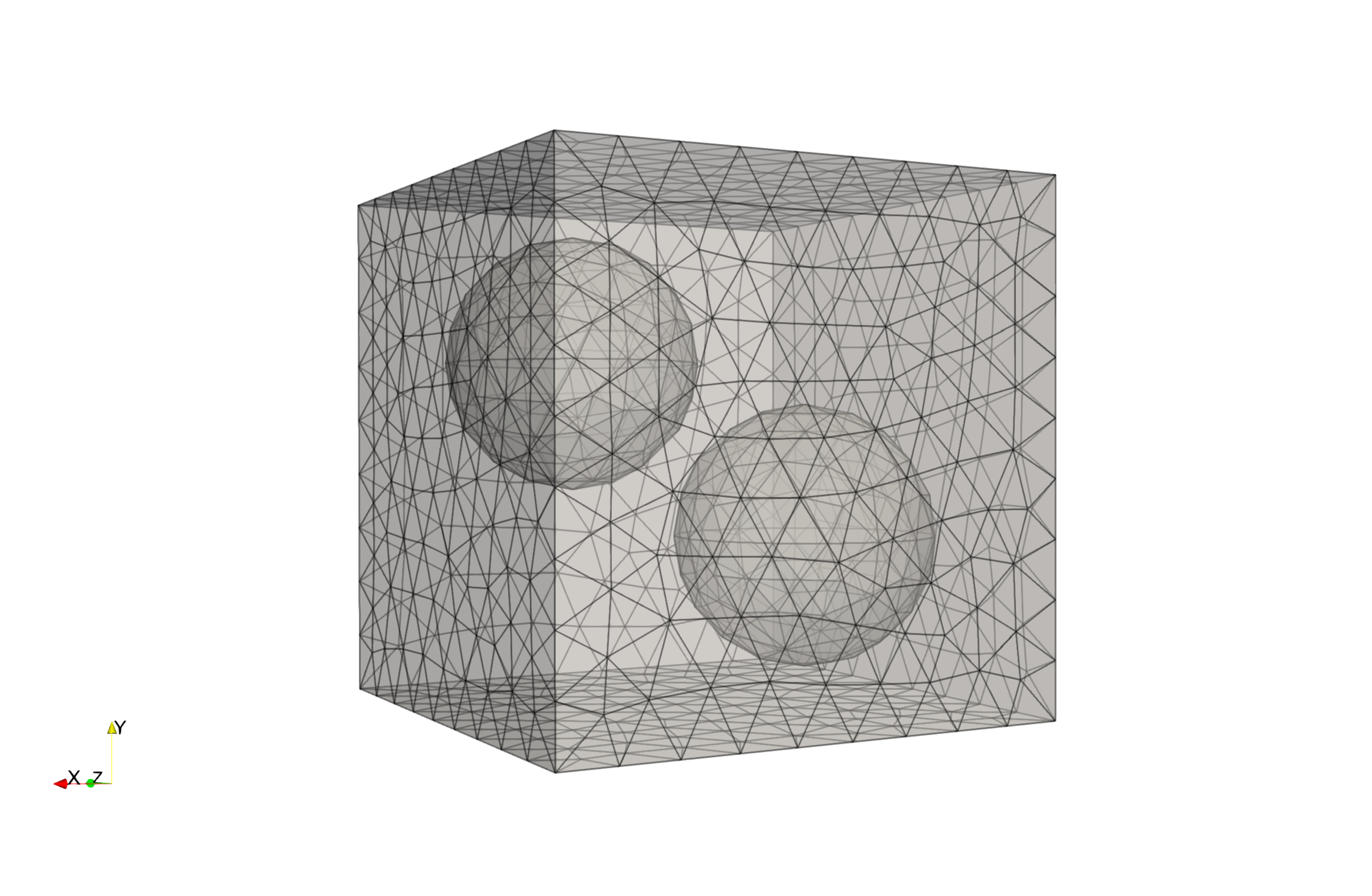}
    \captionof{figure}{Discretisation of artificial porous RVE with quadratic tetrahedral elements.}
    \label{fig:RVE_discretised}
\end{minipage}%
\hspace{0.5cm}
\begin{minipage}[t]{.45\textwidth}
    \centering
    \includegraphics[trim=16cm 2cm 12cm 6cm,clip,scale=0.2]{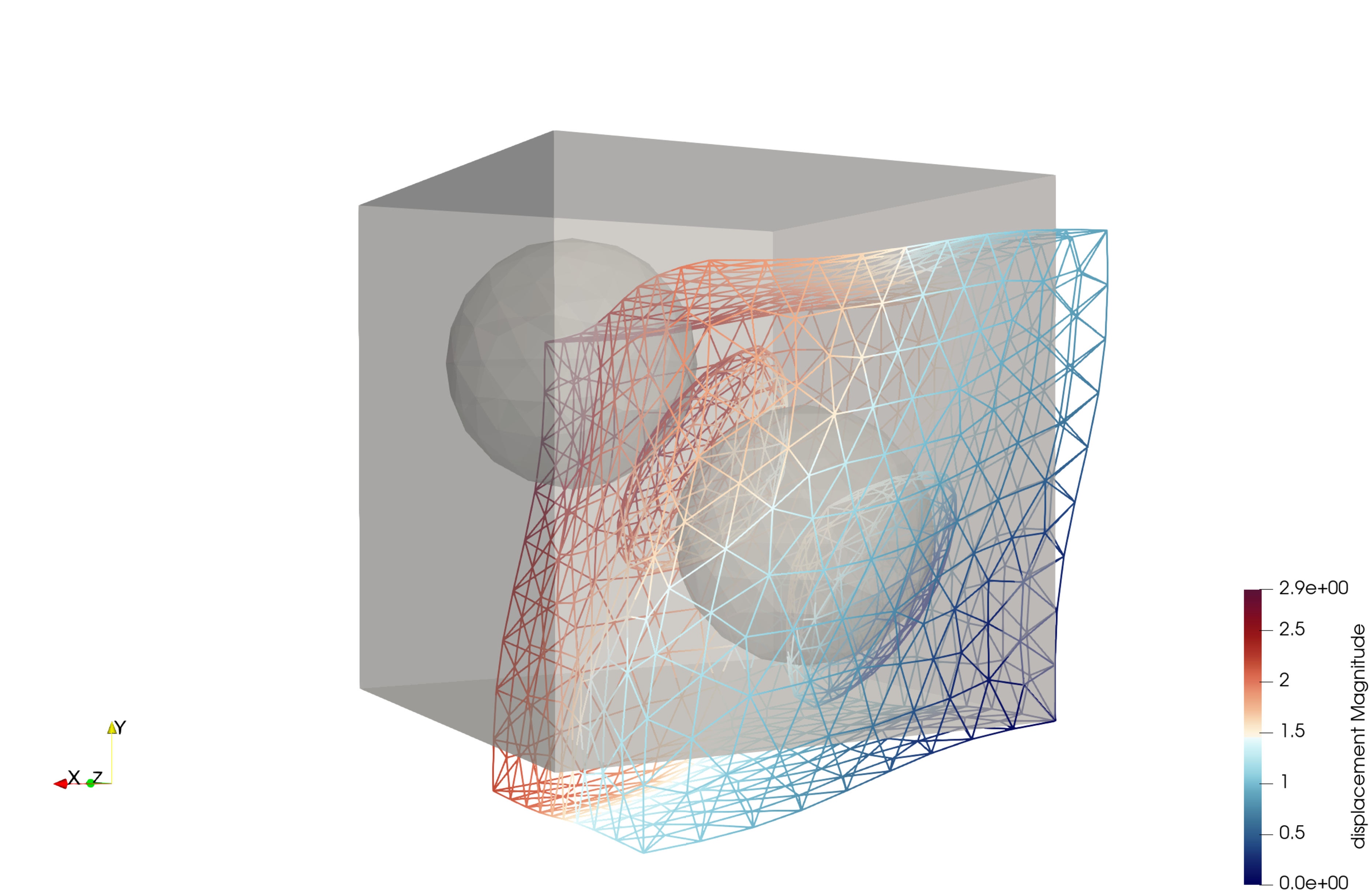}
    \captionof{figure}{Example deformation in RVE at the end of a load path.}
    \label{fig:RVE_deformed}
\end{minipage}
\end{figure}

In this Section, we test the MOR methods outlined above on an exemplary RVE problem.
The artificial RVE used to this end features two spherical pores, as illustrated in Fig.~\ref{fig:RVE_discretised}. The pores are modelled as traction-free voids, such that the RVE features two interior boundaries with homogeneous Neumann boundary conditions. Note that the homogenisation of stresses and stiffnesses becomes slightly more complicated in this special case due to the presence of interior boundaries, but the periodic boundary conditions outlined above satisfy Hill's macro homogeneity condition nonetheless~\cite{Sch:2000:hnk,Sch:2014:nth}. For the investigations outlined below, we use a coarse discretisation with $1388$ elements and $2610$ nodes ($7830$ degrees of freedom, problem A), as well as a moderately fine discretisation with $3998$ elements and $6973$ nodes ($20,919$ degrees of freedom, problem B). The latter is illustrated in Fig.~\ref{fig:RVE_discretised}. This ultimately yields $D=6096$ and $D=17,151$ independent degrees of freedom, with dependent degrees of freedom on the periodic boundaries not considered. Quadratic tetrahedral elements are utilised in both problem A and problem B.
The behaviour of the matrix material is defined via the neo-Hookean stored energy function in Eq.~\eqref{eq:nH}, with a Young's modulus $E=1000$ and Poisson's ratio $\nu=0.2$. Tab.~\ref{tab:general_params} summarises the parameters relevant to the problem definition. 

\begin{figure}[h!]
\centering
\begin{minipage}[t]{.45\textwidth}
    \centering
    \includegraphics[scale=0.2,trim={20cm 5cm 15cm 5cm},clip]{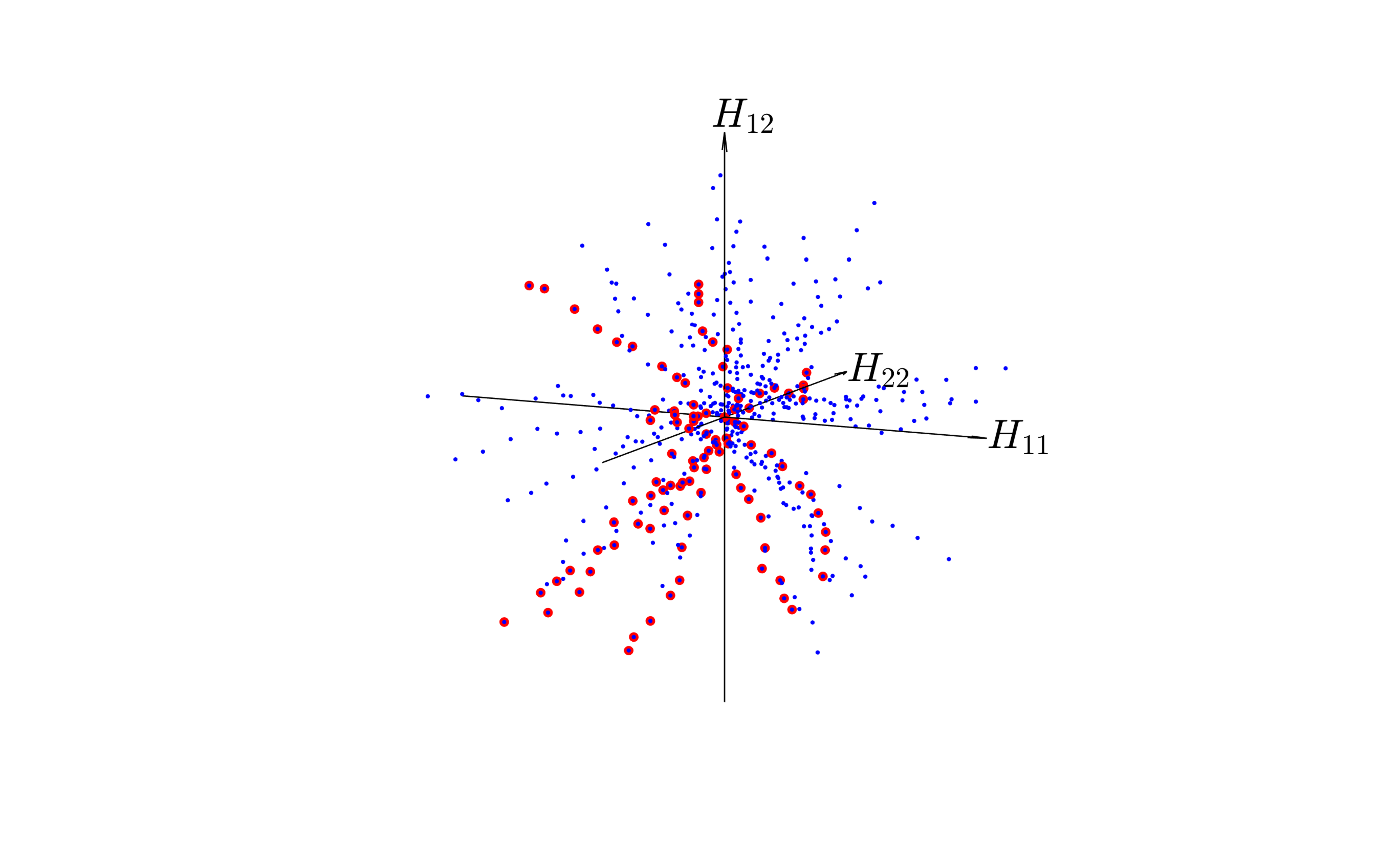}
    \captionof{figure}{Three entries of macroscopic displacement gradient $\ten{\bar{H}}$ for training (red) and validation (blue) load paths.}
    \label{fig:RVE_loadpaths}
\end{minipage}%
\hspace{0.5cm}
\begin{minipage}[t]{.45\textwidth}
    \centering
    \includegraphics[scale=0.85]{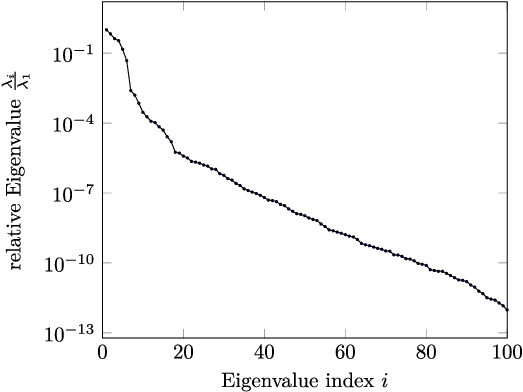}
    \captionof{figure}{Eigenvalue decay of the correlation matrix $\ten{C}$ associated with the example RVE problem.}
    \label{fig:EV_decay}
\end{minipage}
\end{figure}

To emulate the use case in the FE\textsuperscript{2} method, the RVE is subjected to periodic boundary conditions. 
Different boundary value problems are defined representing specific macroscopic loading scenarios via a macroscopic displacement gradient $\ten{\bar{H}}$. These values of the deformation gradient are defined in an artificial manner and do not correspond to any previous simulation but represent possible load paths of a macroscopic point represented by the RVE.
Each load path is defined via $10$ load steps. Increments $\Delta \ten{\bar{H}}$ between successive values of $\ten{\bar{H}}$ along each load path follow a randomly sampled normalised load direction $\ten{N}_{\text{LP}}$ which remains constant along the load path with step length $\Delta H_{\text{LP}}=0.03$. A perturbation following a randomly sampled normalised load direction $\ten{N}_{\text{LS}}$ updated in each load step with step length $\Delta H_{\text{LS}}=0.015$ is additionally applied, such that
\begin{equation*}
    \Delta \ten{\bar{H}} = \Delta H_{\text{LP}} \ten{N}_{\text{LP}} + \Delta H_{\text{LS}} \ten{N}_{\text{LS}}\,, \quad \text{and} \quad
    \ten{\bar{H}} \leftarrow \ten{\bar{H}} + \Delta\ten{\bar{H}}\,.
\end{equation*}
Three example components of displacement gradient load paths generated this way are shown in Fig.~\ref{fig:RVE_loadpaths}. A resulting example deformation in the RVE at the end of one of these load paths is shown to scale in Fig.~\ref{fig:RVE_deformed}.

\begin{table}[h!]
    \centering
    \begin{tabular}{ c c c }
    \hline
    parameter & variable & value  \\
    \hline
    Young's modulus & $E$ & 1000 Nmm\textsuperscript{-2} \\
    Poisson's ratio & $\nu$ & 0.2 \\
    RVE edge length & & 6 mm \\
    pore radius & & 1.5 mm \\
    centre pore 1 ($x,y,z$) &  & (2,2,2) mm \\
    centre pore 2 ($x,y,z$) &  & (4,4,4) mm \\
    snapshot number & $s$ & 100 \\
    validation set size & & 500 \\
    random seed & & 42 \\
    step size & $\Delta {H}_{\text{LP}}$ & 0.03 \\
    perturbation size & $\Delta {H}_{\text{LS}}$ & 0.015 \\
    \hline
    low-fidelity: & & \\
    independent DOFs & $D$ & 6096 \\
    element number & $n_\text{elem}$ & 1388 \\
    node number & $n_\text{node}$ & 2610 \\
    \hline 
    higher-fidelity & & \\
    independent DOFs & $D$ & 17,151  \\ 
    element number & $n_\text{elem}$ & 3988\\
    node number & $n_\text{node}$ & 6973 \\
    \hline
\end{tabular}
    \caption{Problem and simulation parameters for the numerical experiments outlined below. The origin of the coordinate system within the RVE lies in a corner, such that $x_\text{min}=y_\text{min}=z_\text{min}=0$ and $x_\text{max}=y_\text{max}=z_\text{max}=6$.}
    \label{tab:general_params}
\end{table}

The boundary value problem defined above is then used to generate RVE solutions via full-FEM computations. As discussed below, a subset of these is used as snapshots to construct reduced order models and the remainder to validate the results:
first, the nodal displacement $\Vec{u}$ and displacement fluctuation $\tilde{\Vec{u}}$ solutions are computed for each displacement gradient value $\ten{\bar{H}}$ along $50$ load paths, leading to $500$ solutions. Solutions $\Vec{u}_i^\text{TRAIN}, \tilde{\Vec{u}}_i^\text{TRAIN}, i \in [1,100]$ for the $10$ load steps along $10$ of these load paths (which are shown in red in Fig~\ref{fig:RVE_loadpaths}) are later utilised as training data for the ROMs. Solutions $\Vec{u}_i^\text{VAL}, \tilde{\Vec{u}}_i^\text{VAL}, i \in [1,500]$ for the $10$ load steps along the remaining $40$ load paths (which are shown in blue in Fig~\ref{fig:RVE_loadpaths}) are used for validation, alongside the original $10$ load paths. While there are sophisticated sampling strategies to generate ROM training data~\cite{BenGugWil:2015:spm,BhaMat:2020:ndr,FriKun:2018:tdh}, we simply sample far from exhaustively in a random fashion. Note that, in Fig.~\ref{fig:RVE_loadpaths}, there are significant unsampled areas in between the red training load paths, with only three components of the nine-dimensional displacement gradient space shown. If the whole nine-dimensional displacement gradient space could be shown, this whitespace of unsampled regions between the training load paths would appear even more extensive. While not optimal in obtaining accurate ROMs, this ad-hoc strategy is intended to test the robustness of the algorithms outlined above.

A snapshot matrix $\ten{U} \in \mathbb{R}^{D\times s}, s=100$ is assembled from the nodal displacement fluctuation solutions $\tilde{\Vec{u}}_i^\text{TRAIN}, i \in [1,100]$ along the $10$ training load paths. 
As shown in Fig.~\ref{fig:EV_decay}, the Eigenvalues of the associated snapshot correlation matrix $\ten{C}$ (see Eq.~\eqref{eq:POD_EV}) exhibit a rather gradual decay. This suggests that the snapshot data $\ten{U}$, and thus the $\delta=6$-dimensional solution manifold $\mathcal{M}_{\vec{u}}$, does not lie in or close to a low-dimensional linear subspace of the solution space $\mathbb{R}^D$. Hence, it may prove beneficial to utilise nonlinear approximation spaces when performing MOR on this RVE problem. Note that, if $\delta$ were not known a priori, an estimate could be obtained via the correlation dimension, as outlined in ~\ref{s:corr_dim}.

The snapshot data $\ten{U}$ is used to train POD, LPOD, LEM, and LLE ROMs, where the same set of $10$ load paths is used to train each method. 
Note that for the LEM and the LLE, we employ once the global linearisation from~\cite{PytAbe:2016:nmr} (abbreviated as `global lin.' below) and once the local linearisation outlined in Section~\ref{s:loclin} (abbreviated as `local lin.' below). 
Using these ROMs, displacement solutions $\Vec{\bar{u}}i^\text{ROM},i\in[1,500]$ are calculated along all validation load paths.
The displacement results obtained via the ROMs are compared to those obtained from full FEM reference simulations $\Vec{u}_i^\text{VAL},i\in[1,500]$, and maximal as well as mean relative modeling errors computed, i.e.
\begin{equation}
    E_\text{max} = \max_i \left ( \frac{\|\Vec{\bar{u}}_i^\text{ROM}-\vec{u}_i^\text{VAL}\|}{\|\Vec{u}_i^\text{VAL}\|} \right )\,, \quad \text{and} \quad   E_\text{mean} = \frac{1}{500} \sum_{i=1}^{500} \frac{\|\Vec{\bar{u}}_i^\text{ROM}-\vec{u}_i^\text{VAL}\|}{\|\Vec{u}_i^\text{VAL}\|}\,.\label{eq:error}
\end{equation}

Furthermore, parameters for the LEM, LLE, and LPOD, which are used throughout all simulations, except where the contrary is explicitly stated, are shown in Tab.~\ref{tab:LEM_params}, Tab.~\ref{tab:LLE_params}, and Tab.~\ref{tab:LPOD_params}, respectively. The model parameters used for the LEM and LLE outlined therein were motivated by a priori considerations: e.g., the neighbours considered in graph building and local linearisation are chosen to capture neighbourhood information on the solution manifold, the dimension $\delta$ of which can be estimated from the training data as outlined in ~\ref{s:corr_dim}. In the case of the LPOD, the parameters were chosen so as to produce low reproduction errors between the original $\ten{U}$ and reconstructed snapshot data $\ten{\bar{U}}$ for the examples investigated in the following. Detailed parameter optimisation might yield improved results, but in order for this to yield robust results, a broader range of test cases would be required.

\begin{table}[h!]
    \centering
    \begin{tabular}{ c c c }
    \hline
    parameter & variable & value  \\
    \hline
    graph method & & symmetric $k$ nearest neighbours \\
    linearisation method & & local linearisation \\
    orthonormalisation & & orthonormalisation \\
    embedding dimension & $d$ & 15 \\
    graph neighbours & k & 30 \\
    Gaussian weight & $t$ & $\infty$ \\
    linearisation neighbours & n & 20 \\
    \hline
    graph connectivity: & & \\
    min connectivity & & 31 \\
    lqt connectivity & & 32 \\
    median connectivity & & 37 \\
    uqt connectivity & & 56 \\
    max connectivity & & 86 \\
    \hline
\end{tabular}
    \caption{Standard parameters for the \textbf{LEM}. Unless the contrary is explicitly stated, these parameters are used for all numerical experiments outlined below.}
    \label{tab:LEM_params}
\end{table}

\begin{table}[h!]
    \centering
    \begin{tabular}{ c c c }
    \hline
    parameter & variable & value  \\
    \hline
    graph method & & symmetric $k$ nearest neighbours \\
    linearisation method & & local linearisation \\
    orthonormalisation & & orthonormalisation \\
    embedding dimension & $d$ & 15 \\
    graph neighbours & k & 30 \\
    conditioning coefficient & $\Delta$ & 0.001 \\
    linearisation neighbours & n & 20 \\
    \hline
    graph connectivity: & & \\
    min connectivity & & 31 \\
    lqt connectivity & & 32 \\
    median connectivity & & 37 \\
    uqt connectivity & & 56 \\
    max connectivity & & 86 \\
    \hline
\end{tabular}
    \caption{Standard parameters for the \textbf{LLE}. Unless the contrary is explicitly stated, these parameters are used for all numerical experiments outlined below.}
    \label{tab:LLE_params}
\end{table}

\begin{table}[h!]
    \centering
    \begin{tabular}{ c c c }
    \hline
    parameter & variable & value  \\
    \hline
    clustering method & & Lloyd's algorithm \\
    embedding dimension & $d$ & 15 \\
    cluster number & $k$ & 6 \\
    softening ratio & $r$ & 1 \\
    min. core cluster size & $|\mathcal{C}|_\text{core,min}$ & 7 \\
    min. cluster size & $|\mathcal{C}|_\text{min}$ & 30 \\
    max. cluster size & $|\mathcal{C}|_\text{max}$ & 50 \\
    \hline
\end{tabular}
    \caption{Standard parameters for the \textbf{LPOD}. Unless the contrary is explicitly stated, these parameters are used for all numerical experiments outlined below.}
    \label{tab:LPOD_params}
\end{table}

\subsection{Single-stage MOR}

\begin{figure}[h!]
\centering
\begin{subfigure}[t]{.45\textwidth}
  \centering
  \includegraphics[scale=0.9]{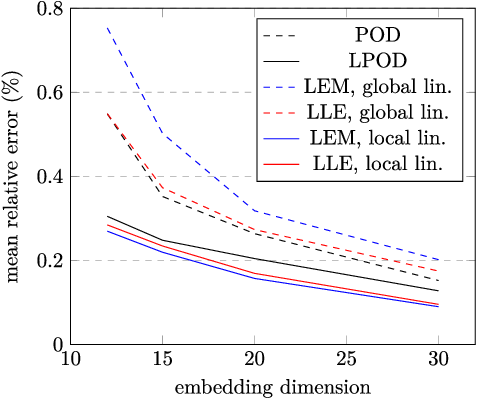}
  \caption{Mean percentage modeling error $E_{\text{mean}}$ over embedding dimension $d$.}
  \label{fig:RVE_ed1}
\end{subfigure}%
\hspace{0.5cm}
\begin{subfigure}[t]{.45\textwidth}
  \centering
  \includegraphics[scale=0.9]{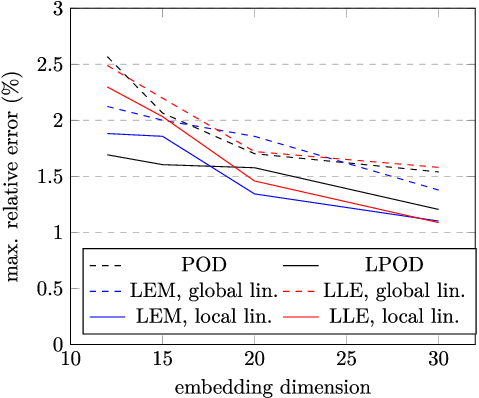}
  \caption{Maximal percentage modeling error $E_{\text{max}}$ over embedding dimension $d$. }
  \label{fig:RVE_ed2}
\end{subfigure}
\caption{\textbf{Basic results with orthonormalisation:} Mean and maximal $L2$ norm of the modeling error in the displacement field produced by reduced FEM simulations with different embedding dimensionalities $d$ with respect to reference displacement fields produced by full FEM simulations, over all validation load paths. Results are shown for MOR via POD (black, dashed lines) and LPOD (black, solid lines) as well as globally (dashed lines) and locally (solid lines) linearised, orthonormalised LEM (blue lines) and LLE (red lines). The lower-fidelity discretisation variant was used for both the full-FEM and the reduced FEM simulations.}
\label{fig:RVE_ed}
\end{figure}

In Fig.~\ref{fig:RVE_ed}, the mean and maximal relative modeling error $E_{\text{mean}}$ and $E_{\text{max}}$ in the displacement field computed via Eq.~\eqref{eq:error} are shown over a range of values for the embedding dimensionality $d$, for the POD, LPOD, and globally as well as locally linearised LEM and LLE (with orthonormalisation). 

In terms of the mean relative modeling error, the globally linearised manifold learning methods fail to outperform the POD for all model sizes. This is perhaps unsurprising, since the POD yields the optimal linear subspace based on the available snapshot data in terms of the reconstruction error, contrary to the linear subspaces obtained when globally linearising the LEM or LLE. The LPOD outperforms the POD by a moderate amount, achieving relative reductions in the modeling error in the order of magnitude of around 20\% to 30\% for various model sizes. The locally linearised, orthonormalised LEM and LLE achieve a further improvement, relatively outperforming the POD by approximately 35\% on this measure, over all model sizes. At higher $d$, the manifold learning techniques outperform the LPOD more noticeably. The mean relative modeling error levels achieved by the POD, LPOD, LEM, and LLE in this example -- in the case of the LLE, for example, between 0.3\% and 0.1\% for model sizes $d$ between 12 and 30, with only 10 training load paths -- are generally encouraging.

Alternatively, performance can be assessed in terms of the model dimensionality required to achieve a set level of accuracy. To reach a mean modeling error level of 0.2\%, for example, the POD requires a model size $d$ of approximately 25, the LPOD requires 21, and the locally linearised LEM and LLE approximately 16. This indicates that the locally linearised manifold learning techniques indeed enable a closer parameterisation of the solution manifold and might ultimately enable smaller and thus faster reduced models.

The maximal relative modeling error levels achieved by the ROMs are similarly encouraging, varying in the range of 1\% to 2.5\%. As shown in Fig.~\ref{fig:RVE_ed2}, the error reduction with increased model size $d$ proceeds at less uniform a rate. Furthermore, the maximal modeling error levels over all model sizes lie more closely together than the mean levels, relatively speaking. The POD and the globally linearised LEM and LLE perform similarly on this metric, in this example, while the locally linearised LEM and LLE are outperformed by the LPOD at low model dimensionalities $d=12$ and $d=15$, but outperform it at $d=20$ and $d=30$. LPOD, LEM, and LLE generally outperform the globally linear MOR methods, by margins of around 35\%. Meanwhile, to attain a maximal relative modeling error level of $1.5\%$, the POD requires a model size of $d>30$, the LPOD $d\approx 22$, the LLE $d\approx 20$, and the LEM $d\approx 18$, again highlighting how manifold learning techniques enable the construction of smaller reduced models.

\begin{figure}[h!]
\centering
\begin{subfigure}[t]{.45\textwidth}
  \centering
  \includegraphics[scale=0.9]{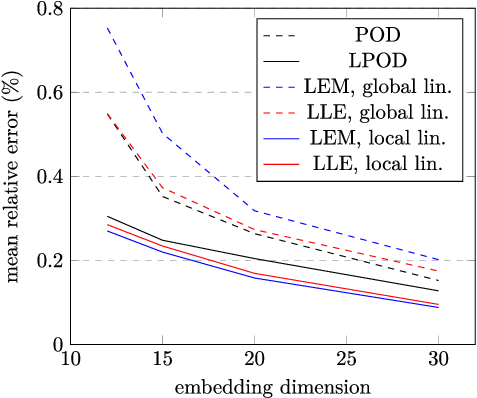}
  \caption{Mean percentage modeling error $E_{\text{mean}}$ over embedding dimension $d$.}
  \label{fig:RVE_noQR_ed1}
\end{subfigure}%
\hspace{0.5cm}
\begin{subfigure}[t]{.45\textwidth}
  \centering
  \includegraphics[scale=0.9]{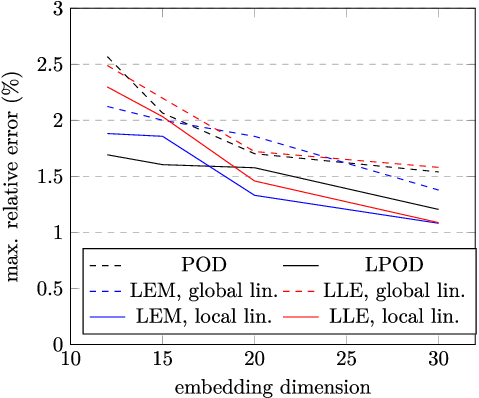}
  \caption{Maximal percentage modeling error $E_{\text{max}}$ over embedding dimension $d$. }
  \label{fig:RVE_noQR_ed2}
\end{subfigure}
\caption{\textbf{Basic results without orthonormalisation:} Mean and maximal $L2$ norm of the modeling error in the displacement field produced by reduced FEM simulations with different embedding dimensionalities $d$ with respect to reference displacement fields produced by full FEM simulations, over all validation load paths. Results are shown for MOR via POD (black, dashed lines) and LPOD (black, solid lines) as well as globally (dashed lines) and locally (solid lines) linearised, but not orthonormalised LEM (blue lines) and LLE (red lines). The lower-fidelity discretisation variant was used for both the full-FEM and the reduced FEM simulations.}
\label{fig:RVE_noQR_ed}
\end{figure}

Equivalent results for the locally linearised LEM and LLE without orthonormalisation are shown in Fig.~\ref{fig:RVE_noQR_ed}. Again, the POD, LPOD, and the globally linearised LEM and LLE are shown for reference. The results are almost identical to those obtained with orthonormalisation in Fig.~\ref{fig:RVE_ed}. In the example considered here, the solutions obtained via the ROMs appear not to be strongly influenced by whether an orthonormal projection matrix as in Eq.~\eqref{eq:loclin_newton} or the approximate tangent to the reconstruction as in Eq.~\eqref{eq:loclin_red} is used for reduction. In the case of the LLE, the use of a QR decomposition appears to slightly reduce the number of iterations the Newton scheme requires to reach these solutions. More research is required to investigate the impact of this phenomenon on the overall online cost. In Section~\ref{s:parameter_study}, some preliminary timing results are shown.

Overall, these results suggest that there is a measurable benefit to approximating the solution manifold of this example RVE problem more closely via manifold learning techniques. The nonlinear MOR techniques outlined in Sections~\ref{s:ManLMOR},~\ref{s:LEM},~\ref{s:LLE},and~\ref{s:loclin} generally require smaller ROMs to obtain a set modeling error level and obtain higher levels of accuracy for a given model size.

\begin{figure}[h!]
\centering
\begin{subfigure}[t]{.45\textwidth}
    \centering
    \includegraphics[scale=0.2,trim={18cm 5cm 18cm 5cm},clip]{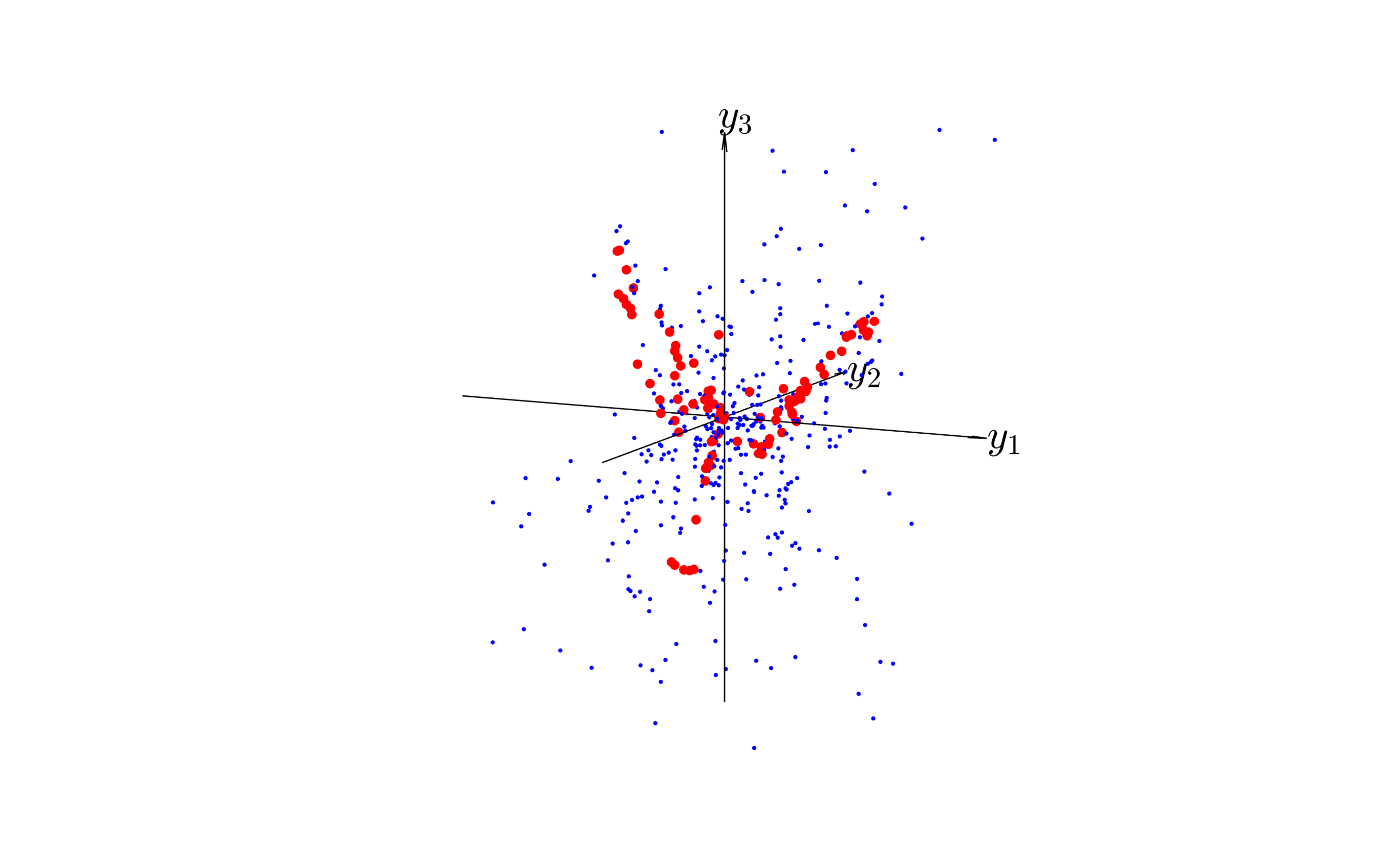}
    \caption{First three components of embedding obtained via LEM and first three components of reduced solutions obtained using locally linearised LEM for model reduction.}
    \label{fig:LEM_embed}
\end{subfigure}%
\hspace{0.5cm}
\begin{subfigure}[t]{.45\textwidth}
    \centering
    \includegraphics[scale=0.2,trim={18cm 5cm 18cm 5cm},clip]{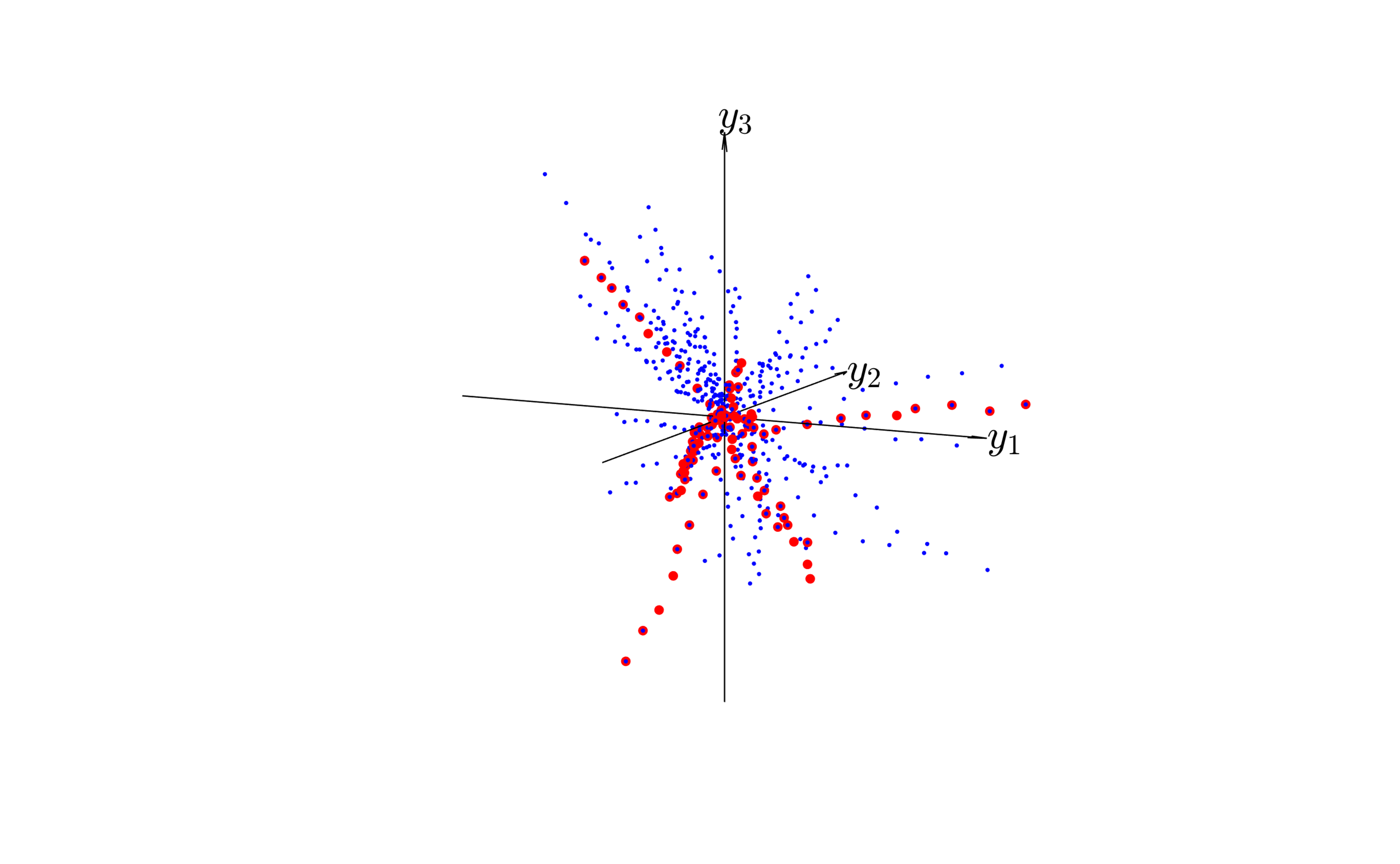}
    \caption{First three components of embedding obtained via LLE, and first three components of reduced solutions obtained using locally linearised LLE for model reduction.}
    \label{fig:LLE_embed}
\end{subfigure}
\caption{First three components of embedding $\ten{Y} \in \mathbb{R}^{d\times s}$ for displacement snapshots $\ten{U} \in \mathbb{R}^{D\times s}$ obtained via manifold learning techniques (in red). Additionally, the first three components of the reduced solutions $\Vec{y} \in \mathbb{R}^d$ obtained via the manifold learning ROMs are shown (in blue).}
\label{fig:embed}
\end{figure}

\begin{figure}[h!]
\centering
\begin{subfigure}[t]{.45\textwidth}
    \centering
    \includegraphics[scale=0.4]{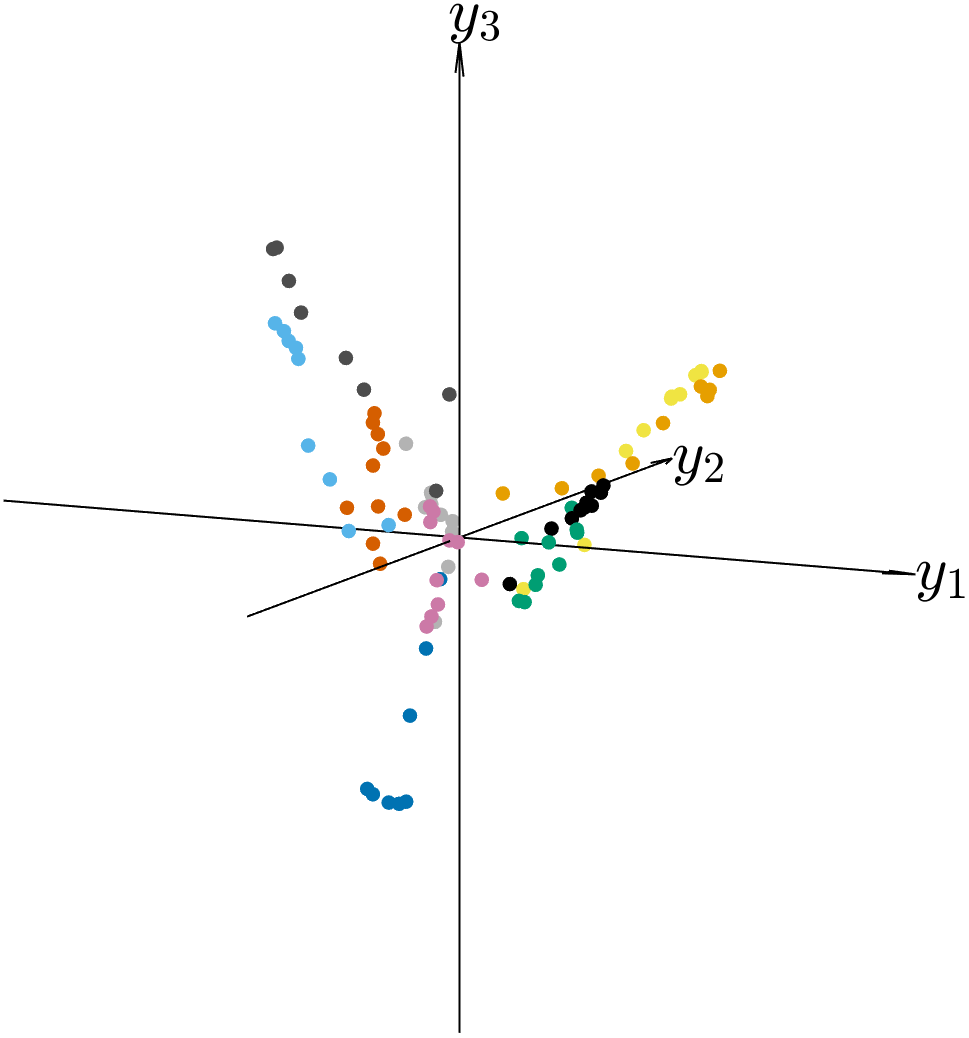}
    \caption{First three components of embedding obtained via LEM.}
    \label{fig:LEM_embed_lp}
\end{subfigure}%
\hspace{0.5cm}
\begin{subfigure}[t]{.45\textwidth}
    \centering
    \includegraphics[scale=0.4]{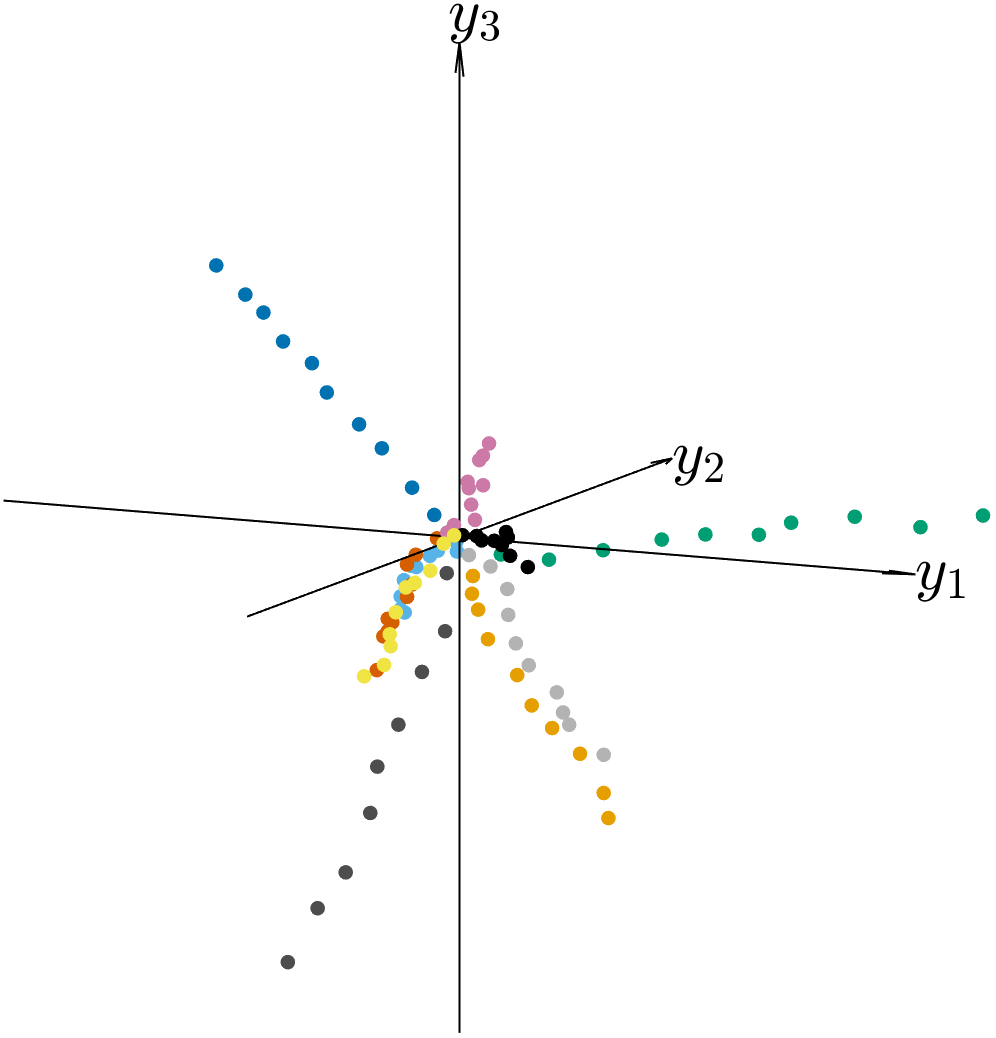}
    \caption{First three components of embedding obtained via LLE.}
    \label{fig:LLE_embed_lp}
\end{subfigure}
\caption{First three components of embedding $\ten{Y} \in \mathbb{R}^{d\times s}$ for displacement snapshots $\ten{U} \in \mathbb{R}^{D\times s}$ obtained via manifold learning techniques. All points belonging to the same load path are shown in the same color.}
\label{fig:embed_lp}
\end{figure}

Finally, Fig.~\ref{fig:embed} displays the first three components of the embeddings obtained via the LEM and the LLE, in red. Both of these are spectral methods which return additional embedding coordinates in order of decreasing information content, meaning that the first three components are the three most relevant to fulfilling the embedding criteria around which the respective manifold learning technique is constructed. Furthermore, Fig.~\ref{fig:embed} shows the first three components of the reduced solutions obtained via the LEM and LLE in blue. In addition, Fig.~\ref{fig:embed_lp} displays the first three components of the embeddings obtained via LEM and LLE, with each load path shown in a different colour. As seen in Fig.~\ref{fig:LLE_embed} and Fig.~\ref{fig:LLE_embed_lp}, the LLE produces an embedding $\ten{Y}$ for the snapshot data $\ten{U}$ which clearly captures the load path structure of the data -- see Fig.~\ref{fig:RVE_loadpaths} for comparison -- even though the parametric inputs shown in Fig.~\ref{fig:RVE_loadpaths} are not available to the manifold learning method. Similarly, the reduced solutions $\Vec{y}$ found via the LLE, which are shown in blue in Fig.~\ref{fig:LLE_embed}, feature a clear load path structure. In contrast, the LEM produces an embedding featuring no clearly visible load path structure, and the resulting reduced solutions do not share such a structure, either. Note, however, that the LEM does not perform significantly worse than the LLE and mostly outperforms the LPOD in terms of the accuracy obtained for different embedding dimensionalities, suggesting that the embedding nevertheless captures structural features of the solution manifold critical to the solution procedure. The unintuitive embedding may be due firstly to the LEM ensuring that neighbouring snapshots remain close, but not that distant snapshots remain distant. Fig.~\ref{fig:LEM_embed_lp} indicates that this is indeed the case as load steps belonging to the same load path generally cluster closely, but different clusters also feature considerable overlaps. For further illustrations of this phenomenon, see e.g. the examples in~\cite{LeeVer:2007:ndr}. Furthermore, an unweighted graph was used for these simulation runs (see Tab.~\ref{tab:LEM_params}), meaning that only proximity and not distance information are available to the LEM. In the parameter study in Section~\ref{s:parameter_study}, alternative Gauss weighting factors, as well as variations of further parameters, are investigated briefly.

\subsection{Two-stage MOR scheme}

\begin{figure}[h!]
\centering
\begin{subfigure}[t]{.45\textwidth}
  \centering
  \includegraphics[scale=0.9]{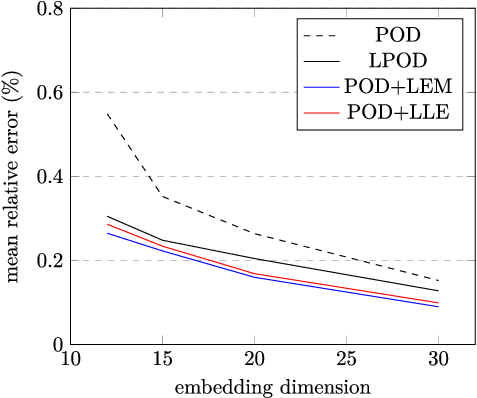}
  \caption{Mean percentage modeling error $E_{\text{mean}}$ over embedding dimension $d$.}
  \label{fig:RVE_POD_ed1}
\end{subfigure}%
\hspace{0.5cm}
\begin{subfigure}[t]{.45\textwidth}
  \centering
  \includegraphics[scale=0.9]{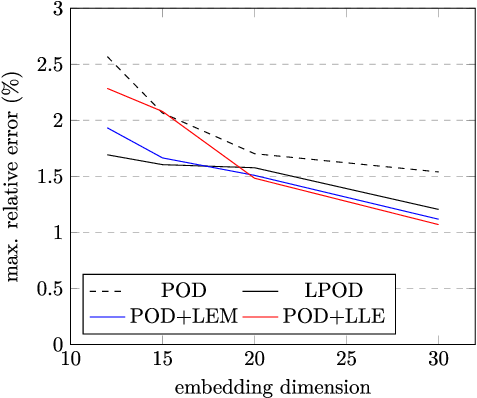}
  \caption{Maximal percentage modeling error $E_{\text{max}}$ over embedding dimension. }
  \label{fig:RVE_POD_ed2}
\end{subfigure}
\caption{\textbf{Results for two-stage MOR scheme:} Mean and maximal $L2$ norm of the modeling error in the displacement field produced by reduced FEM simulations with different embedding dimensionalities $d$ with respect to reference displacement fields produced by full FEM simulations, over all validation load paths. Results are shown for MOR via POD (black, dashed lines) and LPOD (black, solid lines) as well as locally linearised LEM (blue lines) and LLE (red lines), with a first POD MOR stage. The lower-fidelity discretisation variant was used for both the full-FEM and the reduced FEM simulations.}
\label{fig:RVE_POD_ed}
\end{figure}

Fig.~\ref{fig:RVE_POD_ed} displays the results obtained via the two-stage MOR scheme introduced in Section~\ref{s:two_stage} for the locally linearised LEM and LLE, with the POD and LPOD again shown for reference. Both in terms of the mean and the maximal relative modeling error obtained for given model dimensionalities, the performance trends are almost identical to those observed with a single-stage MOR scheme. 
While the LPOD outperforms the POD by around 20\% to 30\%, the two-stage MOR schemes achieve reductions of approximately 35\% in terms of the modeling error achieved using a given model size. In terms of the model dimensionality required to achieve a mean modeling error level of 0.2\%, the POD and the LPOD require a model sizes $d$ of approximately 25 and 21, and the two-stage schemes require approximately 16.
This suggests that, in this example, little information is lost in the first, linear data compression stage performed via the POD. Consequently, components of the nonlinear MOR algorithms outlined in Section~\ref{s:ManLMOR} such as the local linearisation and the QR decomposition may be formulated with respect to the intermediate space, leading to an improved scaling behaviour of $\mathcal{O}(\bar{d})$ for the QR decomposition and the local linearisation with little cost in terms of accuracy.

\section{Parameter investigation}\label{s:parameter_study}

This Section outlines results of a simple parameter study investigating the influence of some algorithm and problem parameters on the performance of the nonlinear MOR techniques proposed in this article. In this proof-of-concept investigation, we are interested in the qualitative influence of these parameters rather than performance optimisation. For simplicity, we therefore vary one factor at a time. The parameters outlined in Tab.~\ref{tab:general_params},~\ref{tab:LEM_params},~\ref{tab:LLE_params}, and~\ref{tab:LPOD_params} which were selected via a priori considerations, are used as reference values wherever the contrary is not explicitly stated.

\begin{figure}[h!]
\centering
\begin{subfigure}[t]{.45\textwidth}
  \centering
  \includegraphics[scale=0.9]{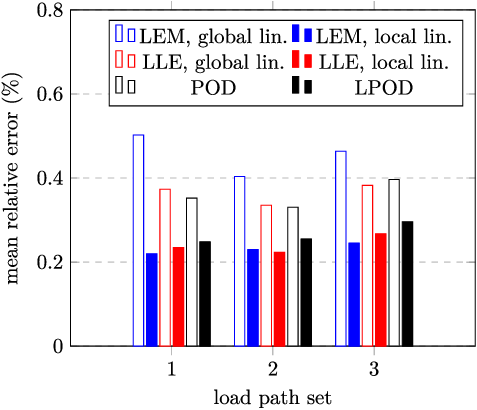}
  \caption{Mean percentage modeling error $E_{\text{mean}}$ for different load path sets.}
  \label{fig:RVE_ers1}
\end{subfigure}%
\hspace{0.5cm}
\begin{subfigure}[t]{.45\textwidth}
  \centering
  \includegraphics[scale=0.9]{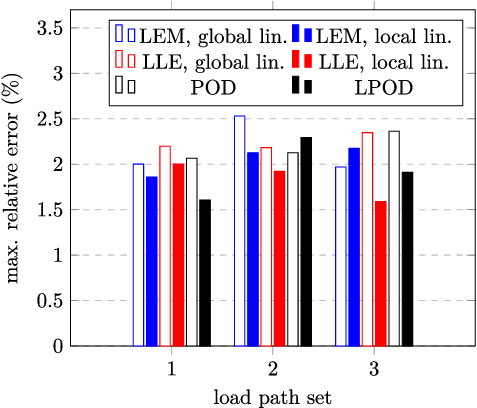}
  \caption{Maximal percentage modeling error $E_{\text{max}}$ for different load path sets.}
  \label{fig:RVE_ers2}
\end{subfigure}
\caption{\textbf{Results for different load path sets:} Mean and maximal $L2$ norm of the modeling error in the displacement field produced by reduced FEM simulations for different training and validation load path sets with respect to reference displacement fields produced by full FEM simulations, over all validation load paths. Results shown for MOR via POD (black, unfilled bars) and LPOD (black, filled bars) as well as globally (unfilled bars) and locally (filled bars) linearised LEM (blue bars) and LLE (red bars). The lower-fidelity discretisation variant was used for both the full-FEM and the reduced FEM simulations. While the modeling error values vary between load path sets, the observed trends appear stable.}
\label{fig:RVE_ers}
\end{figure}

Firstly, different random realisations of the training and testing load paths parameterised via the macroscopic displacement gradient $\ten{\bar{H}}$ are considered. The results for three such new, independent realisations (and $d=15$) are shown in Fig.~\ref{fig:RVE_ers}. Furthermore, the mean value and the range of the mean and maximal modeling error values obtained using different load path sets are summarised in Tab.~\ref{tab:mean_e_rs} and in Tab.~\ref{tab:max_e_rs}, respectively. While the mean relative modeling error values obtained using all methods shift slightly between random realisations, the relative trends between the performance of different methods remain stable. The locally linearised LEM and LLE outperform the POD and globally linearised LEM and LLE by a relative margin of around 35\%, with the LPOD producing slightly worse results. As is to be expected, the maximal relative modeling error varies more strongly and the results obtained using different methods vary less significantly. However, the locally linearised LLE consistently yields lower maximal modeling error values than the globally linear methods. The locally linearised LEM seems more beset by variance and less robust, which might be due to the limited preservation of global structure in the embeddings it produces, as discussed in Section~\ref{s:examples}. Similarly, the maximal modeling errors produced by the LPOD for different realisations of the load paths vary quite strongly.

\begin{table}[h!]
    \centering
    \begin{tabular}{ p{4cm}p{4cm}p{4cm}  }
    \hline
    Method & Mean of $E_\text{mean}$ $(\%)$ & Range of $E_\text{mean}$ $(\%)$ \\
    \hline
    POD & 0.3598 & 0.06582 \\
    LEM, global lin. & 0.4566 & 0.09866 \\
    LLE, global lin. & 0.3636 & 0.04730 \\
    LPOD & 0.2664 & 0.04772 \\
    LEM, local lin. & 0.2316 & 0.02577 \\
    LLE, local lin. & 0.2418 & 0.04394 \\
    \hline
\end{tabular}
    \caption{Mean and range of $E_\text{mean}$ (see Fig.~\ref{fig:RVE_ers1}) obtained via all manifold learning methods for different random seeds.}
    \label{tab:mean_e_rs}
\end{table}
\begin{table}[h!]
    \centering
    \begin{tabular}{ p{4cm}p{4cm}p{4cm}  }
    \hline
    Method & Mean of $E_\text{max}$ $(\%)$ & Range of $E_\text{max}$ $(\%)$ \\
    \hline
    POD & 2.185 & 0.2981 \\
    LEM, global lin. & 2.168 & 0.5622 \\
    LLE, global lin. & 2.243 & 0.1645 \\
    LPOD & 1.936 & 0.6882 \\
    LEM, local lin. & 2.054 & 0.3186 \\
    LLE, local lin. & 1.838 & 0.4128 \\
    \hline
\end{tabular}
    \caption{Mean and range of $E_\text{max}$ (see Fig.~\ref{fig:RVE_ers2}) obtained via all manifold learning methods for different random seeds.}
    \label{tab:max_e_rs}
\end{table}

\begin{figure}[h!]
\centering
\begin{subfigure}[t]{.45\textwidth}
  \centering
  \includegraphics[scale=0.9]{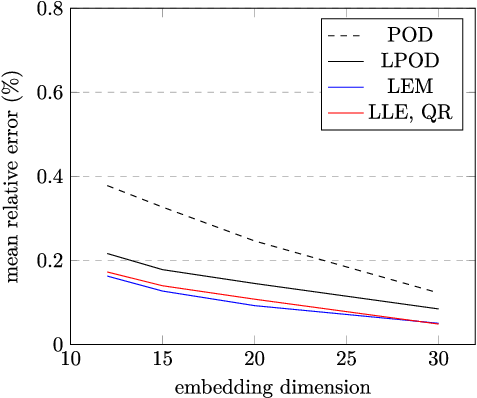}
  \caption{Mean percentage modeling error $E_{\text{mean}}$ over embedding dimension $d$.}
  \label{fig:RVE_ed1_tr04}
\end{subfigure}%
\hspace{0.5cm}
\begin{subfigure}[t]{.45\textwidth}
  \centering
  \includegraphics[scale=0.9]{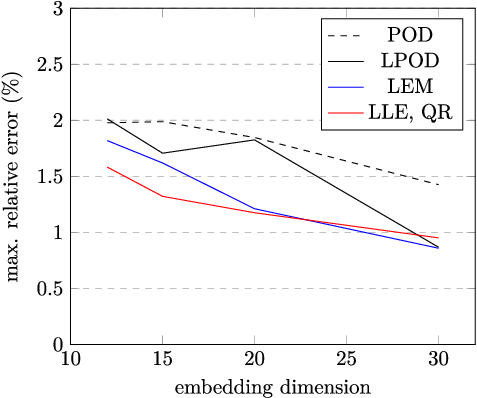}
  \caption{Maximal percentage modeling error $E_{\text{max}}$ over embedding dimension $d$.}
  \label{fig:RVE_ed2_tr04}
\end{subfigure}
\caption{\textbf{Results for more training data:} Mean and maximal $L2$ norm of the modeling error in the displacement field produced by reduced FEM simulations with different embedding dimensionalities $d$ with respect to reference displacement fields produced by full FEM simulations, over all validation load paths. Results are shown for MOR via POD (black, dashed lines) and LPOD (black, solid lines) as well as locally linearised LEM (blue lines) and locally linearised, orthonormalised LLE (red lines). The lower-fidelity discretisation variant was used for both the full-FEM and the reduced FEM simulations. $200$ snapshots were used to train all reduced order models, rather than the $100$ snapshots used elsewhere. While the modeling error values are somewhat lower than in the case with fewer snapshots, the observed trends appear stable.}
\label{fig:RVE_ed_tr04}
\end{figure}

Secondly, results obtained using a larger training data set consisting of $20$ rather than $10$ load paths are shown in Fig.~\ref{fig:RVE_ed_tr04}. As might be expected, the mean and maximal relative modeling error obtained via all investigated techniques over all model sizes $d$ are reduced. The mean relative modeling error is generally reduced more significantly: for model sizes between $d=12$ and $d=30$, the POD yields a modeling error of between $0.38\%$ and $0.12\%$, the LPOD between $0.22\%$ and $0.09\%$, and the LEM and LLE between $0.17\%$ and $0.05\%$. This is intuitive since the mean relative modeling error is tied more closely to the quality of interpolation well within the subset of the solution manifold sampled by the testing data, which scales with the training data density. In contrast, the maximum relative modeling error is due to the quality of extrapolation to regions of the solution manifold not sampled in the training data, which does not scale as strongly with data density. 
Generally, in this example the locally linearised LEM and LLE seem to benefit more from the additional training data than the LPOD does: the convergence behaviour in both the mean and the maximal relative modeling error with increased model sizes becomes more consistent and uniform, and both the maximal and the mean relative modeling error are reduced more significantly in relative terms. One possible explanation is that the more continuously nonlinear approximation of the solution manifold obtained via LEM and LLE improves more continuously with added data, as the local approximation improves with every data point, while the LPOD encounters more discrete thresholds above which the use of additional clusters is warranted.

\begin{figure}[h!]
    \centering
    \includegraphics[scale=0.4,trim={8cm 4.5cm 8cm 3cm},clip]{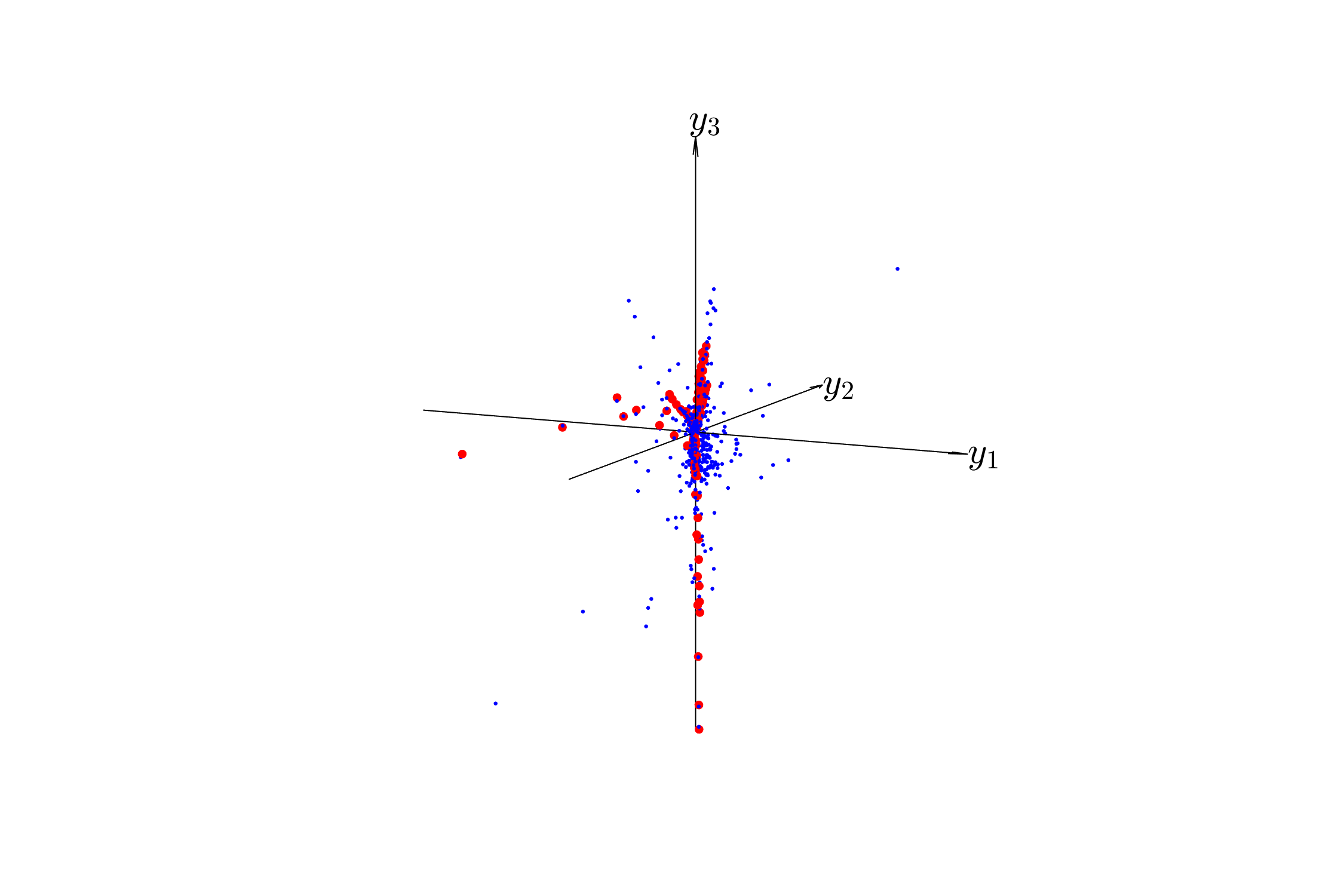}
    \caption{First three components of embedding obtained via LEM with a Gauss weight of $t=5$ (in red), and first three components of reduced solutions obtained using locally linearised LEM with $t=5$ for model reduction (in blue).}
\label{fig:LEM_w5_embed}
\end{figure}

Furthermore, the influence of the Gaussian weight $t$ on the mean and maximal relative modeling error obtained via the locally linearised LEM is shown in Tab.~\ref{tab:LEM_weights}. $t$ appears to have a moderate influence on both metrics; the mean relative modeling error varies in a relative range of approximately 20\% and the maximal relative modeling error in a relative range of approximately 25\%. The performance achieved with an unweighted graph (which was used in the examples in Section~\ref{s:examples}) lies toward the top of this interval.
Performance improvements can thus be expected from the use of a judiciously chosen Gauss weight. Likely, this stems from improved local and global structure retention in the mapping provided by the LEM, as local distance information is captured in the weighted graph edges. However, as shown in Fig.~\ref{fig:LEM_w5_embed}, the introduction of a weighted graph does not lead to an embedding which retains the load path structure of the snapshot data to a much greater extent than an unweighted graph (see Fig.~\ref{fig:LEM_embed}). While it is thus possible to obtain slight performance improvements via a weighted graph, this does not lead to extreme changes in the qualitative behaviour of the MOR scheme.

Note that the optimal choice of $t$ is problem-dependent; the Gauss weight has to account for variations in the order of magnitude of the unknown degrees of freedom as well as for their number $D$. Fortunately, appropriate weight values can be estimated from information available in the offline graph-building step: to distinguish between a wide range of distances, $t$ should be chosen such that the nonzero entries of $\ten{W}$ scatter neither around 0 nor 1, but are distributed more homogeneously.

\begin{table}[h!]
    \centering
    \begin{tabular}{ p{4cm}p{4cm}p{4cm}  }
    \hline
    Gaussian weight $t$ & mean error ($\%$) & maximal error ($\%$) \\
    \hline
    1 & 0.2156 & 1.466 \\
    2 & 0.2132 & 1.378 \\
    5 & 0.1958 & 1.376 \\
    15 & 0.2071 & 1.737 \\
    25 & 0.2053 & 1.868 \\
    $\infty$ & 0.2196 & 1.858 \\
    \hline
\end{tabular}
    \caption{Mean and maximal $L2$ norm of the modeling error in the displacement field produced by reduced FEM simulations with respect to reference displacement fields produced by full FEM simulations, over all validation load paths. For reduction, \textbf{LEM} was used with orthonormalised local linearisation. Results shown for different values of the Gaussian weight $t$.}
    \label{tab:LEM_weights}
\end{table}

Moreover, the influence of the number of neighbours $n$ utilised in the data-based local linearisation step is highlighted by Tab.~\ref{tab:LEM_lin} (for the LEM) and Tab.~\ref{tab:LLE_lin} (for the LLE). Similar trends are apparent for both methods: if too few neighbours are chosen (e.g. $n=15$ for $d=15$), the linearisation is not sufficiently robust and prohibitively high modeling errors result. If too many neighbours are chosen (i.e. $n\rightarrow s$), the local linearisation tends towards the less fruitful global linearisation. In these data-poor examples the optimal number of neighbours lies relatively close to the dimension of the reduced space $d$. A sufficient number of neighbours is required for robustness, but smaller $n$ approximate the tangent to the solution manifold more closely.

\begin{table}[h!]
    \centering
    \begin{tabular}{ p{4cm}p{4cm}p{4cm}  }
    \hline
    linearisation & & \\ neighbours $n$ & mean error ($\%$) & maximal error ($\%$) \\
    \hline
    15 & 4.502 & 14.00 \\
    20 & 0.2196 & 1.858 \\
    25 & 0.2253 & 1.922 \\
    30 & 0.2289 & 1.685 \\
    40 & 0.2441 & 1.819 \\
    50 & 0.2683 & 1.851 \\
    \hline
\end{tabular}
    \caption{Mean and maximal $L2$ norm of the modeling error in the displacement field produced by reduced FEM simulations with respect to reference displacement fields produced by full FEM simulations, over all validation load paths. For reduction, \textbf{LEM} was used with orthonormalised local linearisation. Results shown for different values of the number of neighbours $n$ used for linearisation.}
    \label{tab:LEM_lin}
\end{table}

\begin{table}[h!]
    \centering
    \begin{tabular}{ p{4cm}p{4cm}p{4cm}  }
    \hline
    linearisation & & \\ neighbours $n$ & mean error ($\%$) & maximal error ($\%$) \\
    \hline
    15 & 3.776 & 13.24 \\
    20 & 0.2345 & 2.034 \\
    25 & 0.2416 & 1.859 \\
    30 & 0.2467 & 1.928 \\
    40 & 0.2560 & 2.006 \\
    50 & 0.2666 & 1.919 \\
    \hline
\end{tabular}
    \caption{Mean and maximal $L2$ norm of the modeling error in the displacement field produced by reduced FEM simulations with respect to reference displacement fields produced by full FEM simulations, over all validation load paths. For reduction, \textbf{LLE} was used with orthonormalised local linearisation. Results shown for different values of the number of neighbours $n$ used for linearisation.}
    \label{tab:LLE_lin}
\end{table}

Next, Tab.~\ref{tab:LEM_kNN} and Tab.~\ref{tab:LLE_kNN} highlight the influence of the graph construction methodology and the resulting graph connectivity on the results achieved with the LEM and the LLE, respectively. The symmetric and mutual $k$-nearest neighbour approaches are investigated and the parameter $k$ is varied (see e.g. Section~\ref{s:LEM}). The median and the interquartile range (IQR) of the graph connectivity, i.e., the number of neighbours to which each node is connected, are shown for reference.

In the case of the LEM, low values of $k$ for the symmetric $k$-nearest neighbour approach seem to yield ineffective embeddings, with a high modeling error level for $k=25$ and diverging simulations for lower values. The mutual $k$-nearest neighbour approach appears more robust, even at low connectivities, and generally enables slightly lower mean and maximal modeling errors.
In the case of the LLE, the influence of the graph algorithm and the resulting connectivity seems less pronounced; once more, lower mean and maximal modeling error values can be achieved using a mutual $k$-nearest neighbour approach. 

Generally, trends in the influence of the graph connectivity are less pronounced and less uniform than those observable in the influence of other parameters explored above. Overall, a relatively low graph connectivity seems desirable, as relevant neighbourhood relationships are prioritised, while excessively low values of $k$ may lead to relevant neighbourhood relationships being neglected. The mutual $k$-nearest neighbour approach may be preferable on account of prioritising essential neighbourhood relationships, while the symmetric alternative may over-connect nodes toward the centre of the snapshot data in particular.

\begin{table}[h!]
    \centering
    \begin{tabular}{ p{2.5cm}p{2.5cm}p{2.5cm}p{2.5cm}p{1.75cm}p{1.75cm}  }
    \hline
    graph & connected & connectivity & connectivity & mean & max. \\
    algorithm & neighbours $k$ & median & IQR & error ($\%$) &  error ($\%$)  \\
    \hline
    symmetric & 15 & 20 & 10 & -- & -- \\
    symmetric & 20 & 26 & 17 & -- & -- \\
    symmetric & 25 & 31 & 22 & 0.2562 & 2.674 \\
    symmetric & 30 & 37 & 14 & 0.2196 & 1.858 \\
    symmetric & 35 & 43 & 23 & 0.2194 & 1.635 \\
    \hline 
    mutual & 25 & 12 & 11 & 0.2021 & 1.677 \\
    mutual & 30 & 14 & 17 & 0.1999 & 1.279 \\
    mutual & 40 & 19 & 28 & 0.2077 & 1.487 \\
    mutual & 50 & 33 & 36 & 0.2067 & 1.640 \\
    mutual & 60 & 52 & 38 & 0.2115 & 1.808 \\
    \hline
\end{tabular}
    \caption{Mean and maximal $L2$ norm of the modeling error in the displacement field produced by reduced FEM simulations with respect to reference displacement fields produced by full FEM simulations, over all validation load paths. For reduction, \textbf{LEM} was used with orthonormalised local linearisation. Results shown for different values of the number of neighbours $n$ used for graph building. The mean and interquartile range of the graph connectivity are also shown. Simulations using $k=15$ and $k=20$ failed to converge.}
    \label{tab:LEM_kNN}
\end{table}

\begin{table}[h!]
    \centering
    \begin{tabular}{ p{2.5cm}p{2.5cm}p{2.5cm}p{2.5cm}p{1.75cm}p{1.75cm}  }
    \hline
    graph & connected & connectivity & connectivity & mean & max. \\
    algorithm & neighbours $k$ & median & IQR & error ($\%$) &  error ($\%$)  \\
    \hline
    symmetric & 15 & 20 & 10 & 0.2215 & 1.854 \\
    symmetric & 20 & 26 & 17 & 0.2367 & 2.059 \\
    symmetric & 25 & 31 & 22 & 0.2368 & 2.051 \\
    symmetric & 30 & 37 & 14 & 0.2345 & 2.034 \\
    symmetric & 35 & 43 & 23 & 0.2370 & 1.841 \\
    \hline 
    mutual & 25 & 12 & 11 & 0.2091 & 1.689 \\
    mutual & 30 & 14 & 17 & 0.2130 & 1.444 \\
    mutual & 40 & 19 & 28 & 0.2289 & 1.745 \\
    mutual & 50 & 33 & 36 & 0.2399 & 2.160 \\
    mutual & 60 & 52 & 38 & 0.2315 & 1.655 \\
    \hline
\end{tabular}
    \caption{Mean and maximal $L2$ norm of the modeling error in the displacement field produced by reduced FEM simulations with respect to reference displacement fields produced by full FEM simulations, over all validation load paths. For reduction, \textbf{LLE} was used with orthonormalised local linearisation. Results shown for different values of the number of neighbours $n$ used for graph building. The median and interquartile range of the graph connectivity are also shown.}
    \label{tab:LLE_kNN}
\end{table}

\begin{figure}[h!]
\centering
\begin{subfigure}[t]{.45\textwidth}
  \centering
  \includegraphics[scale=0.9]{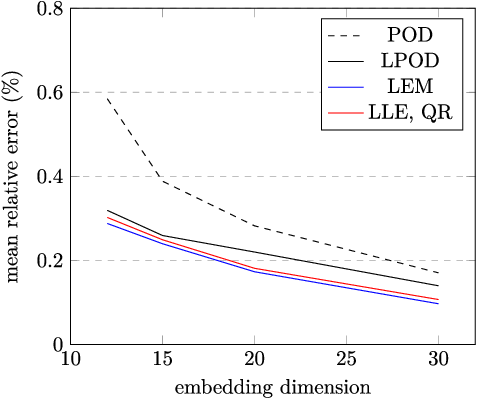}
  \caption{Mean percentage modeling error $E_{\text{mean}}$ over embedding dimension $d$.}
  \label{fig:RVE_ed_fine1}
\end{subfigure}%
\hspace{0.5cm}
\begin{subfigure}[t]{.45\textwidth}
  \centering
  \includegraphics[scale=0.9]{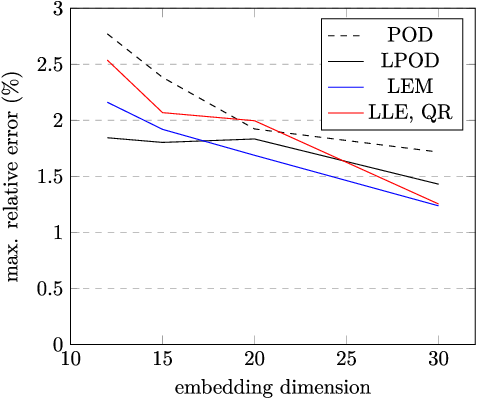}
  \caption{Maximal percentage modeling error over embedding dimension $d$. }
  \label{fig:RVE_ed_fine2}
\end{subfigure}
\caption{\textbf{Results for finer discretisation:} Mean and maximal $L2$ norm of the modeling error in the displacement field produced by reduced FEM simulations with different embedding dimensionalities $d$ with respect to reference displacement fields produced by full FEM simulations, over all validation load paths. Results shown for MOR via POD (black, dashed lines) and LPOD (black, solid lines) as well as locally linearised LEM (blue lines) and locally linearised, orthonormalised LLE (red lines). The higher-fidelity discretisation variant was used for both the full-FEM and the reduced FEM simulations. 
}
\label{fig:RVE_ed_fine}
\end{figure}

In addition, results obtained using the higher-fidelity discretisation of the RVE problem with $D=17,151$ are shown in Fig.~\ref{fig:RVE_ed_fine}. The trends broadly mirror those observed for the lower-fidelity variant in Fig.~\ref{fig:RVE_ed}. As is to be expected with the increased number of entries in the vector of primary unknown variables, the mean and maximal modeling error values obtained by all methods are slightly higher overall. Nevertheless, the LEM and the LLE outperform the POD by some margin (around 40\%) in terms of the mean relative modeling error. Especially at higher model sizes, a significant improvement (around 20\%) over the LPOD is also achieved. 
Conversely, the LEM and LLE outperform the POD and LPOD in terms of the model size required to achieve a given modeling error level: to achieve an error level of e.g. $0.2\%$, the LEM and LLE require a model size of $d\approx 18$, the LPOD requires $d\approx 23$, and the POD $d\approx 28$.
Again, the methods yield closer values in the maximal relative modeling error, which is more closely tied to extrapolation rather than interpolation characteristics. Furthermore, the LEM outperforms the LPOD at higher model dimensionalities ($d>15)$, while the LLE yields slightly worse results here. 

\begin{figure}[h!]
\centering
\begin{subfigure}[t]{.45\textwidth}
  \centering
  \includegraphics[scale=0.9]{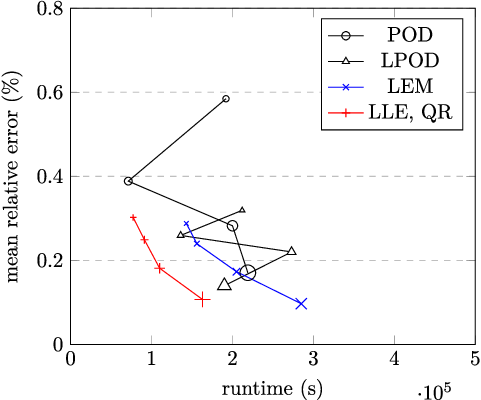}
  \caption{Mean percentage modeling error $E_{\text{mean}}$ over embedding dimension $d$.}
  \label{fig:RVE_et_fine1}
\end{subfigure}%
\hspace{0.5cm}
\begin{subfigure}[t]{.45\textwidth}
  \centering
  \includegraphics[scale=0.9]{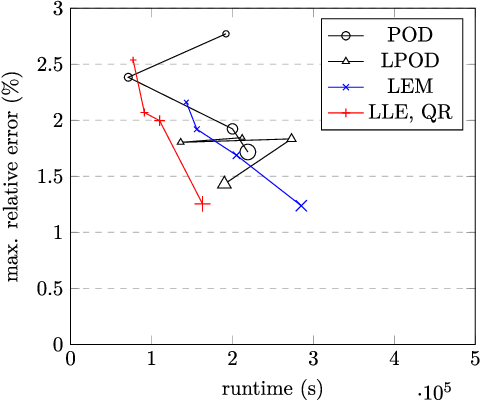}
  \caption{Maximal percentage modeling error $E_{\text{max}}$ over embedding dimension $d$. }
  \label{fig:RVE_et_fine2}
\end{subfigure}
\caption{\textbf{Results for finer discretisation:} Mean and maximal $L2$ norm of the modeling error in the displacement field produced by reduced FEM simulations with runtime $t$ for different embedding dimensionalities $d$ with respect to reference displacement fields produced by full FEM simulations, over all validation load paths. Results shown for MOR via POD (black circles) and LPOD (black triangles) as well as locally linearised LEM (blue $+$ symbols) and locally linearised, orthonormalised LLE (red $\times$ symbols). The marker size in each case is chosen to be proportional to the underlying embedding dimension $d$. The higher-fidelity discretisation variant was used for both the full-FEM and the reduced FEM simulations. 
}
\label{fig:RVE_et_fine}
\end{figure}

Finally, Fig.~\ref{fig:RVE_et_fine} highlights some basic timing results. Performance optimisations are not an emphasis of this proof-of-concept investigation; we did not consider, for example, hyper-reduction techniques to accelerate the costly assembly of stiffness matrices and residua. Such measures are necessary for real-time online performance in projection-based MOR methods, but are beyond the scope of this investigation. 
Nevertheless, all methods were implemented in the same FE framework and the algorithms relevant for the online solution procedure differ only in the methodology used to obtain the projection matrices required for reducing the linear system of equations resulting from a Newton-Raphson scheme. Thus, runtime variations are mostly due to the complexity of the procedure used to obtain these projection matrices, as well as the number of iterations required to achieve convergence\footnote{Note that all methods were implemented in our in-house Python FEM and MOR frameworks, which are primarily tools for algorithmic development.}. While the absolute runtime values are therefore not particularly meaningful, preliminary observations about relative performance trends may be drawn.

In Fig.~\ref{fig:RVE_et_fine}, the mean and maximal relative modeling error obtained for all model sizes $d$ are laid out over the runtime required for simulations along all load paths to converge. The marker size in each case is chosen to be proportional to $d$. While the LEM and LLE behave as expected, with smaller models yielding lower runtimes, the variation in runtime exhibited by the LPOD does not simply scale with the model size, as discussed in more detail below. The POD yields lower runtimes for smaller models for $d=30$, $d=20$, and $d=15$, but sees an uptick at $d=12$. This is because of the increased number of iterations required for the reduced Newton scheme to converge in many load steps, which may be due to the small, linear model not sufficently capturing the solution manifold.

The locally linearised, orthonormalised LLE Pareto-dominates the other methods in the trade-off between mean relative modeling error and runtime, and also appears the best choice overall in terms of the maximal relative modeling error obtainable in a range of runtimes. The LEM, though it obtains similarly low values in the modeling errors, requires considerably more time to do so -- which is almost exclusively due to the higher number of iterations required for convergence. This might be a consequence of the reduced degree to which the LEM with $t\rightarrow\infty$ retains global structural information in its embedding (see Fig.~\ref{fig:LEM_embed}) which leads the Newton scheme to follow less straightforward a path along the approximation of the solution manifold. 

Meanwhile, the LPOD is subject to a considerable variation in runtime which does not scale simply with changing model dimension. This appears to be due to the behaviour of the method around cluster transition zones: when a solution lies close to the transition zone between two or more local ROMs with this method, the search for solutions may proceed alternatingly in the reduced spaces defined by the one, and then the other ROM. Even when no modeling errors are created in the cluster transition itself, and even if repeated transitions do not require projections from one reduced space to the other, constraining the search for solutions alternatingly to one of two potentially quite different subspaces may lead to an undesirable zig-zagging behaviour. Perhaps additional modifications to the cluster transition methodology beyond these suggested in~\cite{AmsZahWas:2015:flr} could improve the performance of the LPOD in the context of quasi-static solid-mechanical problems to be solved with a Newton-Raphson scheme. For example, employing a basis enrichment procedure when an appropriate zig-zagging criterion is met may allay some of these issues.

Overall, the results of this exploratory timing investigation are encouraging. In terms of the level of accuracy achievable in a given amount of time, the LPOD yields an improvement over the POD, and the locally linearised LLE yields a further improvement yet. The LEM is less competitive on account of the number of iteration required for convergence in the Newton scheme.

\section{Summary and Conclusions}\label{s:conclusion}

In this proof-of-concept investigation, we propose a manifold learning approach to nonlinear projection-based MOR for parameterised quasi-static problems, which addresses some of the shortcomings of the POD and LPOD.
The POD may require significant model dimensionalities $d$ to successfully capture a nonlinear solution manifold resulting from a quasi-static solid-mechanical problem. The LPOD from~\cite{AmsZahFar:2012:nmo} has been generally successful in addressing this deficiency by making use of local, rather than global, POD bases. However, this approach comes with some drawbacks, especially in the data-poor case: cluster-wise linear approximations do not parameterise the solution manifold as closely as possible, cluster transitions are precarious and may yield instabilities and inefficiencies in the solution scheme, and the quality of the obtained ROM scales with the quality of the underlying clusters. 
In contrast, we apply manifold learning techniques to obtain a continuous and nonlinear, rather than discontinuous and locally linear, approximation space. A locally linear mapping between the reduced and original solution spaces is computed in the online phase.

After the manifold learning approach is motivated in the Introduction, Section~\ref{s:FEM} covers some essentials of FE modelling for quasi-static solid-mechanical problems.
Section~\ref{s:problem} then outlines our conceptual approach to nonlinear projection-based MOR. Section~\ref{s:state_of_art} recalls the POD and the LPOD, and Section~\ref{s:ManLMOR} introduces our alternative approach. Two manifold learning methods are covered in Subsections~\ref{s:ManL},~\ref{s:LEM}, and~\ref{s:LLE}, and the linearisation scheme in Subsection~\ref{s:loclin}. Section~\ref{s:two_stage} finally lays out a two-stage MOR scheme with which a scaling behaviour independent from the original problem size $D$ can be achieved. Numerical experiments on a simple artificial RVE are detailed in Section~\ref{s:examples}, and Section~\ref{s:parameter_study} features an ad-hoc parameter study which characterises the qualitative influence of algorithmic parameters.

In the examples explored in Sections~\ref{s:examples} and~\ref{s:parameter_study}, the locally linearised LEM and LLE yield more of a performance boost over the POD than the LPOD, when considering the level of accuracy achievable with a given model size. Conversely, the LEM and LLE achieve a set modeling error level with smaller model sizes, meaning that smaller and potentially more efficient models can be used to attain a target level of accuracy. The two-stage MOR scheme, in which the POD is used for a nearly lossless compression before the manifold leaning step, encouragingly yields almost identical results. This indicates that, in the investigated example, the local linearisation as well as the QR decomposition can be performed with respect to the lower-dimensional intermediate space rather the solution space, yielding an improved scaling behaviour with little penalty in terms of accuracy. Performance trends appear stable for independent random realisations of the training and testing load paths, a different amount of training data, a different underlying FE model fidelity, and a range of reasonable model parameters. While the investigations summarised in Section~\ref{s:examples} were conducted with parameters motivated by simpler preliminary investigations, the parameter study outlined in Section~\ref{s:parameter_study} indicates that slight performance improvements can readily be realised via further parameter tuning. Via a judiciously chosen Gauss weight or the selection of a mutual $k$-nearest neighbour graph building algorithm, for example, the mean modeling error level achieved by the locally linearised LEM with $d=15$ can be reduced by approximately $10\%$. At the same time, the performance of the locally linearised manifold learning algorithms does not vary too strongly when model parameters are varied within reasonable bounds, suggesting that the methods are sufficiently robust in the face of parameter changes.

No in-depth timing experiments were conducted, as performance optimisation and hyper-reduction were not an objective of this proof-of-concept investigation. However, preliminary runtime results suggest that the locally linearised, orthonormalised LLE is able to produce results with a given level of accuracy in a noticeably shorter time than the POD and the LPOD{, or to achieve a higher accuracy in a given budget of computational time; i.e. the locally linearised LLE Pareto-dominates these existing methods on the considered example}. Both the LPOD and the locally linearised LEM suffer from slow convergence in some instances. In the case of the LPOD, {we noted that} this is likely due to alternating iterations within distinct subspaces in cluster transition regions. In the case of the LEM, we hypothesised that this is due to the reduced degree to which the LEM retains global structural information, causing the Newton scheme to follow less straightforward a path along the approximation of the solution manifold.

Overall, the results of this proof-of-concept investigation are encouraging: the locally linearised LLE (as well as the LEM, providing convergence issues can be addressed) appears to be able to outperform the state-of-the-art LPOD in reducing a quasi-static RVE problem with a nonlinear solution manifold that proves more challenging for the POD.

\section{Outlook}\label{s:outlook}

Given the encouraging results of this proof-of-concept investigation, further research on nonlinear projection based MOR via graph-based manifold learning techniques in quasi-static solid mechanics appears warranted. 
Firstly, the implementation of a hyper-reduction scheme such as one of those proposed in~\cite{Ryc:2009:hrm,ChaSor:2010:nmr,CarFar:2011:lcg,NegManAms:2015:emr,JaiTis:2019:hnm} is necessary to achieve competitive online performance. 

Secondly, the resulting nonlinear MOR framework ought to be applied to a range of problems, beyond the simple artificial RVE computations investigated here as proof-of-concept verification. Performance comparisons to alternative approaches such as the LPOD, autoencoder applications, and Ansatz manifold parameterisations for nonlinear projection-based MOR, but also alternative MOR techniques such as surrogate models are particularly intriguing in this context.
This would serve to highlight for which classes of problems projection-based MOR schemes based on manifold learning techniques can be applied profitably. It would also help diagnose open issues and identify suitable directions for further development.
The handling of spatial localisation phenomena such as those resulting from plasticity or damage in solids might be one such research direction. 
In parallel, further developments of the LPOD, e.g. considering adaptivity and additional cluster transition safeguards, might improve its performance on quasi-static problems in solid mechanics.

If the further development outlined above proves successful, nonlinear projection-based MOR methods using manifold learning techniques may carve out a niche in applications to parametric problems where little data is available to train a ROM and little human-time available to craft a specialised model, but comparatively high levels of accuracy are required. 
Such applications maybe found in fields ranging from multiscale modelling over inverse problems to optimisation.
Beyond use as a standalone MOR scheme, nonlinear projection-based MOR methods could be applied to manufacture additional training data for data-hungry surrogate models which excel in terms of online computation time, but may feature prohibitively expensive offline phases.

\section*{CRediT authorship contribution statement}

\noindent
\textbf{Lisa Scheunemann}: Conceptualization, Methodology, Investigation, Writing - Review \& Editing, Supervision, Funding acquisition. 
\textbf{Erik Faust}: Conceptualization, Methodology, Software, Validation, Formal analysis, Investigation, Data Curation, Writing - Original Draft, Visualization

\section*{Data availability}

\noindent
For access to the GitHub repositories in which implementations of the methods and examples outlined above are archived, please contact the authors.

\section*{Acknowledgements}

\noindent
{We thank Ramses Sala, Felix Steinmetz, and Rozan Rosandi for their helpful comments.}
Funded by the Deutsche Forschungsgemeinschaft (DFG) within the TRR375/1, subproject A01, project number 511263698.


\bibliographystyle{elsarticle-num} 
\bibliography{sources.bib}







\pagebreak

\appendix

\section{Derivation of global linearisation}\label{a:globlin}

A linear projection matrix $\ten{\phi}\in \mathbb{R}^{D\times d}$ is sought to optimally map between the original snapshot matrix $\ten{U}\in \mathbb{R}^{D\times s}$ and the corresponding reduced coordinates $\ten{Y}\in\mathbb{R}^{d\times s}$ in the least-squares sense. A rigorous derivation can be found in~\cite{Pyt:2018:muo}; here, we provide a simple sketch using linear algebra.

The objective for the least-squares problem can be written in index notation as
\begin{equation*}
    \min_{\ten{\phi}} f(\ten{\phi}) = \min_{\ten{\phi}} \sum_i \sum_j ( U_{ij} - \sum_k \phi_{ik} Y_{kj} )^2\,.
\end{equation*}
The necessary conditions for an optimum are given by
\begin{equation*}
    \frac{\partial f}{\partial \phi_{ab}} = 0\,.
\end{equation*}
Differentiation yields
\begin{equation*}
    -\sum_j 2 Y_{bj} ( U_{aj} - \sum_k \phi_{ak} Y_{kj} ) = 0
\end{equation*}
and
\begin{equation*}
    \sum_j \sum_k Y_{bj} Y_{kj} \phi_{ak} = \sum_j Y_{bj} U_{aj}\,. 
\end{equation*}
In matrix notation, this is equivalent to
\begin{equation*}
    \ten{Y} \ten{Y}^T \ten{\phi}^T = \ten{Y} \ten{U}^T\,.
\end{equation*}
such that $\ten{\phi}$ is given by
\begin{equation*}
    \ten{\phi}^T = (\ten{Y} \ten{Y}^T )^{-1} \ten{Y} \ten{U}^T\,, \quad \text{and} \quad
    \ten{\phi} = \ten{U} \ten{Y}^T ( \ten{Y} \ten{Y}^T )^{-1}\,.
\end{equation*}

\section{Correlation dimension}\label{s:corr_dim}

The dimensionality $\delta$ of a solution manifold $\mathcal{M}_{\vec{u}}$ is not always known a priori. Luckily, it is possible to estimate $\delta$ from the snapshot data $\Vec{u}_i \in \mathbb{R}^D, i = [1,..,s]$ which lies on $\mathcal{M}_{\vec{u}}$ via the correlation dimension originally utilised by~\cite{GraPro:1983:mss}. For the sake of completeness, we provide a simple derivation here.

To this end, it is useful to consider the probability $P_\text{CD}(\varepsilon)$ that two arbitrary snapshots $i$ and $j$ lie at a distance of less than some length scale $\varepsilon$ from each other, i.e.~\cite[p.53]{LeeVer:2007:ndr}
\begin{equation*}
    P_\text{CD}(\varepsilon) = P(\|\Vec{u}_i-\Vec{u}_j\|_2\leq \varepsilon)\,.
\end{equation*}
It is possible to estimate this probability from the snapshot data by computing the distances between all pairs of points $\Vec{u}_i$ and $\Vec{u}_j$, and determining the proportion of distances smaller than $\varepsilon$, i.e.
\begin{equation*}
    \tilde{P}_\text{CD}(\varepsilon) = \frac{1}{s(s-1)} \sum_{i=1}^s\sum_{j=i+1}^s H(\varepsilon-\|\Vec{u}_i-\Vec{u}_j\|)\,,
\end{equation*}
where $H$ is the Heaviside function~\cite[p.54]{LeeVer:2007:ndr}. Assuming the data to be sufficiently dense and homogeneously spaced, in the limit as $\varepsilon \rightarrow 0$, $\tilde{P}_\text{CD}$ grows as a length for a  one-dimensional manifold, as a surface for a two-dimensional manifold, and so on~\cite[p.54]{LeeVer:2007:ndr}, i.e. $\tilde{P}_\text{CD}(\varepsilon) = k \varepsilon^\delta$, or
\begin{equation*}
    \delta = \lim_{\varepsilon\rightarrow 0} \frac{\log(\frac{1}{k}\tilde{P}_\text{CD}(\varepsilon))}{\log(\varepsilon)} = \lim_{\varepsilon\rightarrow 0} \frac{\log(\tilde{P}_\text{CD}(\varepsilon)) - \log(k)}{\log (\varepsilon)} = \lim_{\varepsilon\rightarrow 0} \frac{\log(\tilde{P}_\text{CD}(\varepsilon))}{\log(\varepsilon)}
\end{equation*}
with the proportionality factor $k$ disappearing in the limit. As both numerator and denominator tend to $-\infty$, we may apply l'Hospital
\begin{equation*}
    \delta = \lim_{\varepsilon\rightarrow 0} \nicefrac{\frac{\partial \log(\tilde{P}_\text{CD}(\varepsilon))}{\partial \varepsilon}}{\frac{\partial \log(\varepsilon)}{\partial \varepsilon}}\,,
\end{equation*}
and use the definition of the derivative with $\varepsilon_2>\varepsilon_1$
\begin{equation*}
    \delta = \lim_{\varepsilon_1\rightarrow 0} \lim_{\varepsilon_2\rightarrow\varepsilon_1} \frac{ \log(\tilde{P}_\text{CD}(\varepsilon_2))- \log(\tilde{P}_\text{CD}(\varepsilon_1))}{\varepsilon_2-\varepsilon_1}\frac{\varepsilon_2-\varepsilon_1}{ \log(\varepsilon_2)-\log(\varepsilon_1)}\,,
\end{equation*}
to yield
\begin{equation*}
    \delta = \lim_{\varepsilon_1\rightarrow 0} \lim_{\varepsilon_2\rightarrow\varepsilon_1} \frac{ \log(\tilde{P}_\text{CD}(\varepsilon_2))- \log(\tilde{P}_\text{CD}(\varepsilon_1))}{ \log(\varepsilon_2)-\log(\varepsilon_1)}\,.
\end{equation*}

As the limits cannot be realised exactly with finite data, the dimensionality $\delta$ of $\mathcal{M}_{\vec{u}}$ can instead be practically estimated via a scale-dependent correlation dimension $\delta_\text{SD}$ defined over a range of length scales $\varepsilon$ and via finite increments $\Delta \varepsilon$ as
\begin{equation*}
    \delta_\text{SD} = \frac{ \log(\tilde{P}_\text{CD}(\varepsilon +\Delta \varepsilon))- \log(\tilde{P}_\text{CD}(\varepsilon))}{ \log(\varepsilon +\Delta \varepsilon)-\log(\varepsilon)}\,.
\end{equation*}
Under the aforementioned assumptions of sufficiently dense, homogeneous data and at relatively low $\varepsilon$, this scale-dependent correlation dimension provides an estimate of $\delta$ which only relies on the snapshot data $\Vec{u}_i \in \mathbb{R}^D, i = [1,..,s]$ and does not require a priori knowledge about $\mathcal{M}_{\vec{u}}$. 

\begin{figure}[h!]
    \centering
    \includegraphics{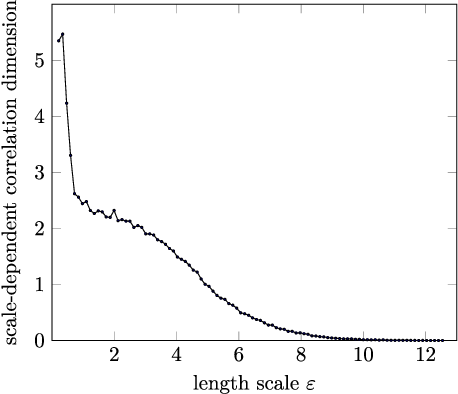}
    \caption{Correlation dimension for the snapshot data $\ten{U}$ associated with the example RVE problem investigated in Section~\ref{s:examples}.}
    \label{fig:corr_dim}
\end{figure}

Note that practically, the parameters $\varepsilon$ and $\Delta \varepsilon$ must be chosen such that a non-negligible number of points lie at a distance of less than $\varepsilon+\Delta \varepsilon$, but more than $\varepsilon$ from each other. If $\Delta \varepsilon$ is chosen to be excessively small, $\delta_\text{SD}$ is prone to noise, while for excessively large $\Delta \varepsilon$, it becomes difficult to isolate the desired limit of $\delta_\text{SD}$. For this analysis, we varied $\varepsilon$ between $\varepsilon_\text{min} = 0$ and $\varepsilon_\text{max} =\max_{i,j}(\|\Vec{u}_i-\Vec{u}_j\|_2)$, and selected $\Delta \varepsilon= \frac{\varepsilon_\text{max} - \varepsilon_\text{min}}{100}$.
In Fig.~\ref{fig:corr_dim}, the scale-dependent correlation dimension $\delta_\text{CD}$ associated with the snapshot data for the example discussed in Section~\ref{s:examples} is shown over this range of length scales $\varepsilon$. At low $\varepsilon$, we approximately recover the known a priori manifold dimensionality $\delta=6$.

\pagebreak

\section{Pseudocode}\label{s:code}

\appendix


\begin{algorithm}[h!]
\caption{Solve a nonlinear mechanical problem with periodic boundary conditions via the FEM}
\KwIn{$\ten{X}_{\text{glob}},\ten{E}_{\text{nodes}},\vec{E}_{\text{mat}},\text{ELEM},\text{MAT},\text{SHAPE},\text{param},\text{BC info}$}
\tcp{Initialise global displacement fluctuation}
$\Tilde{\vec{u}} \gets \Vec{0}$\;

\tcp{Incremental time-stepping}
$t_n \gets \Delta t$\;
\While{$t_n<=1$}{
\tcp{Obtain current load}
$\ten{\bar{H}}_n = \ten{\bar{H}} (t_n)$\;

\tcp{Iterative solution for nonlinear problem}
$\text{res} \gets 10^{10}$\;
$\text{res}_{\text{old}} \gets 10^{11}$\;
\While{$\text{res}>\text{res}_{\text{max}}$}{
\tcp{Assemble stiffness matrix and reaction vector}
$\ten{K},\vec{r} \gets \text{ASSEMBLE}(\ten{X}_{\text{glob}},\ten{E}_{\text{nodes}},\vec{E}_{\text{mat}},\Tilde{\Vec{u}},\text{ELEM},\text{MAT},\text{SHAPE},\text{param},\ten{\bar{H}_n})$\;
\tcp{Apply BCs}
$\ten{K}_{\text{BC}},\vec{r}_{\text{BC}} \gets \text{BC}(\ten{K},\Vec{r},\text{BC info})$\;
\tcp{Determine residual (trivial without external loads)}
$\Vec{g}_{\text{BC}} \gets \Vec{r}$\;
\tcp{Solve linearly for reduced displacement fluctuation increment}
$\Delta \tilde{\Vec{u}}_{\text{BC}} \gets \text{SOLVE} \left ( \ten{K}_{\text{BC}} \Delta \tilde{\vec{u}}_{\text{BC}} = - \vec{g}_{\text{BC}} \right )$\;
\tcp{Sort into whole displacement fluctuation increment vector}
$\Delta \Tilde{\Vec{u}} \gets 0$\;
$\Delta \Tilde{\Vec{u}}_{I_\text{idep}} \gets \Delta \Tilde{\Vec{u}}_{\text{BC}}$\;
\tcp{Increment displacement fluctuation}
$\Tilde{\Vec{u}} \gets \Tilde{\Vec{u}} + \Delta \Tilde{\Vec{u}}$\;

\tcp{Get residual for convergence check}
$\vec{r} \gets \text{ASSEMBLE}(\ten{X}_{\text{glob}},\ten{E}_{\text{nodes}},\vec{E}_{\text{mat}},\Tilde{\Vec{u}},\text{ELEM},\text{MAT},\text{SHAPE},\text{param},\ten{\bar{H}_n})$\;
$\vec{r}_{\text{BC}} \gets \text{BC}(\text{BC info})$\;
$\Vec{g}_{\text{BC}} \gets \Vec{r}_{\text{BC}}$\;
$\text{res} \gets \text{max}(\text{abs}(\Vec{g}_{\text{BC}}))$\;
}

\tcp{Increment time}
$t_n \gets t_n + \Delta t$\;
}

\label{alg:FEM}
\end{algorithm}



\begin{algorithm}[h!]
\caption{Use the POD to find the best approximating linear subspace for snapshot data}
\tcp{Input snapshot vectors and information ratio threshold}
\KwIn{$\Vec{u}_i, i \in [1,..,s], r_{\text{info}}$}
\tcp{Make snapshot matrix out of snapshot vectors}
$\ten{U} \gets [ \Vec{u}_i ], \quad i \in [1,..s]$\;
\tcp{Compute snapshot covariance matrix}
$\ten{C} \gets \frac{1}{s-1} \ten{U}^T \ten{U}$\;
\tcp{Solve Eigenvalue problem}
$\lambda_i, \Vec{v}_i \gets \text{EV}_i \left ( \ten{C} \right ), \quad i \in [1,..s]$\;
\tcp{Sort Eigenvalues and Eigenvectors in descending order of Eigenvalues}
$\lambda_i, \Vec{v}_i \gets \text{sort}_{\lambda_i}^- \left (\lambda_i, \Vec{v}_i \right )$\;
\tcp{Compute vector of cumulative information content in the first $i$ Eigenvectors/modes}
$I_i \gets \nicefrac{\sum_{j=1}^{i}\lambda_j}{\sum_{j=1}^{s}\lambda_j}$\;
\tcp{Determine number of Eigenvectors necessary to reproduce snapshots with specified information content}
$d \gets \text{find} \left ( I_i > r_{\text{info}} \right )$\;
\tcp{Get spatial modes/orthonormal mapping matrix}
$\ten{\psi} \gets [\nicefrac{\ten{U} \Vec{v_i}}{\| \ten{U} \Vec{v_i} \|}], \quad i \in [1,..,d]$\;
\KwOut{$\ten{\psi}$}
\label{alg:POD_offline}
\end{algorithm}

\begin{algorithm}[h!]
\caption{Solve a nonlinear mechanical problem with periodic boundary conditions via the FEM, using the POD as a MOR method}
\KwIn{$\ten{X}_{\text{glob}},\ten{E}_{\text{nodes}},\vec{E}_{\text{mat}},\text{ELEM},\text{MAT},\text{SHAPE},\text{param},\text{BC info}, \ten{\psi}$}
\tcp{Initialise global unreduced and reduced displacement fluctuation}
$\Tilde{\vec{u}} \gets \Vec{0}$, 
$\Vec{y} \gets \ten{\psi}^T \Tilde{\vec{u}}$\;

\tcp{Incremental time-stepping}
$t_n \gets \Delta t$\;
\While{$t_n<=1$}{
\tcp{Obtain current load}
$\ten{\bar{H}}_n = \ten{\bar{H}} (t_n)$\;

\tcp{Iterative solution for nonlinear problem}
$\text{res} \gets 10^{10}$\;
$\text{res}_{\text{old}} \gets 10^{11}$\;
\While{$\text{res}>\text{res}_{\text{max}}$}{
\tcp{Assemble stiffness matrix and reaction vector}
$\ten{K},\vec{r} \gets \text{ASSEMBLE}$\;
\tcp{Apply BCs}
$\ten{K}_{\text{BC}},\vec{r}_{\text{BC}} \gets \text{BC}$\;
\tcp{Determine residual (trivial without external loads)}
$\Vec{g}_{\text{BC}} \gets \Vec{r}$\;
\tcp{Reduce stiffness matrix and residuum}
$\ten{K}_{\text{red}} \gets \psi^T \ten{K}_{\text{BC}} \ten{\psi}$,
$\vec{g}_{\text{red}} \gets \psi^T \Vec{g}_{\text{BC}}$\;
\tcp{Solve linearly for reduced displacement fluctuation increment}
$\Delta \Vec{y} \gets \text{SOLVE} \left ( \ten{K}_{\text{red}} \Delta \vec{y} = - \vec{g}_{\text{red}} \right )$\;
\tcp{Compute independent displacement fluctuation increment DOFs}
$\Delta \Tilde{\vec{u}}_{\text{BC}} \gets \ten{\psi} \Delta \vec{y}$\;
\tcp{Sort into whole displacement fluctuation increment vector}
$\Delta \Tilde{\Vec{u}} \gets 0$\;
$\Delta \Tilde{\Vec{u}}_{I_\text{idep}} \gets \Delta \Tilde{\Vec{u}}_{\text{BC}}$\;
\tcp{Increment displacement fluctuation}
$\Tilde{\Vec{u}} \gets \Tilde{\Vec{u}} + \Delta \Tilde{\Vec{u}}$\;

\tcp{Get residual for convergence check}
$\vec{r} \gets \text{ASSEMBLE}$\;
$\vec{r}_{\text{BC}} \gets \text{BC}$\;
$\Vec{g}_{\text{BC}} \gets \Vec{r}_{\text{BC}}$\;
$\vec{g}_{\text{red}} \gets \psi^T \Vec{g}_{\text{BC}}$\;
$\text{res} \gets \text{max}(\text{abs}(\Vec{g}_{\text{red}}))$\;
}

\tcp{Increment time}
$t_n \gets t_n + \Delta t$\;
}

\label{alg:POD_online}
\end{algorithm}



\begin{algorithm}[h!]
\caption{Use the LPOD to find best approximating linear subspaces for clusters of snapshot data}
\label{alg:LPOD_offline}
\tcp{Input snapshot vectors, information ratio threshold, number of clusters, cluster enlargement ratio, and robustness parameters}
\KwIn{$\Vec{u}_i, i \in [1,..,s], r_{\text{info}}, k, r, |\mathcal{C}|_{\text{min}}, |\mathcal{C}|_{\text{max}}, |\mathcal{C}|_{\text{core,min}}$}

\tcp{Only accept clustering with minimum core cluster size}
\While{$\text{any}_j(|\mathcal{C}_j|<|\mathcal{C}|_{\text{core,min}})$}{
\tcp{$k$-means clustering using Lloyd's algorithm}
\tcp{Initialise cluster centroids}
$\Vec{u}_j^c \gets \Vec{u}_i, i \gets \text{rand}([1,..,s]), j \in [1,..,k]$\;
\While{$\Vec{u}_j^c$, $j \in [1,..,k]$ not converged}{
\tcp{Update cluster association}
$c_i^u \gets \text{argmin}_j d(\vec{u}_i,\Vec{u}_j^c), i \in [1,...,s]$\;
$\mathcal{C}_j \gets \{ i: c_i = j\}$\;
\tcp{Compute cluster centroids}
$\Vec{u}_j^c \gets \frac{1}{|\mathcal{C}_j|} \sum_{i \in \mathcal{C}_j} \Vec{u}_i, j \in [1,...,k]$\;
}
}

\tcp{Enlarge clusters for robustness}
\For{$j \gets [1,...,k]$}{
\tcp{Compute target (enlarged) cluster size}
$|\mathcal{C}|_{\text{new}} \gets \max \big (|\mathcal{C}|_{\text{min}},\min(|\mathcal{C}_j|+\text{ceil}(r|\mathcal{C}_j|),|\mathcal{C}|_{\text{max}}) \big)$\;
\tcp{Enlarge cluster to target}
\While{$|\mathcal{C}|_j<|\mathcal{C}|_\text{new}$}{
\tcp{Find closest solution not in cluster}
$i_\text{new} \gets \text{argmin}_i d(\vec{u}_j^c,\Vec{u}_i), i \in [1,...,s], i \notin \mathcal{C}_j$\;
\tcp{Add solution to cluster}
$\mathcal{C}_j \gets \mathcal{C}_j \cup i_\text{new}$\;
}
}

\tcp{Perform local POD for each cluster}
\For{$j \gets [1,...,k]$}{
\tcp{Center cluster snapshots using centroid}
$\Vec{u}_i^s \gets \Vec{u}_i - \Vec{u}_j^c\,, i \in \mathcal{C}_j$\;
\tcp{Make snapshot matrix and perform snapshot POD}
$\ten{U}_j^c \gets [\vec{u}_{i}^s], i\in \mathcal{C}_j$\;
$\ten{\psi}_j^c \gets \text{POD}(\ten{U}_j^c,r_\text{info})$\;
}

\KwOut{$\ten{\psi}_j^c, \Vec{u}_j^c, j \in [1,..,k]$}
\end{algorithm}

\begin{algorithm}[h!]
\caption{Solve a nonlinear mechanical problem with periodic boundary conditions via the FEM, using the LPOD as a MOR method}
\label{alg:LPOD_online}
\KwIn{$\ten{X}_{\text{glob}},\ten{E}_{\text{nodes}},\vec{E}_{\text{mat}},\text{ELEM},\text{MAT},\text{SHAPE},\text{param},\text{BC info}, \ten{\psi}_j^c$, $\Vec{u}_j^c$, $j \in [1,..,k]$}
\tcp{Initialise global unreduced and reduced displacement fluctuation}
$\Tilde{\vec{u}} \gets \Vec{0}$, 
$c \gets \text{argmin}_j d(\vec{u},\Vec{u}_j^c)$,
$\Vec{y} \gets \ten{\psi}_c^{cT} \Tilde{\vec{u}}$\;

\tcp{Incremental time-stepping}
$t_n \gets \Delta t$\;
\While{$t_n<=1$}{
\tcp{Modify current load via load multiplier}
$\ten{\bar{H}}_n = \ten{\bar{H}} t_n$\;

\tcp{Iterative solution for nonlinear problem}
$\text{res} \gets 10^{10}$\;
$\text{res}_{\text{old}} \gets 10^{11}$\;
\While{$\text{res}>\text{res}_{\text{max}}$}{
\tcp{Assemble stiffness matrix and reaction vector}
$\ten{K},\vec{r} \gets \text{ASSEMBLE}$\;
\tcp{Apply BCs}
$\ten{K}_{\text{BC}},\vec{r}_{\text{BC}} \gets \text{BC}$\;
\tcp{Determine residual (trivial without external loads)}
$\Vec{g}_{\text{BC}} \gets \Vec{r}$\;
\tcp{Identify current cluster}
$a \gets \text{argmin}_j d(\vec{u},\Vec{u}_j^c)$\;
\tcp{Reduce stiffness matrix and residuum}
$\ten{K}_{\text{red}} \gets \ten{\psi}_a^{cT} \ten{K}_{\text{BC}} \ten{\psi}_a^{c}$,
$\vec{g}_{\text{red}} \gets \ten{\psi}_a^{cT} \Vec{g}_{\text{BC}}$\;
\tcp{Solve linearly for reduced displacement fluctuation increment}
$\Delta \Vec{y} \gets \text{SOLVE} \left ( \ten{K}_{\text{red}} \Delta \vec{y} = - \vec{g}_{\text{red}} \right )$\;
\tcp{Compute independent displacement fluctuation increment DOFs}
$\Delta \Tilde{\vec{u}}_{\text{BC}} \gets \ten{\psi}_a^{c} \Delta \vec{y}$\;
\tcp{Sort into whole displacement fluctuation increment vector}
$\Delta \Tilde{\Vec{u}} \gets 0$\;
$\Delta \Tilde{\Vec{u}}_{I_\text{idep}} \gets \Delta \Tilde{\Vec{u}}_{\text{BC}}$\;
\tcp{Increment displacement fluctuation}
$\Tilde{\Vec{u}} \gets \Tilde{\Vec{u}} + \Delta \Tilde{\Vec{u}}$\;

\tcp{Get residual for convergence check}
$\ten{r} \gets \text{ASSEMBLE}$\;
$\vec{r}_{\text{BC}} \gets \text{BC}$\;
$\Vec{g}_{\text{BC}} \gets \Vec{r}_{\text{BC}}$\;
$a \gets \text{argmin}_j d(\vec{u},\Vec{u}_j^c)$\;
$\vec{g}_{\text{red}} \gets \ten{\psi}_a^{cT} \Vec{g}_{\text{BC}}$\;
$\text{res} \gets \text{max}(\text{abs}(\Vec{g}_{\text{red}}))$\;
}

\tcp{Increment time}
$t_n \gets t_n + \Delta t$\;
}
\end{algorithm}



\begin{algorithm}[h!]
\caption{Use the LEM to compute a local distance preserving embedding of snapshot vectors}
\tcp{Input graph weight matrix and target embedding dimension}
\KwIn{$\ten{W} \in \mathbb{R}^{s\times s},d$}
\tcp{Compute diagonal edge weight matrix}
$d_i \gets \sum_j W_{ij}$, $\ten{D} \gets \text{diag}(\Vec{d})$\;
\tcp{Compute graph Laplacian}
$\ten{L} \gets \ten{D} - \ten{W}$\;
\tcp{Scale graph Laplacian}
$\ten{\bar{L}} \gets \ten{D}^{-1} \ten{L}$\;
\tcp{Solve Eigenvalue problem}
$\lambda_i, \Vec{v}_i \gets \text{EV}_i \left ( \ten{\bar{L}} \right ), \quad i \in [1,..,s]$\;
\tcp{Eigenvectors in ascending order of Eigenvalues}
$\lambda_i, \Vec{v}_i \gets \text{sort}_{\lambda_i}^+ \left (\lambda_i, \Vec{v}_i \right )$\;
\tcp{Select second to $d$-th Eigenvectors as embedding}
$\ten{Y} \gets [\Vec{v}_i]^T,\quad i \in [2,..,d+1]$\;
\KwOut{$\ten{Y}$}
\label{alg:LEM}
\end{algorithm}



\begin{algorithm}[h!]
\caption{Use LLE to compute a local angle preserving embedding of snapshot vectors}
\tcp{Input adjacency matrix, snapshot matrix, target embedding dimension, and invertibility constant}
\KwIn{$\ten{G} \in \mathbb{R}^{s\times s}, \ten{U} \in \mathbb{R}^{D\times s},d, \delta $}
\tcp{Initialise reconstruction weight matrix}
$\ten{W} \gets \ten{0}$\;
\tcp{Compute local reconstruction weights for each node on graph}
\For{$i \in [1,..,2]$}{
\tcp{Current point on graph}
$\Vec{u}_N \gets \Vec{U}_i$\;
\tcp{Neighbours of current point}
$I_N \gets \text{find}_j (G_{ij} \text{and} \text{not} i==j) $\;
\tcp{Neighbours snapshot matrix}
$\ten{U}_N \gets [\Vec{U}_j], \quad j \in I_N$\;
\tcp{Compute local covariance matrix}
$\ten{C}_N \gets ( \vec{u}_N - \ten{U}_N)^T ( \vec{u}_N - \ten{U}_N)$\;
\tcp{Ensure invertibility}
$n_N \gets \text{length}(I_N)$\;
\If{$n_N>D$}{
$\ten{C}_N \gets \ten{C} + \nicefrac{\delta^2}{n_N} \ten{I}$\;
}
\tcp{Compute current row of weight matrix}
$\Vec{w}_N \gets \text{SOLVE}(\ten{C}_N \Vec{w} = \vec{\bar{1}})$\;
\tcp{Normalise}
$\vec{w}_N \gets \nicefrac{\Vec{w}}{\sum_j w_i} $\;
\tcp{Sort into reconstruction weight matrix}
$W_{i I_N} \gets \Vec{w}_N$\;
}

\tcp{Compute matrix defining reconstruction Eigenproblem}
$\ten{M} \gets (\ten{I}-\ten{W})^T (\ten{I}-\ten{W})$\;
\tcp{Solve Eigenvalue problem}
$\lambda_i, \Vec{v}_i \gets \text{EV}_i \left ( \ten{M} \right ), \quad i \in [1,..s]$\;
\tcp{Eigenvectors in ascending order of Eigenvalues}
$\lambda_i, \Vec{v}_i \gets \text{sort}_{\lambda_i}^+ \left (\lambda_i, \Vec{v}_i \right )$\;
\tcp{Select second to $d$-th Eigenvectors as embedding}
$\ten{Y} \gets [\Vec{v}_i]^T,\quad i \in [2,..,d+1]$\;
\KwOut{$\ten{Y}$}
\label{alg:LLE}
\end{algorithm}



\begin{algorithm}[h!]
\caption{Compute approximate global (least-squares) mapping between embedded and original snapshot vectors}
\tcp{Input original and reduced snapshot vectors (including zero vector)}
\KwIn{$\ten{U} \in \mathbb{R}^{D \times s}, \ten{Y} \in \mathbb{R}^{d\times s}, \Vec{y}_0 \in \mathbb{R}^d$}
\tcp{Centre reduced snapshot vectors around embedding of zero vector}
$\ten{Y} \gets \ten{Y} - \Vec{y}_0$\;
\tcp{Compute projection operator via right Moore-Penrose inverse}
$\ten{\phi} \gets \frac{1}{D} \ten{U} \ten{Y}^T ( \ten{Y} \ten{Y}^T )^{-1}$\;
\tcp{Orthonormalise}
${\ten{\phi}}_\perp,\ten{R} \gets \text{QR}(\ten{\phi})$\;
\KwOut{${\ten{\phi}}_\perp$}
\label{alg:globlin}
\end{algorithm}



\begin{algorithm}[h!]
\caption{Find nearest snapshots to a new point in reduced space}
\tcp{Input new point in reduced space, snapshot vectors in reduced and original spaces, and number of neighbours to find}
\KwIn{$\Vec{y} \in \mathbb{R}^d,\ten{Y} \in \mathbb{R}^{d\times s},\ten{U} \in \mathbb{R}^{D\times s},n$}
\tcp{Compute distances from new to previous points in reduced space (reduced snapshot vectors)}
$d_i \gets \sqrt{\sum_j (y_i-Y_{ij})^2} \quad i \in [1,..,s]$\;
\tcp{Get indices of $n$ closest points in $\ten{Y}$ to $\Vec{y}$}
$\Vec{I_N} \gets \text{n\_argmin}_i \left ( d_i \right )$\;
\tcp{Collect corresponding reduced and original coordinates}
$\ten{Y}_N \gets [\Vec{Y}_i], \quad i \in I_N$\;
$\ten{U}_N \gets [\Vec{U}_i], \quad i \in I_N$\;
\KwOut{$\ten{Y}_N,\ten{U}_N$}
\label{alg:neighbours}
\end{algorithm}


\begin{algorithm}[h!]
\caption{Use local linearisation to approximate the mapping to a local tangent to the solution manifold}
\tcp{Input current point in reduced space, reduced as well as original snapshot vectors, and number of neighbours to use for linearisation}
\KwIn{$\Vec{y},\ten{Y},\ten{U},n$}
\tcp{Find $n$ nearest neighbours of $\Vec{y}$ on graph}
$\ten{Y}_N,\ten{U}_N \gets \text{Neighbours}\left ( \Vec{y},\ten{Y},\ten{U},n \right )$\;
\tcp{Compute local linear mapping matrix with right Moore-Penrose inverse (least-squares solution to overdetermined problem, allowing for constant offset)}
$\ten{W} \gets \ten{I} - \nicefrac{1}{n} \ten{1}$\;
$\ten{\varphi}_{\text{loc}} \gets \ten{U}_N \ten{W} \ten{Y}_N^T \left ( \ten{Y} \ten{W} \ten{Y}^T \right)^{-1}$\;
\tcp{Return local mapping matrix}
\KwOut{$\ten{\varphi}_{\text{loc}}$}
\label{alg:loclin}
\end{algorithm}

\begin{algorithm}[h!]
\caption{Solve a nonlinear mechanical problem with periodic boundary conditions via the FEM, using graph-based manifold learning methods for MOR, with orthonormalised local linearisation}
\KwIn{$\ten{X}_{\text{glob}},\ten{E}_{\text{nodes}},\vec{E}_{\text{mat}},\text{ELEM},\text{MAT},\text{SHAPE},\text{param},\text{BC info}, \ten{U}, \ten{Y}$}
\tcp{Initialise global unreduced and reduced displacement fluctuation}
$\Tilde{\vec{u}} \gets \Vec{0}$, 
$\Vec{y} \gets \Vec{y}_0$\;

\tcp{Incremental time-stepping}
$t_n \gets \Delta t$
\While{$t_n<=1$}{
\tcp{Modify current load via load multiplier}
$\ten{\bar{H}}_n = \ten{\bar{H}} (t_n)$\;

\tcp{Iterative solution for nonlinear problem}
$\text{res} \gets 10^{10}$, $\text{res}_{\text{old}} \gets 10^{11}$\;
\While{$\text{res}>\text{res}_{\text{max}}$}{
\tcp{Assemble stiffness matrix and reaction vector}
$\ten{K},\vec{r} \gets \text{ASSEMBLE}$\;
\tcp{Apply BCs}
$\ten{K}_{\text{BC}},\vec{r}_{\text{BC}} \gets \text{BC}$\;
\tcp{Determine residual (trivial without external loads)}
$\Vec{g}_{\text{BC}} \gets \Vec{r}$\;
\tcp{Locally linearise graph approximation of solution manifold}
$\ten{\varphi} \gets \text{LOCLIN}(\Vec{y},\ten{Y},\ten{U},k_{NN})$\;
\tcp{Orthonormalise with reduced QR factorisation}
$\ten{\varphi}_\perp,\ten{R}_\perp \gets \text{QR}(\ten{\varphi})$\;
\tcp{Reduce stiffness matrix and residuum}
$\ten{K}_{\text{red}} \gets \ten{\varphi}_\perp^T \ten{K}_{\text{BC}} \ten{\varphi}_\perp$,
$\vec{g}_{\text{red}} \gets \ten{\varphi}_\perp^T \Vec{g}_{\text{BC}}$\;
\tcp{Solve linearly for reduced displacement fluctuation increment}
$\Delta \Vec{y}_\perp \gets \text{SOLVE} \left ( \ten{K}_{\text{red}} \Delta \vec{y} = - \vec{g}_{\text{red}} \right )$\;
$\Delta \Vec{y} \gets \ten{R}^{-1} \Delta y_\perp$\;
\tcp{Compute independent displacement fluctuation increment DOFs}
$\Delta \Tilde{\vec{u}}_{\text{BC}} \gets \ten{\varphi} \Delta \vec{y}$\;
\tcp{Sort into whole displacement fluctuation increment vector}
$\Delta \Tilde{\Vec{u}} \gets 0$, 
$\Delta \Tilde{\Vec{u}}_{I_\text{idep}} \gets \Delta \Tilde{\Vec{u}}_{\text{BC}}$\;
\tcp{Increment displacement fluctuation and reduced variables}
$\Tilde{\Vec{u}} \gets \Tilde{\Vec{u}} + \Delta \Tilde{\Vec{u}}$\,, $\vec{y} \gets \vec{y} + \Delta \vec{y}$\;

\tcp{Get residual for convergence check}
$\vec{r} \gets \text{ASSEMBLE}$\;
$\vec{r}_{\text{BC}} \gets \text{BC}$\;
$\Vec{g}_{\text{BC}} \gets \Vec{r}_{\text{BC}}$\;
$\ten{\varphi} \gets \text{LOCLIN}(\Vec{y},\ten{Y},\ten{U},k_{NN})$\;
$\ten{\varphi}_\perp,\ten{R}_\perp \gets \text{QR}(\ten{\varphi})$\;
$\vec{g}_{\text{red}} \gets \ten{\varphi}_\perp^T \Vec{g}_{\text{BC}}$\;
$\text{res} \gets \text{max}(\text{abs}(\Vec{g}_{\text{red}}))$\;
}

\tcp{Increment time}
$t_n \gets t_n + \Delta t$\;
}
\label{alg:manl_online}
\end{algorithm}


\begin{algorithm}[h!]
\caption{Solve a nonlinear mechanical problem with periodic boundary conditions via the FEM, using graph-based manifold learning methods for MOR, with orthonormalised local linearisation and intermediate POD reduction}
\KwIn{$\ten{X}_{\text{glob}},\ten{E}_{\text{nodes}},\vec{E}_{\text{mat}},\text{ELEM},\text{MAT},\text{SHAPE},\text{param},\text{BC info}, \ten{U}, \ten{Y}, \ten{\bar{Y}}, \ten{\psi}$}
\tcp{Initialise global unreduced and reduced displacement fluctuation}
$\Tilde{\vec{u}} \gets \Vec{0}$, 
$\Vec{y} \gets \Vec{y}_0$\;

\tcp{Incremental time-stepping}
$t_n \gets \Delta t$
\While{$t_n<=1$}{
\tcp{Modify current load via load multiplier}
$\ten{\bar{H}}_n = \ten{\bar{H}} (t_n)$\;

\tcp{Iterative solution for nonlinear problem}
$\text{res} \gets 10^{10}$, $\text{res}_{\text{old}} \gets 10^{11}$\;
\While{$\text{res}>\text{res}_{\text{max}}$}{
\tcp{Assemble stiffness matrix and reaction vector}
$\ten{K},\vec{r} \gets \text{ASSEMBLE}$\;
\tcp{Apply BCs}
$\ten{K}_{\text{BC}},\vec{r}_{\text{BC}} \gets \text{BC}$\;
\tcp{Determine residual (trivial without external loads)}
$\Vec{g}_{\text{BC}} \gets \Vec{r}$\;
\tcp{Locally linearise graph approximation of solution manifold}
$\ten{\varphi} \gets \text{LOCLIN}(\Vec{y},\ten{Y},\ten{\bar{Y}},k_{NN})$\;
\tcp{Orthonormalise with reduced QR factorisation}
$\ten{\varphi}_\perp,\ten{R}_\perp \gets \text{QR}(\ten{\varphi})$\;
\tcp{Reduce stiffness matrix and residuum}
$\ten{K}_{\text{red}} \gets \ten{\varphi}_\perp^T \ten{\psi}^T \ten{K}_{\text{BC}} \ten{\psi} \ten{\varphi}_\perp$,
$\vec{g}_{\text{red}} \gets \ten{\varphi}_\perp^T \ten{\psi}^T\Vec{g}_{\text{BC}}$\;
\tcp{Solve linearly for reduced displacement fluctuation increment}
$\Delta \Vec{y}_\perp \gets \text{SOLVE} \left ( \ten{K}_{\text{red}} \Delta \vec{y} = - \vec{g}_{\text{red}} \right )$\;
$\Delta \Vec{y} \gets \ten{R}^{-1} \Delta y_\perp$\;
\tcp{Compute independent displacement fluctuation increment DOFs}
$\Delta \Tilde{\vec{u}}_{\text{BC}} \gets \ten{\psi} \ten{\varphi} \Delta \vec{y}$\;
\tcp{Sort into whole displacement fluctuation increment vector}
$\Delta \Tilde{\Vec{u}} \gets 0$, 
$\Delta \Tilde{\Vec{u}}_{I_\text{idep}} \gets \Delta \Tilde{\Vec{u}}_{\text{BC}}$\;
\tcp{Increment displacement fluctuation and reduced variables}
$\Tilde{\Vec{u}} \gets \Tilde{\Vec{u}} + \Delta \Tilde{\Vec{u}}$\,, $\vec{y} \gets \vec{y} + \Delta \vec{y}$\;

\tcp{Get residual for convergence check}
$\vec{r} \gets \text{ASSEMBLE}$\;
$\vec{r}_{\text{BC}} \gets \text{BC}$\;
$\Vec{g}_{\text{BC}} \gets \Vec{r}_{\text{BC}}$\;
$\ten{\varphi} \gets \text{LOCLIN}(\Vec{y},\ten{Y},\ten{U},k_{NN})$\;
$\ten{\varphi}_\perp,\ten{R}_\perp \gets \text{QR}(\ten{\varphi})$\;
$\vec{g}_{\text{red}} \gets \ten{\varphi}_\perp^T \ten{\psi}^T\Vec{g}_{\text{BC}}$\;
$\text{res} \gets \text{max}(\text{abs}(\Vec{g}_{\text{red}}))$\;
}

\tcp{Increment time}
$t_n \gets t_n + \Delta t$\;
}
\end{algorithm}

\end{document}